\documentclass[11pt]{article}
\pdfoutput=1
\usepackage{graphicx,epsfig}
\usepackage{color}
\usepackage[colorlinks,citecolor=red,linkcolor=blue]{hyperref}
\usepackage{slashed}
\usepackage[left=2.5cm, right=2.5cm, top=2cm]{geometry}

\usepackage{epsfig,float,amsmath,amsfonts,amssymb,graphicx,bm,bbm,array,dcolumn}
\usepackage{subfigure}

\newcommand{\Veff}{V_{\rm eff}}
\newcommand{\Ptunl}{P_{\rm tunl}}
\newcommand{\rhoh}{\hat\rho}
\newcommand{\hh}{\hat h}
\newcommand{\Veffh}{\hat{V}_{\rm eff}}
\newcommand{\half}{\frac{1}{2}}

\def\beq{\begin{equation}}
\def\eeq{\end{equation}}
\def\bea{\begin{eqnarray}}
\def\eea{\end{eqnarray}}
\def\bmat{\begin{pmatrix}}
\def\emat{\end{pmatrix}}
\newcommand{\pslash}{\mbox{$\not \hspace{-0.08cm} p$ }}
\newcommand{ \slashchar }[1]{\setbox0=\hbox{$#1$}   
   \dimen0=\wd0                                     
   \setbox1=\hbox{/} \dimen1=\wd1                   
   \ifdim\dimen0>\dimen1                            
      \rlap{\hbox to \dimen0{\hfil/\hfil}}          
      #1                                            
   \else                                            
      \rlap{\hbox to \dimen1{\hfil$#1$\hfil}}       
      /                                             
   \fi}                                             %

\def\to{\rightarrow}

\begin{document}

\title{
\normalsize{\bf Higgs Vacuum Stability with Vector-like Fermions}}

\author{\normalsize{Shrihari Gopalakrishna$^{a,b}$\thanks{shri@imsc.res.in}\ , Arunprasath Velusamy$^a$\thanks{arunprasath@imsc.res.in}}
\\
$^a$~\small{Institute of Mathematical Sciences (IMSc), Chennai 600 113, India.}\\
$^b$~\small{Homi Bhabha National Institute (HBNI), Anushaktinagar, Mumbai 400 094, India.}
}

\maketitle

\begin{abstract}
  
  We present the effects of vector-like fermions (VLF) on the stability of the Higgs electroweak vacuum, 
  using the renormalization group improved Higgs effective potential.
  We review the calculation of the one-loop beta-functions of the standard model couplings paying particular attention
  to the fermion contributions. 
  From this, we derive the VLF contributions to the beta-functions.
  We also include the significant two-loop contributions to the beta-functions.  
  Using these beta-functions, we determine the scale at which
  the effective Higgs quartic-coupling becomes zero and goes negative, signaling vacuum instability. 
  We find that for certain VLF masses and Yukawa couplings, the Higgs quartic stays positive for field values
  all the way up to the Planck scale,
  implying that the meta-stable vacuum of the standard model can be rendered absolutely stable if VLFs are present
  with certain parameters.
  For other values of VLF parameters, the Higgs vacuum is metastable as in the standard model. 
  For cases where the vacuum is metastable,
  we compute the probability of quantum tunneling from the false electroweak vacuum
  into a deeper true vacuum in our Hubble volume
  by numerically solving for the bounce configuration in Euclidean space-time and computing the bounce action for it.
  We compare our numerical solution with the analytical approximation for the bounce action
  commonly used in the literature, and comment on when the latter may be used. 
  
\end{abstract}

\vfill\eject


\vfill\eject

\section{Introduction}

The stability of the electroweak (EW) vacuum can be studied using the Higgs effective potential
(for a review see Ref.~\cite{Sher:1988mj}). 
Recent investigations (see for example Refs.~\cite{Bezrukov:2012sa,Degrassi:2012ry,Buttazzo:2013uya,Bezrukov:2014ina})
at the NNLO level have revealed that the Higgs vacuum is metastable in the standard model~(SM),
with the life-time in the false (EW) vacuum being much much larger than the age of the Universe.
This situation arises because the Higgs quantum effective potential
$V_{\rm eff}(h)$ has a smaller value for $h \sim 10^{10}~$GeV when compared to
its value at the EW vacuum expectation value (VEV) $v\approx 246$~GeV, i.e.,
$V_{\rm eff}(h\! \sim\! 10^{10}~GeV) < V_{\rm eff}(v)$ for the SM.

There are many compelling reasons to expect physics beyond the standard model (BSM). 
These include theoretical reasons such as the gauge hierarchy problem,
and observational reasons, such as neutrino mass generation, dark matter,
and generation of the baryon asymmetry of the Universe.
A plethora of BSM extensions have been proposed to address these shortcomings of the SM.
These inevitably add new particles to the SM particle content.
In particular, resolution of the gauge hierarchy problem necessarily have new states coupled to the Higgs.
In such cases, the above conclusions on Higgs EW vacuum stability must be revisited by including the effects of such new
particles.
Many such BSM extensions include vector-like fermions (VLF) that couple to the Higgs, 
and are often the lightest BSM states.
They therefore have a significant effect on EW vacuum stability.
Some examples of such models that include vector-like fermions are in the following contexts:
the gauge hierarchy problem such as AdS-space/composite-Higgs models
in Refs.~\cite{Agashe:2003zs,Contino:2006qr,Gopalakrishna:2013hua,Zheng:1996hq},
Higgs-portal dark matter models 
in Refs.~\cite{Patt:2006fw,Gopalakrishna:2009yz,Baek:2011aa,Aravind:2015xst,Lu:2017uur},
gauge-coupling unification
in Refs.~\cite{Dermisek:2012ke,Bhattacherjee:2017cxh,EmmanuelCosta:2005nh,Barger:2006fm,Dorsner:2014wva},
neutrino mass generation and vacuum stability
in Refs.~\cite{Chao:2012xt,Masina:2012tz,Kobakhidze:2013pya,Mohapatra:2014qva,Rose:2015fua,Goswami:2018jar},
universal extra dimension model in Ref.~\cite{Datta:2012db},
SM extensions with an additional $U(1)$ gauge symmetry in Refs.~\cite{Anchordoqui:2012fq,Coriano:2014mpa,Coriano:2015sea,Accomando:2016sge}, 
models with extended scalar sector in Refs.~\cite{EliasMiro:2012ay,Lebedev:2012zw},
a combination of these
in Ref.~\cite{Joglekar:2012vc,Hamada:2014xka,Haba:2015nwl}, models of inflation in Refs.~\cite{Kobakhidze:2013tn,Kobakhidze:2014xda},
and effective models
in Refs.~\cite{Chao:2012mx,Altmannshofer:2013zba,Xiao:2014kba}.
Motivated by these considerations, we study the effect of VLFs that are coupled to the Higgs on
EW vacuum stability.\footnote{
  New chiral (4th generation) fermions that get their mass from the Higgs
  are severely disfavored with a single Higgs doublet
  from the recent LHC Higgs cross-section and couplings measurements.
  In contrast, VLFs tend to have milder constraints on them owing to their nice decoupling property.
}
Many of these models may also contain new bosonic states apart from VLFs. 
In such a case, a full conclusion about the stability of the EW vacuum in that model can be reached only after
including the contributions of these bosonic states also.
However, fermions usually have the biggest role in destabilizing the EW vacuum,
and so our analysis here addresses the most crucial ingredient in this problem.
Hence, our goal here is to analyze model-independently the generic effects of VLFs on EW vacuum stability.

We set the stage for our analysis by writing the classical Higgs potential as
\beq
{\cal V} = \frac{m^2_{h}}{2} h^2 + \frac{\lambda}{4} h^4 \ .   
\eeq
Including quantum effects, we can write the quantum Higgs effective potential as 
\beq
    V_{\rm eff}(h) = \frac{m^2_{h\, {\rm eff}}}{2} h^2 + \frac{\lambda_{\rm eff}(h)}{4} h^4 \ ,
    \label{VeffhDefn.EQ}
\eeq
where $\lambda_{\rm eff}$ has a dependence on $h$ of the form $\ln{(h/M)}$ with $M$ being a subtraction scale.
For $h \gg m_h$ (the physical Higgs mass $m_h \approx 125$~GeV), the mass term has a negligible effect,
and thus, to an excellent approximation, we can write
\beq
    \Veff(h) = \frac{\lambda_{\rm eff}(h)}{4} h^4 \ .
    \label{Veffh4.EQ}
\eeq
Denoting the field value as $h\equiv \mu$ and denoting $\lambda_{\rm eff}(\mu)$ as just $\lambda(\mu)$,
it can be shown (see for example Ref.~\cite{Weinberg:1996kr})
that $\lambda(\mu)$ obeys a renormalization group equation (RGE) of the form
\beq
\frac{d\, \lambda(\mu)}{d\ln{\mu}} = \beta_\lambda\left(\lambda(\mu), y_t(\mu), g_3(\mu), g_2(\mu), g_1(\mu), ... \right) \ . 
\eeq
The RGE is interpreted now as an evolution with field value $h=\mu$,
and the $\beta$-function $\beta_\lambda$ is the usual $\beta$-function for the coupling $\lambda$, governed by the RGE.
The $\lambda(\mu)$ obtained by integrating the RGE has the leading logs of the form $\log^n\!{(\mu/M)}$ resummed.  
$\beta_\lambda$ is shown as a function of $\lambda$ itself, and also of the other couplings
that contribute significantly, 
which, in the SM, are the top Yukawa coupling $y_t$ and the $SU(3), SU(2), U(1)$ gauge couplings
$g_a=\{g_3, g_2, g_1 \}$.
All these couplings also evolve with $\mu$ via analogous RGE equations with their corresponding
$\beta$-functions $\beta_{y_t}, \beta_{g_a}$. 
We neglect the contributions of the other SM couplings to the $\beta$-functions as they contribute insignificantly.
From Eq.~(\ref{Veffh4.EQ}) we see that for $h \gg m_h$,
the instability is signalled by the Higgs quartic effective coupling $\lambda(\mu)$ becoming negative.

As we show explicitly later, $\beta_\lambda$ obtains a negative contribution from $y_t$,
while it obtains a positive contribution from $\lambda(\mu)$ itself and from gauge couplings.
Thus, the top quark has the important effect of decreasing $\lambda(\mu)$,
and for $y_t$ as large as in the SM, for the observed $m_h$, it drives $\lambda(\mu)$ negative at higher energies,
signaling vacuum instability.
The effect of fermions coupled to the Higgs is generally to destabilize the electroweak vacuum,
although in this work we show that this statement is not so definite. 
Many extensions beyond the SM (BSM) include new fermions, and the question we address in this work is 
what the effects of new fermions might be on Higgs vacuum stability in light of the observation made above. 

The subject of this work is to include VLF contributions to the $\beta$-functions
and find the consequences for EW vacuum stability and how it is changed from the SM.
We ask if $\lambda(\mu)$ still becomes zero with VLF present, and if so at what $\mu$,
and compare it with the SM case.
If the vacuum is unstable, we compute the tunneling probability to ascertain if it decays within the age of the Universe,
in which case it is unacceptable.
On the contrary, if the lifetime in the EW vacuum is comparatively much larger than the age of the Universe,
it is metastable and phenomenologically acceptable.
To this end, we study some simple VLF extensions of the SM,
where the VLFs are either in the trivial or fundamental representations of $SU(3), SU(2)$ and $U(1)$,
and demonstrate their effects on Higgs vacuum stability.
In particular, the VLFs we add are of two kinds, namely, $SU(3)$ triplet vector-like quarks (VLQ) and
$SU(3)$ singlet vector-like leptons (VLL).

The paper is organized as follows:
In Sec.~\ref{RGE.SEC}  
we list the 1-loop RGE in the SM and include some significant 2-loop corrections from the literature.
We present a derivation of the fermionic contributions to the RGE in Appendix~\ref{betaFcnSM.SEC}. 
We then derive the 1-loop VLF contributions to the RGE and add these to the SM RGE.
We present the calculational details of the VLF contributions in Appendix~\ref{betaFcnVLF.SEC}.
We add the significant two-loop contributions to the beta-functions. 
We integrate the RGE numerically and show the evolution of the couplings as a function of the field value $h\equiv \mu$. 
In Sec.~\ref{TUNL.SEC} we compute the probability that our electroweak vacuum would have tunneled into
a deeper true vacuum in our Hubble volume in the case where the EW vacuum is metastable.
We do so by solving for the bounce configuration numerically and computing the Euclidean action for this. 
In Sec.~\ref{AbsltStab.SEC} we make some remarks for the case when the VLFs render the EW vacuum absolutely stable.
In Sec.~\ref{compSB.SEC} we compare our numerical evaluation of the bounce action to an approximation commonly used in
the literature, and provide a cautionary note on when the approximation can be applied. 
We offer our conclusions in Sec.~\ref{Concl.SEC}.

\section{Renormalization group improved Higgs effective potential}
\label{RGE.SEC}

We have in the SM the Lagrangian density, showing only the terms relevant to our analysis here, 
\beq
{\cal L} \supset \bar{t} i \slashchar{\partial} t - \lambda (H^\dagger H)^2  - (y_t \bar{q}_L \cdot H^* t_R + h.c.) \ ,  
\label{LSMth.EQ}
\eeq
where the $\cdot$ represents the antisymmetric combination in $SU(2)$ space,
$q_L = (t_L\ b_L)^T$ is the $SU(2)$ doublet, 
and, $t=(t_L\ t_R)^T$ and $b=(b_L\ b_R)^T$ are the top-quark and bottom-quark Dirac fermions.
It is sufficient for our purposes to keep only the top Yukawa coupling $y_t$ in the SM as the others are much suppressed.
Next, we present the SM $\beta$-functions and extend them to include the VLF contributions.

\subsection{SM RGE}
\label{SMRGE.SEC}

We first discuss the SM RGE $\beta$-functions at the 1-loop level and include some significant 2-loop effects.
We use the SM RGE to find the Higgs field value at which
$\lambda(\mu)$ becomes zero and compare our results with those in the literature.
Denoting the relevant SM couplings generically as $\kappa_i = \{\lambda, y_t, g_3, g_2, g_1 \}$,
the RGE are of the form  
\beq
\frac{d\, \kappa_i(\mu)}{d\ln{\mu}} = \beta_{\kappa_i}(\kappa_j(\mu)) \ .  
\eeq
We derive the fermion contributions to $\beta_\kappa$ in Appendix~\ref{betaFcnSM.SEC}
since our goal in this work is to extend them to include VLF contributions.
We take the other terms from the literature (see for example Ref.~\cite{Buttazzo:2013uya}).
Putting these together, the 1-loop $\beta$-functions, $\beta^{(1)}_\kappa$, are
\bea
\beta^{(1)}_\lambda &=& \frac{1}{16 \pi^2} \left[ 24 \lambda^2 + 4 N_c y_t^2 \lambda - 2 N_c y_t^4 - 9 g_2^2 \lambda - \frac{9}{5} g_1^2 \lambda
  + \frac{9}{8} \left( g_2^4 + \frac{2}{5} g_2^2 g_1^2 + \frac{3}{25} g_1^4 \right)  \right] \ , \label{RGEbeta1lamSM.EQ}  \\
\beta^{(1)}_{y_t} &=& \frac{y_t}{16 \pi^2} \left[ \frac{(3+2 N_c)}{2} y_t^2 - 8 g_3^2 - \frac{9}{4} g_2^2 - \frac{17}{20} g_1^2  \right] \ , \label{RGEbeta1ygaSM.EQ} \\
\beta^{(1)}_{g_a} &=& \frac{g_a^3 b_a}{16 \pi^2} \ , \label{RGEbeta1gaSM.EQ}
\eea
with $b_a = (-7, -19/6, 41/10 )$ for $g_a = (g_3, g_2, g_1)$ respectively, and $N_c = 3$ for a fermion in the fundamental representation of $SU(3)$.
For $g_1$, we use the $SU(5)$ normalization,
i.e. the SM hypercharge gauge-coupling $g'$ is related to $g_1$ by $g_1 = \sqrt{5/3} g'$. 

The precision of the full 2-loop (or higher order) calculations that are available in the literature
are not required for our purposes since our goal is to analyze BSM physics contributions that involve
as yet experimentally undetermined parameters.
However, to help compare our numerical results to what has been obtained in the literature for the SM,
we will include 2-loop SM contributions to the $\beta$-functions that depend on
$y_t$ and $g_3$ as they are numerically the most significant. 
They are (see for example Ref.~\cite{Buttazzo:2013uya})
\bea
\beta^{(2)}_\lambda &=& \frac{y_t^2}{(16 \pi^2)^2} (30 y_t^4 - 32 g_3^2 y_t^2 + 80 \lambda g_3^2 + ...) \ , \\
\beta^{(2)}_{y_t} &=& \frac{y_t}{(16 \pi^2)^2} \left[ \left(-\frac{404}{3} + \frac{40}{9} n_3^{\rm (SM)} \right) g_3^4 + 36 y_t^2 g_3^2 - 12 y_t^4 + ... \right] \ , \\
\beta^{(2)}_{g_3} &=& \frac{g_3^3}{(16 \pi^2)^2} \left[ \left(-86 + 10 n_3^{\rm(SM)} \right) g_3^2 - 2 y_t^2 + ... \right]  \ ,
\label{SM2loopDom.EQ}
\eea
where $n_3^{\rm (SM)} = 6$ is the number of $SU(3)$-triplets (i.e. quarks) in the SM. 

We use these RGE to determine the Higgs field value $\mu$ at which $\lambda(\mu)$ becomes zero, signalling vacuum instability.
This will be discussed in Sec.~\ref{RGERes.SEC}.
We discuss next the VLF contributions to the RGE. 

\subsection{VLF contributions to the RGE}
\label{VLFRGE.SEC}

We add an $SU(2)$ doublet VLF $\chi = (\chi_1 \ \chi_2)^T$ and an $SU(2)$ singlet VLF $\xi$ and couple it to the Higgs as follows
\beq
    {\cal L} \supset -M_\chi \bar{\chi} \chi - M_\xi \bar{\xi} \xi - (\tilde{y}\, \bar{\chi} \cdot H^* \xi + h.c.) \ ,
    \label{LagrVLF.EQ}
\eeq
where the $\cdot$ represents the antisymmetric combination in $SU(2)$ space.\footnote{If another $SU(2)$ singlet VLF $\zeta$ is added, we can add the terms
${\cal L} \supset - M_\zeta \bar\zeta \zeta - (\tilde{y}_2 \bar\chi H \zeta + h.c.)$.   
After adding the $\zeta$, the one doublet and two singlet VLF structure then mimics the SM quark or lepton structure in a generation.
For keeping the field content minimal, we will omit the $\zeta$ in our work here, and therefore will not include the $\tilde{y}_2$ term. 
}
Extracting the Higgs interactions from this yields
\beq
{\cal L} \supset -\frac{\tilde{y}}{\sqrt{2}} h ( \bar{\chi}_1 \xi + \bar\xi \chi_1  ) \ . 
\eeq
If $\chi$ and $\xi$ have color $N_c'=3$ we call them vector-like quarks (VLQ) and if they are trivial under $SU(3)$, i.e. $N_c' = 1$, we call them
vector-like leptons (VLL).
The $SU(3), SU(2)$ and $U(1)$ gauge interactions are standard and we do not show them explicitly.
We denote the hypercharge of $\chi$ as $Y_\chi$, that of $\xi$ as $Y_\xi$,
and that of the Higgs doublet is $Y_H = 1/2$ as in the SM.
If $N_c' = 3$ the VLQ have gluon interactions, while if $N_c' = 1$ the VLL do not have gluon interactions.

For SM-like choices of $Y_\chi$ and $Y_\xi$, mixed Yukawa couplings between the VLF and the standard model fermions (SMF) can be written down.
However, collider, flavor changing neutral current and other precision constraints restrict how large such couplings can be
(for details, see for example Ref.~\cite{Ellis:2014dza}).
For simplicity, in this work, we do not turn-on such mixed Yukawa couplings; 
an analysis including such mixed couplings will be the subject of future work. 

We derive the 1-loop RGE contributions due to the VLF (see Appendix~\ref{betaFcnVLF.SEC} for the derivation)
and add them to the SM contributions
given above. The VLF contributions to the RGE in Eqs.~(\ref{RGEbeta1lamSM.EQ})-(\ref{RGEbeta1gaSM.EQ}),
and the SM and VLF contributions to the RGE for the new coupling $\tilde{y}$, are
\bea
\beta^{(1)\, {\rm VLF}}_{g_3} &=& \frac{g_3^3}{16 \pi^2} \left( \frac{2}{3} n_3 \right)  \ , \label{betag3VLF.EQ} \\
\beta^{(1)\, {\rm VLF}}_{g_2} &=& \frac{g_2^3}{16 \pi^2} \left( \frac{2}{3} N_c' n_2 \right)  \ , \label{betag2VLF.EQ} \\
\beta^{(1)\, {\rm VLF}}_{g_1} &=& \frac{g_1^3}{16 \pi^2} \left[ \frac{4}{5} N_c' \left( 2 n_2 Y_\chi^2 + n_1 Y_\xi^2 \right) \right]  \ , \label{betag1VLF.EQ} \\
\beta^{(1)\, {\rm VLF}}_\lambda &=& \frac{2 n_F}{16 \pi^2} \left( 4 N_c' \tilde{y}^2 \lambda - 2 N_c' \tilde{y}^4 \right) \ , \label{betalamVLF.EQ} \\
\beta^{(1)\, {\rm VLF}}_{y_t} &=& \frac{n_F}{16 \pi^2} y_t \left( 2 N_c' \tilde{y}^2 \right) \ , \label{betaytVLF.EQ} \\
\beta^{(1)}_{\tilde{y}} &=& \frac{\tilde{y}}{16 \pi^2} \left[ \frac{(3 \tilde{y}^2 +  2 N_c y_t^2 + 4 n_F N_c' \tilde{y}^2)}{2}
  - 8 \hat{n}_F^{VLQ} g_3^2 - \frac{9}{4} g_2^2 - \frac{9}{5} g_1^2 \left( Y_H^2 + 2 Y_\chi Y_\xi \right) \right] \ , \label{betatwyVLF.EQ}
\eea
where  
$n_3$ is the number of colored VLF $SU(3)$ triplets, i.e. VLQs, $n_2$ is the number of $SU(2)$ doublets,
$n_1$ is the number of $SU(2)$ singlets,
$n_F$ is the number of complete VLF families coupled to the Higgs
(a family is a doublet and a singlet {\em both} present), and
$\hat{n}_F^{VLQ} =  1$ if the VLF is a VLQ family or zero otherwise.   
For example, $n_F = 0$ for either VLF singlets or doublets added (but not both),
and $n_F = 1$ for one $SU(2)$ doublet and one singlet VLF added together such that a Yukawa coupling $\tilde y$ can be written down with the Higgs. 
Only VLQs contribute to $\beta_{g_3}$ and VLLs do not. 
For instance, for one VLQ family of $\chi$ and $\xi$, we have $N_c' = 3$, $n_3 = 3$, $n_2 = 1$, $n_1 = 1$, $n_F = 1$
and $\hat{n}_F^{VLQ} =  1$.

To improve precision, we include the dominant 2-loop VLF contributions to the $\beta$-functions obtained from the package `SARAH'~\cite{Staub:2008uz,Staub:2013tta},
which are, 
\bea
\beta^{(2)\, {\rm VLF}}_{g_3} &=& \frac{g_3^3}{(16 \pi^2)^2} \left( 10\, n_3 g_3^2 - 2\times 2 \hat{n}^{VLQ}_F \tilde{y}^2 + ... \right)  \ , \\
\beta^{(2)\, {\rm VLF}}_\lambda &=&  \frac{\tilde{y}^2 n_F }{(16 \pi^2)^2} \left( 2\times 10 N_c^\prime {\tilde y}^4 - 2 \times 32 \hat{n}^{VLQ}_F g_3^2 \tilde{y}^2 + 2 \times 80 \hat{n}^{VLQ}_F \lambda g_3^2 + ...\right) \ , \\
\beta^{(2)\, {\rm VLF}}_{y_t} &=& \frac{y_t}{(16 \pi^2)^2} \left( \frac{40}{9} n_3 g_3^4 - \frac{9}{2} n_F N_c' \tilde{y}^4 - \frac{9}{2} n_F N_c' \tilde{y}^2 y_t^2
  +  40 n^{VLQ}_F g_3^2 \tilde{y}^2 + ... \right) \ , \\
  \beta^{(2)}_{\tilde{y}} &=& \frac{\tilde{y}}{(16 \pi^2)^2} \left[ -\left( 9 N_c' - \frac{3}{2}  \right) \tilde{y}^4 + \hat{n}^{VLQ}_F \left(-\frac{2\times 485}{9} + \frac{40}{9} \left(n_3^{\rm (SM)} + n_3 \right) \right) g_3^4  \nonumber \right.  \\
    & &  \left. + 56 \hat{n}^{VLQ}_F \tilde{y}^2 g_3^2 + 20 g_3^2 y_t^2 -\frac{27}{4} \tilde{y}^2 y_t^2 - \frac{27}{4} y_t^4 + ... \right] \ , 
\label{VLF2loopDom.EQ}
\eea
where, $n^{VLQ}_F$ is the number of colored families, and as noted earlier, $n_3^{\rm (SM)} = 6$.
We have explicitly checked that the above dominant contributions closely reproduce numerically the full 2-loop running from SARAH.

Our goal in this work is to analyze the stability of the electroweak (EW) vacuum for which the behavior of $\Veff$ at large field values is most important.
We have therefore not kept the finite mass effects in the RGE as they are small,
being of the form $(m/\mu)$ for $\mu \gg m$ where $m$ collectively denotes the particle masses.
We include the VLF contributions only for $\mu \geq M_{VL}$, where $M_{VL}$ is the vector-like fermion mass.

\subsection{RGE Numerical integration results}
\label{RGERes.SEC}

We take the input parameters as follows and as compiled in Ref.~\cite{Buttazzo:2013uya},
with the renormalization point taken as the top mass scale $\tilde m_t$:
\begin{itemize}
\item[] The EW VEV: $v=246.2~$GeV, 
\item[] The Higgs quartic: $\tilde \lambda = 0.12710$ (NNLO), 
\item[] The top Yukawa coupling: $\tilde{y}_t = 0.93558$ (partial 3-loop), 
\item[] The $SU(3)_c$ coupling constant: $\tilde{g}_3 = 1.1666$ (partial 4-loop), 
\item[] The $SU(2)_L$ coupling constant: $\tilde{g}_2 = 0.64755$ (NLO), 
\item[] The $U(1)$ coupling constant: $\tilde{g}_1 = \sqrt{5/3}\, g' = \sqrt{5/3}\times 0.35937$ (NLO). 
\end{itemize}
In terms of these inputs, we set the top mass to be $\tilde{m}_t = \tilde{y}_t v/\sqrt{2}$ and the Higgs mass $\tilde{m}_h = \sqrt{2 \tilde\lambda}\, v$.
In this work, since our interest is in analyzing a new physics (VLF) model with unknown parameters, the full precision to which these are defined are not so important, 
and the above specification is more than adequate for our purposes.

The RGE are a coupled set of first order differential equations for the couplings
$\lambda(\mu)$, $y_t(\mu)$, $g_3(\mu)$, $g_2(\mu)$, $g_1(\mu)$. 
We take the inputs given above at $\tilde{m}_t$ and integrate the RGE numerically,
including both the SM contributions in Section~\ref{SMRGE.SEC} and the VLF contributions in Section~\ref{VLFRGE.SEC}.
As already mentioned, we include the VLF contribution only for $\mu \geq M_{VL}$. 

\begin{figure}
  \begin{center}
    \includegraphics[width=0.4\textwidth] {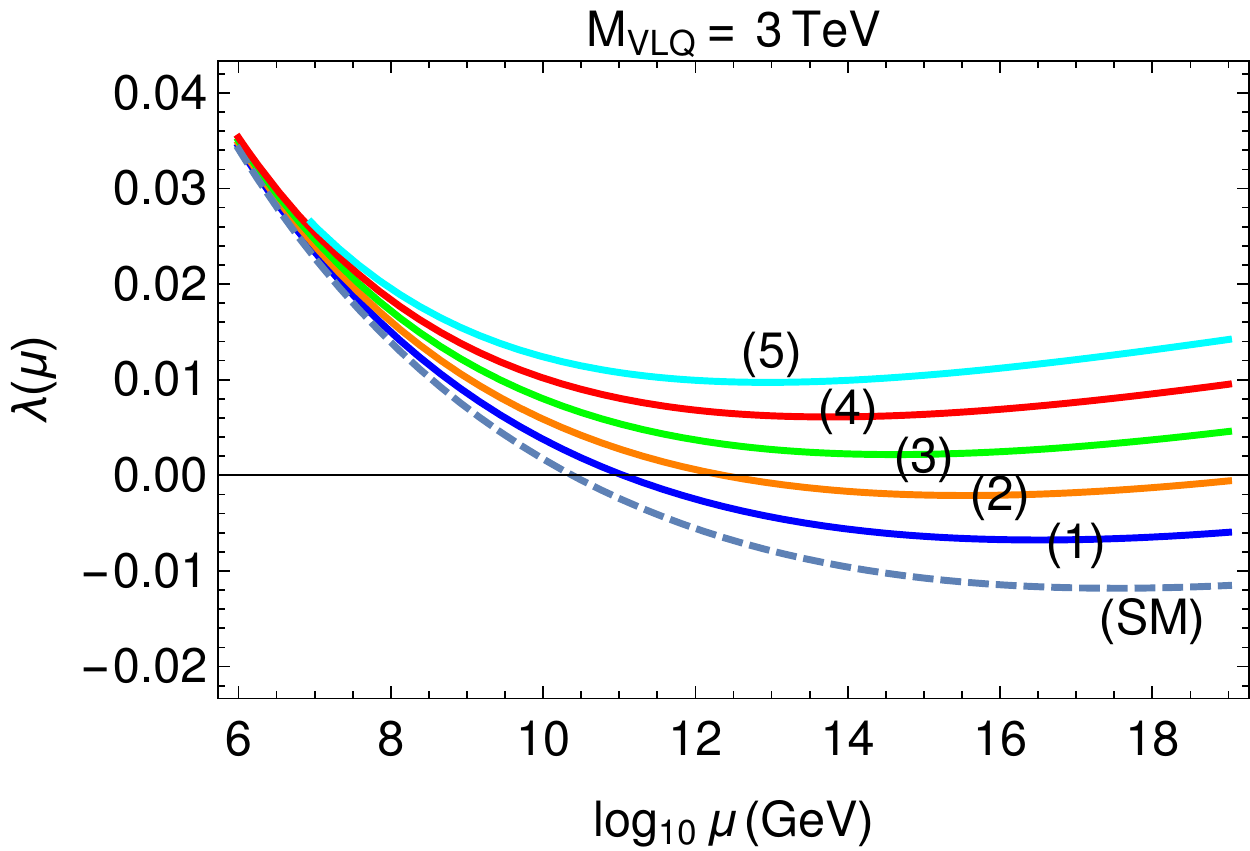} \hspace*{0.25cm}
    \includegraphics[width=0.4\textwidth] {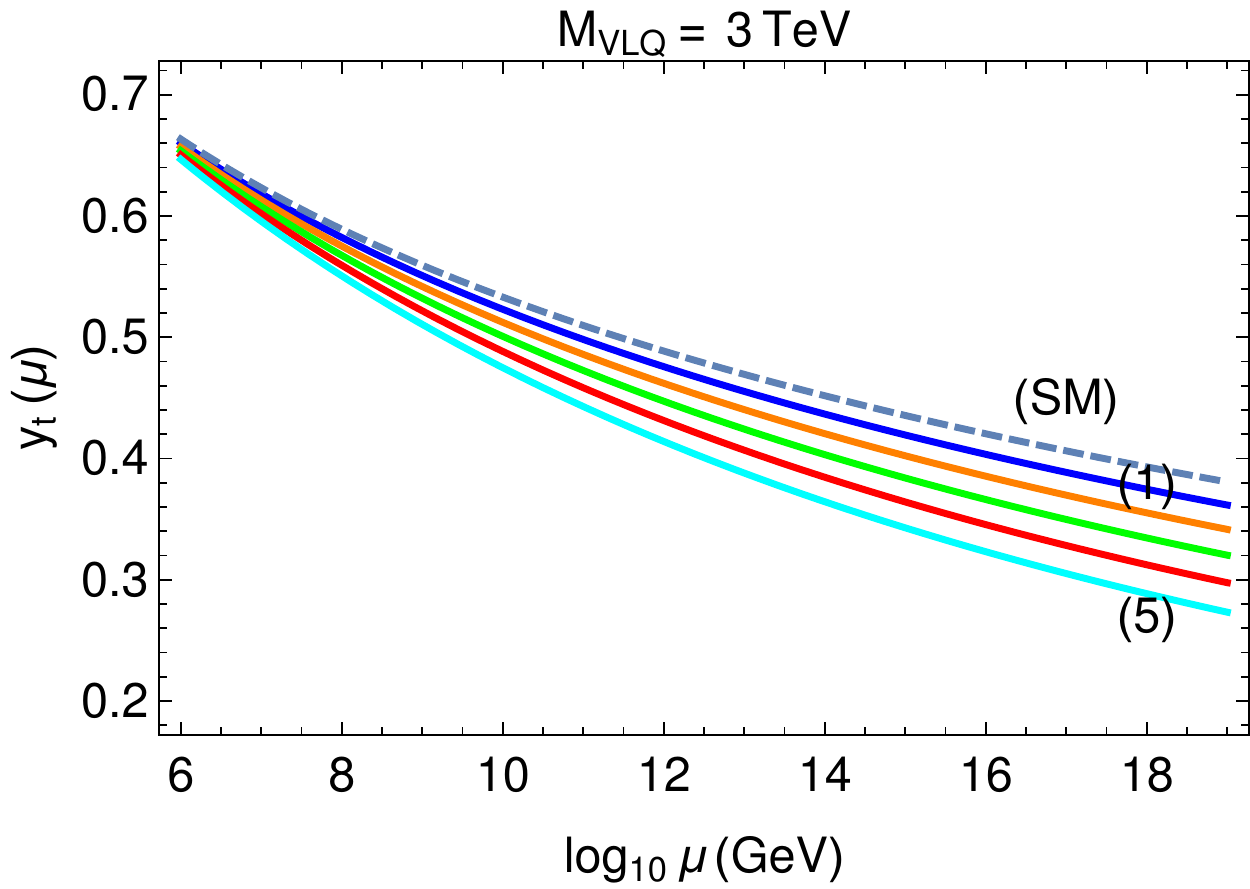} \\
    \vspace*{0.25cm}
    \includegraphics[width=0.4\textwidth] {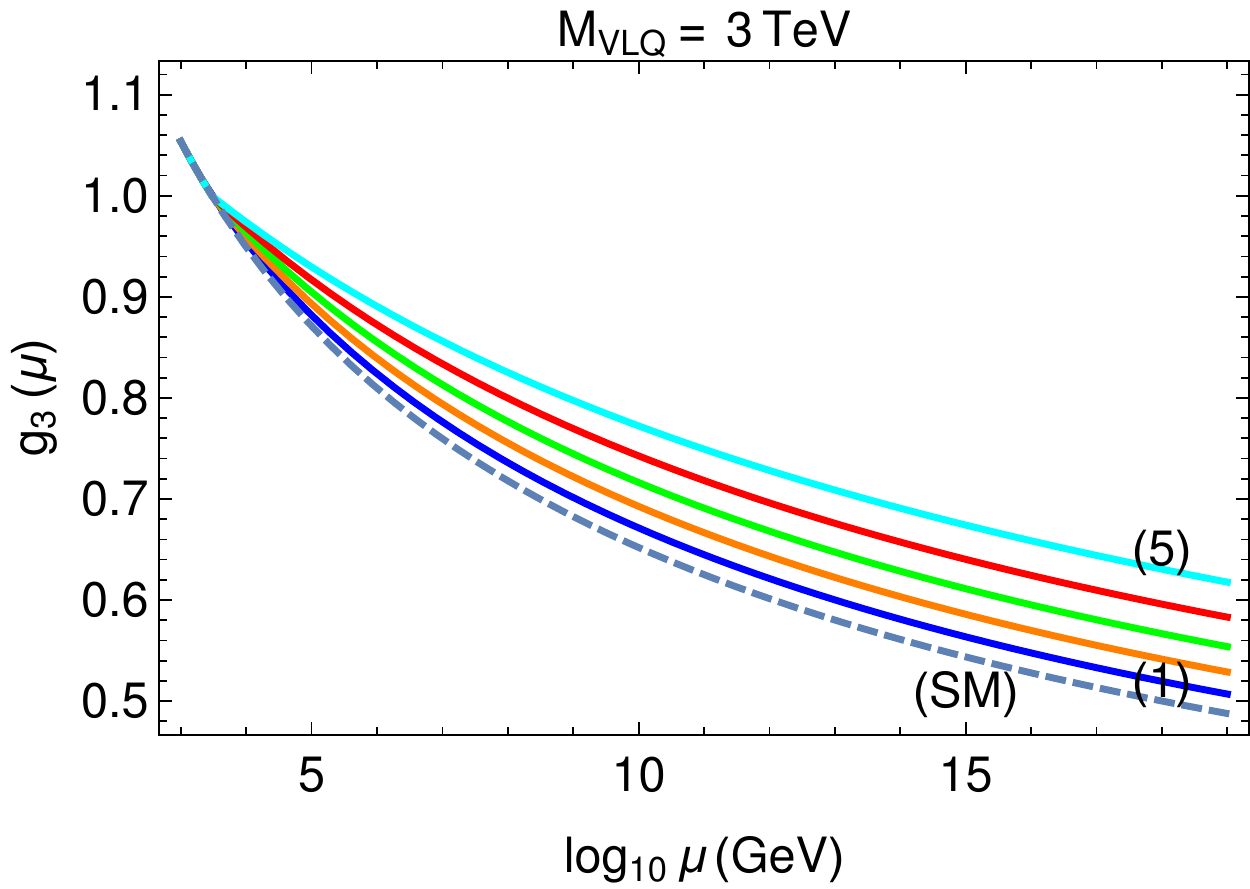} \hspace*{0.25cm}
    \includegraphics[width=0.4\textwidth] {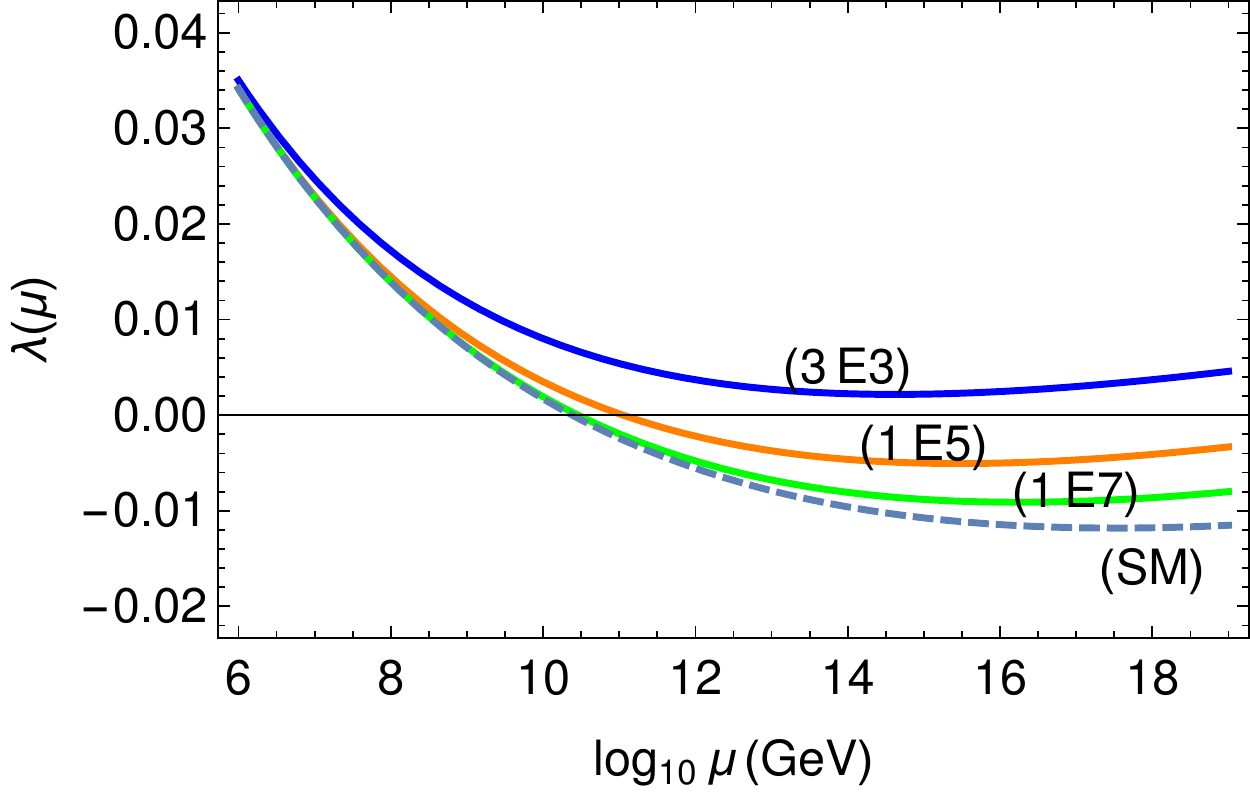}
    \caption{The evolution of $\lambda$, $y_t$, $g_3$ with Higgs field value $\mu$ 
      for $(n)$ number of degenerate $SU(2)$ singlet VLQs of mass 3~TeV (first three plots),
      and $\lambda$ with 3 degenerate singlet VLQs of mass $3\times 10^3$~GeV, $10^5$~GeV and $10^7$~GeV
      shown respectively as $3E3$, $1E5$ and $1E7$ (last plot).  
      \label{VLQsing.FIG}}
  \end{center}
\end{figure}

\begin{figure}
  \begin{center}
    \includegraphics[width=0.4\textwidth] {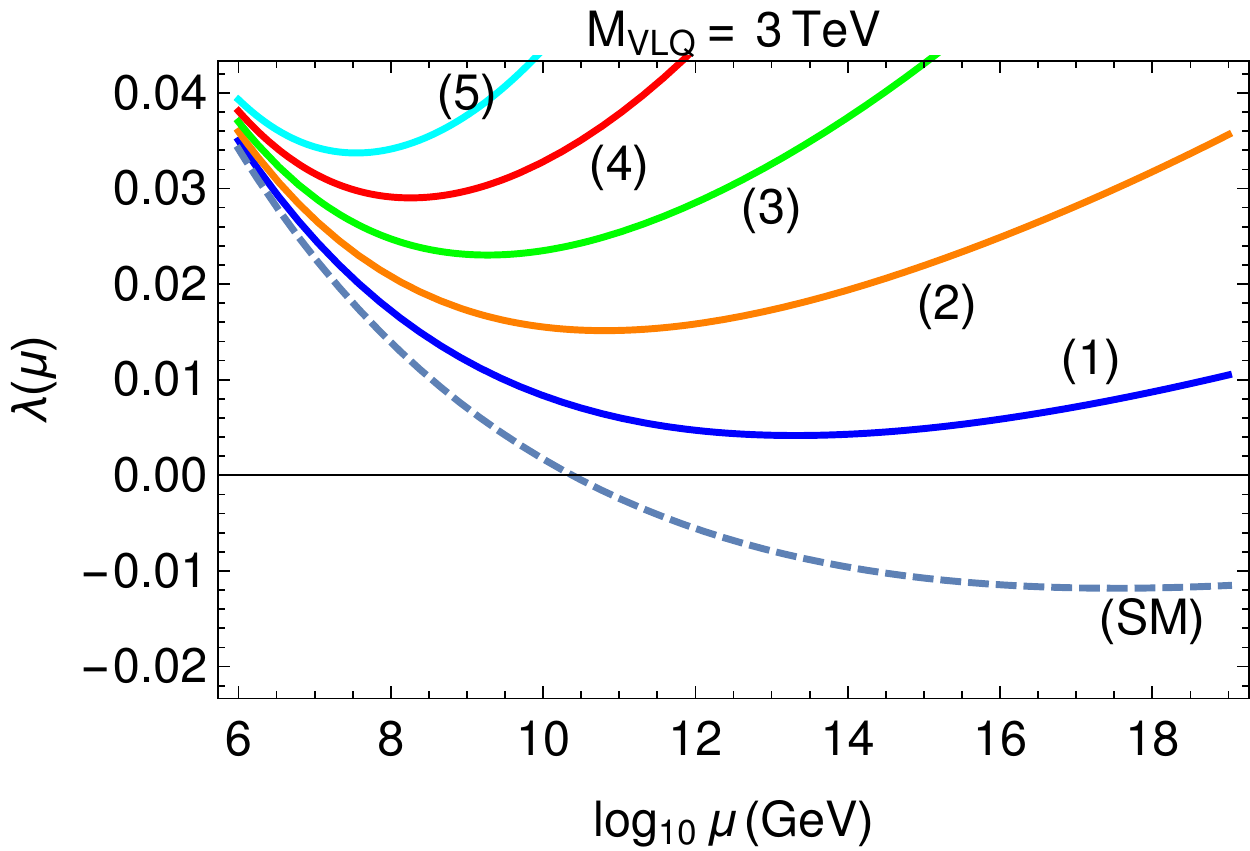} \hspace*{0.25cm}
    \includegraphics[width=0.4\textwidth] {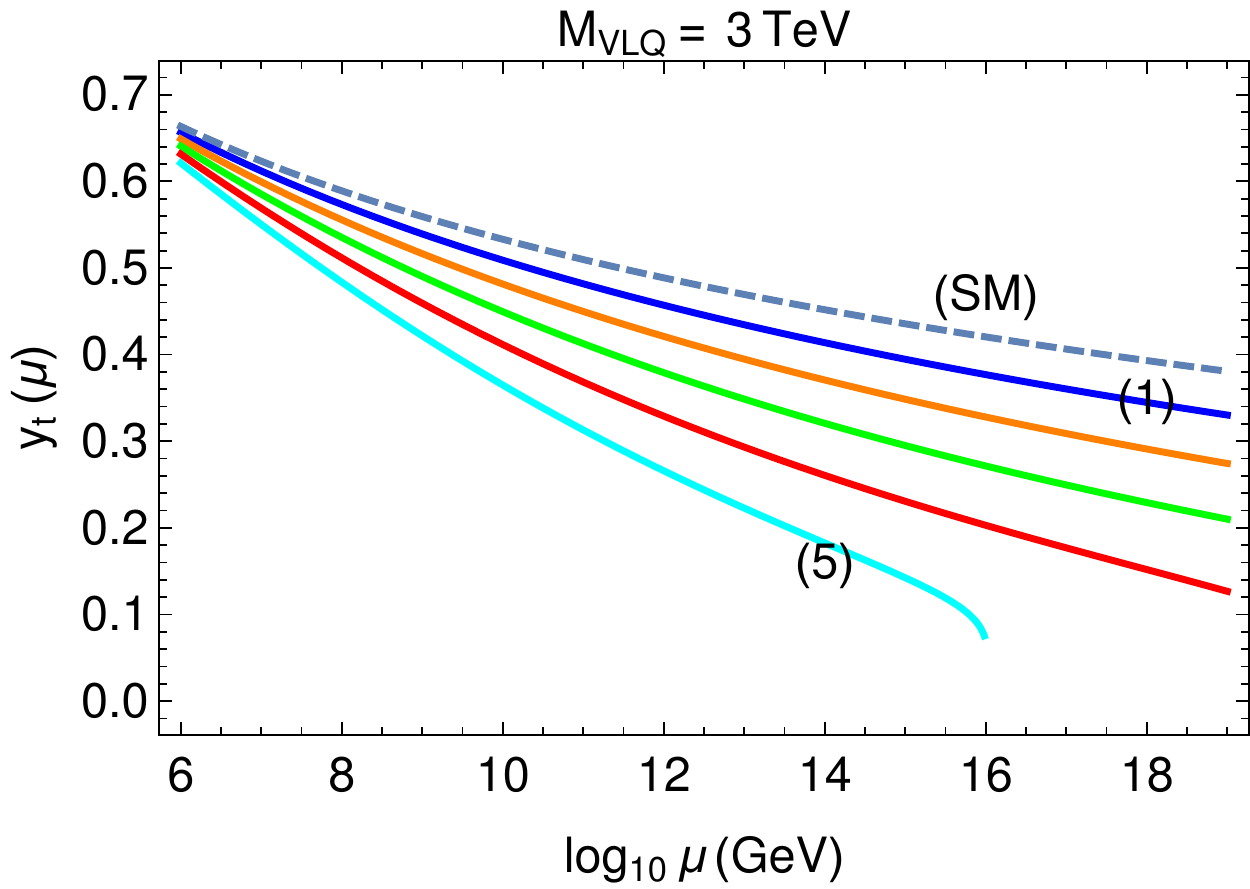} \\
    \vspace*{0.25cm}
    \includegraphics[width=0.4\textwidth] {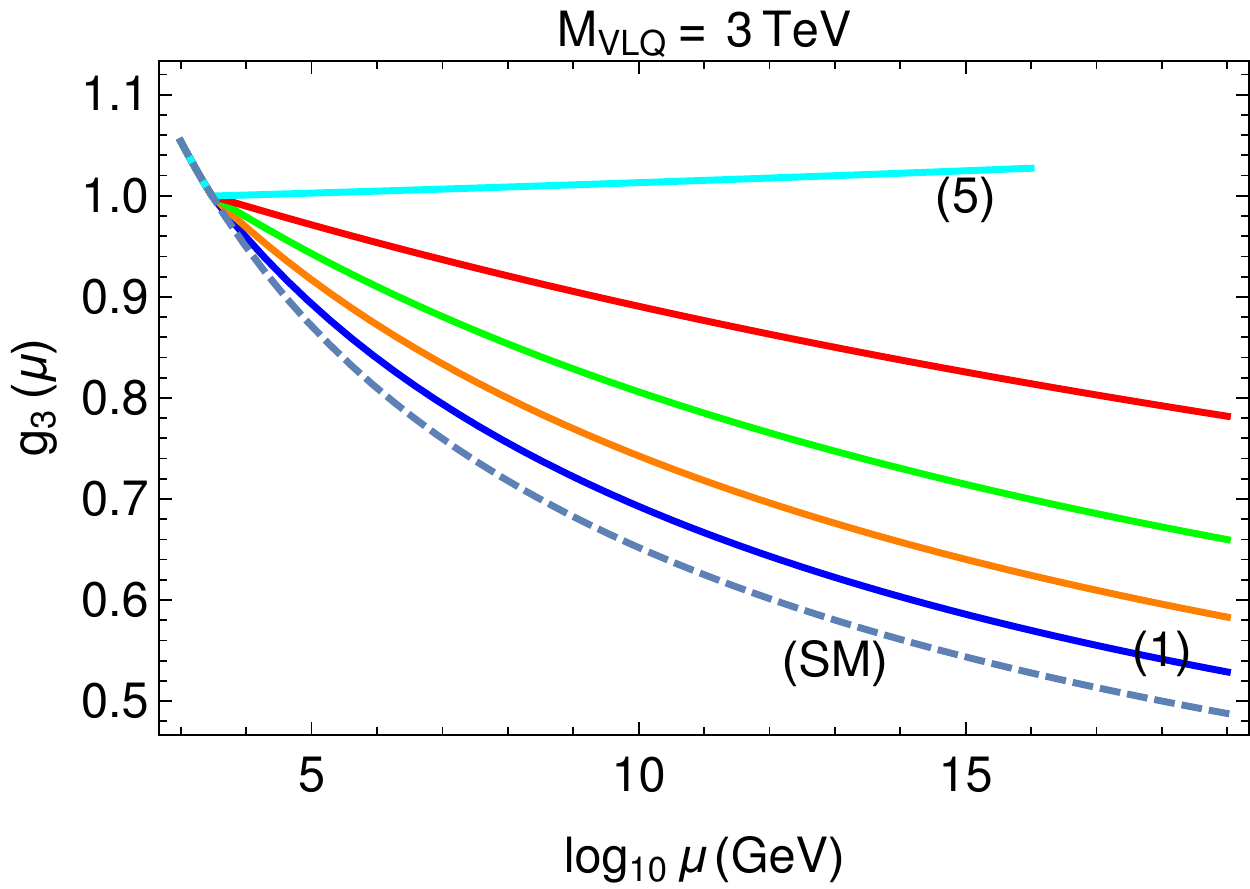} \hspace*{0.25cm}
    \includegraphics[width=0.4\textwidth] {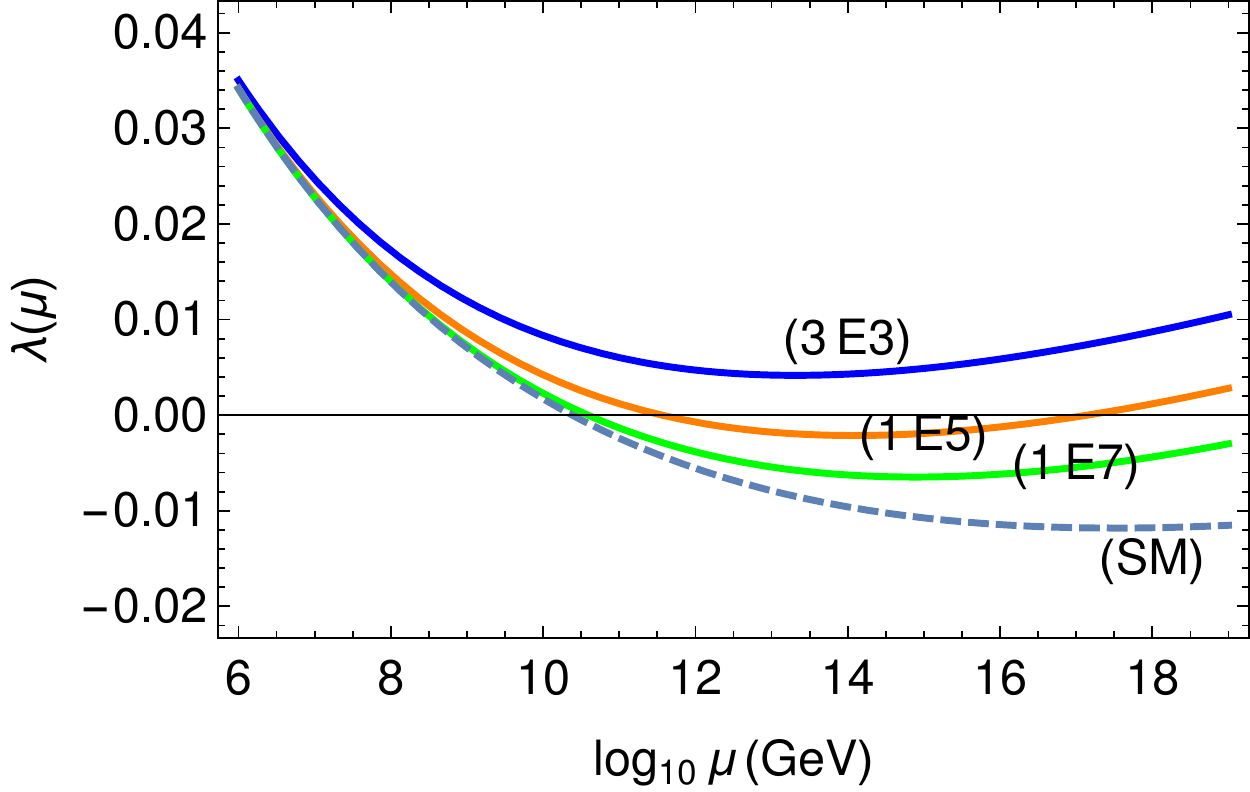}
    \caption{The evolution of $\lambda$, $y_t$, and $g_3$ with Higgs field value $\mu$ 
      for $(n)$ number of degenerate $SU(2)$ doublet VLQs of mass 3~TeV (first three plots),
      and $\lambda$ with a doublet VLQ of mass $3\times 10^3$~GeV, $10^5$~GeV and $10^7$~GeV
      shown respectively as $3E3$, $1E5$ and $1E7$ (last plot).  
      \label{VLQdoub.FIG}}
  \end{center}
\end{figure}

\begin{figure}
  \begin{center}
    \includegraphics[width=0.4\textwidth] {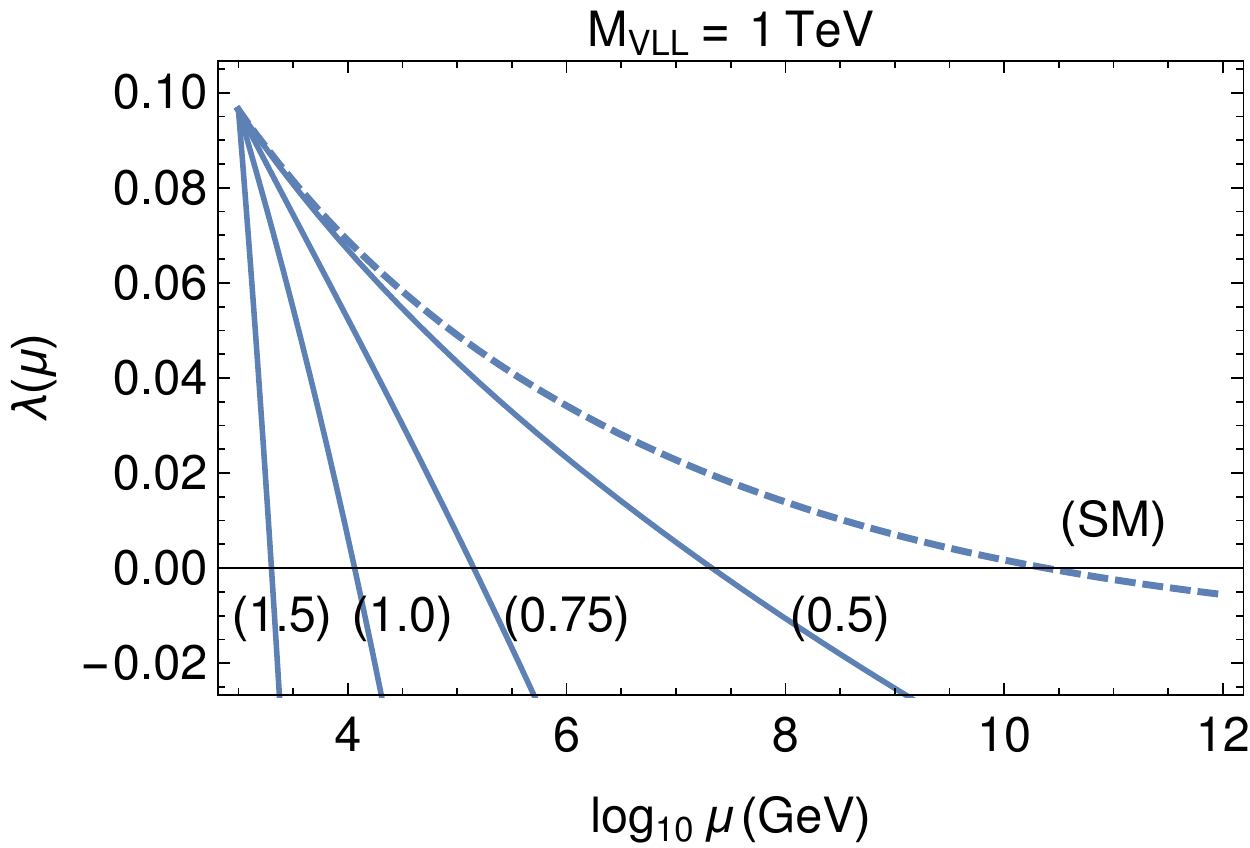} \hspace*{0.25cm}
    \includegraphics[width=0.4\textwidth] {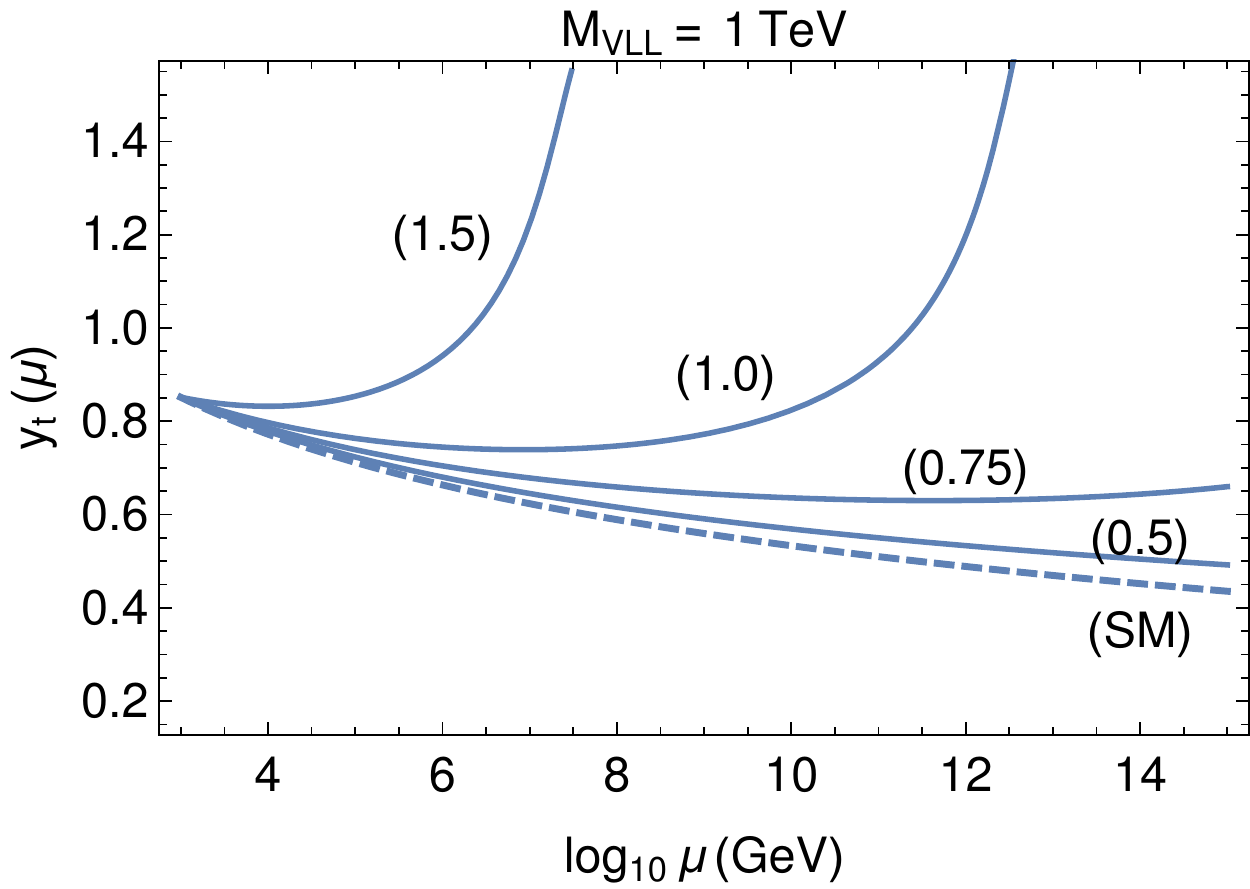} \\
    \vspace*{0.25cm}
    \includegraphics[width=0.4\textwidth] {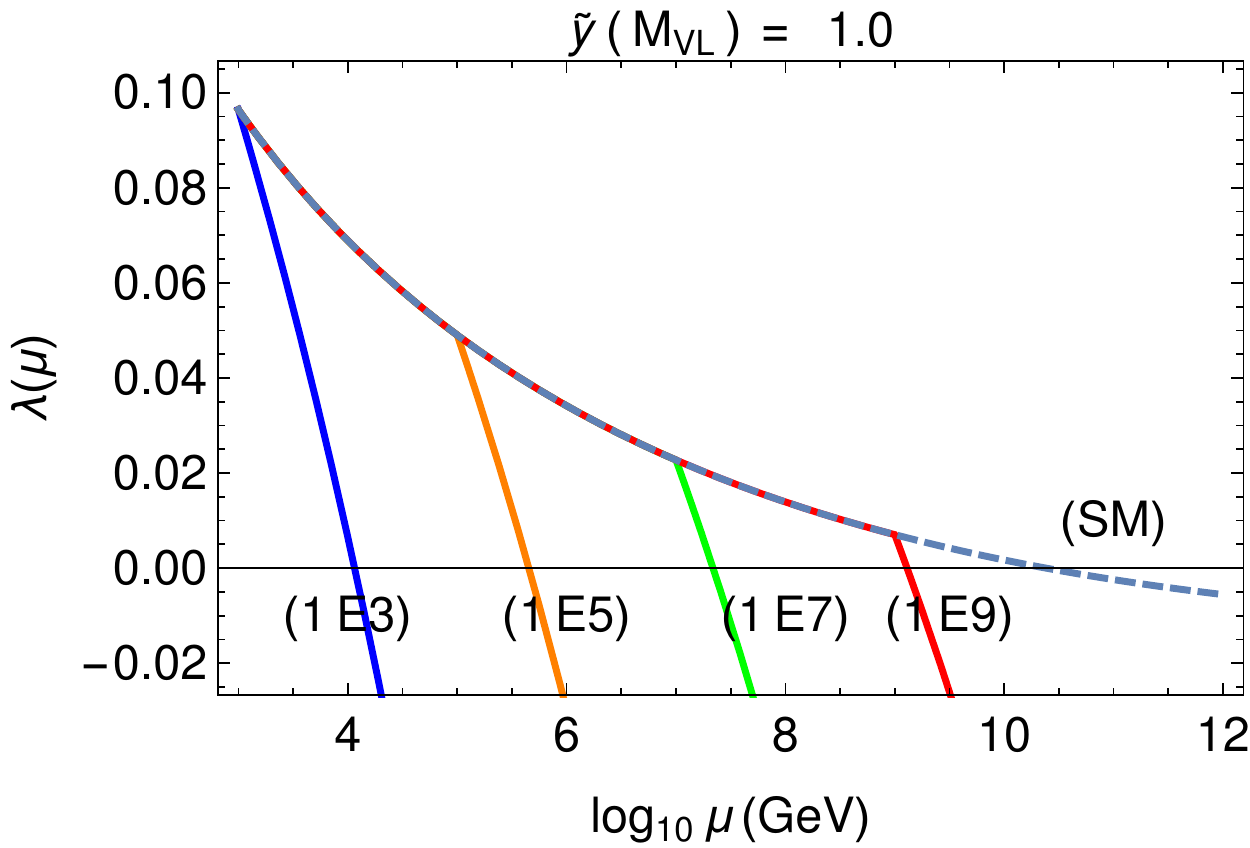} \hspace*{0.25cm}
    \includegraphics[width=0.4\textwidth] {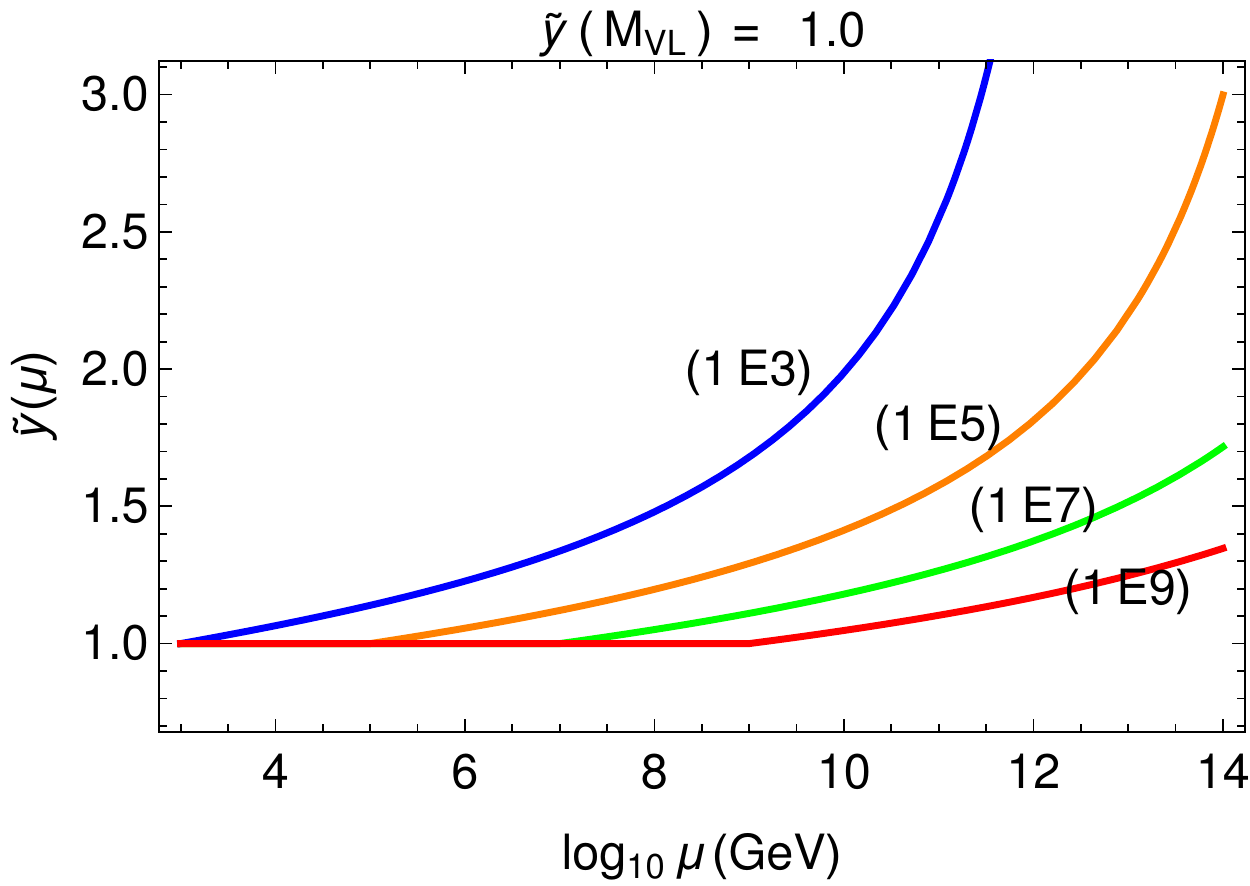}
    \caption{The evolution of $\lambda$, $y_t$, and $\tilde{y}$ with Higgs field value $\mu$
      for a degenerate family of one SU(2) doublet VLL and one singlet VLL, 
      for $M_{VL} = 1$~TeV and various $\tilde{y}$ (first two plots), and
      for $\tilde{y}(M_{VL}) = 1$ and various $M_{VL}$ (in GeV) (last two plots). 
      \label{VLL2211.FIG}}
  \end{center}
\end{figure}

\begin{figure}
  \begin{center}
    \includegraphics[width=0.4\textwidth] {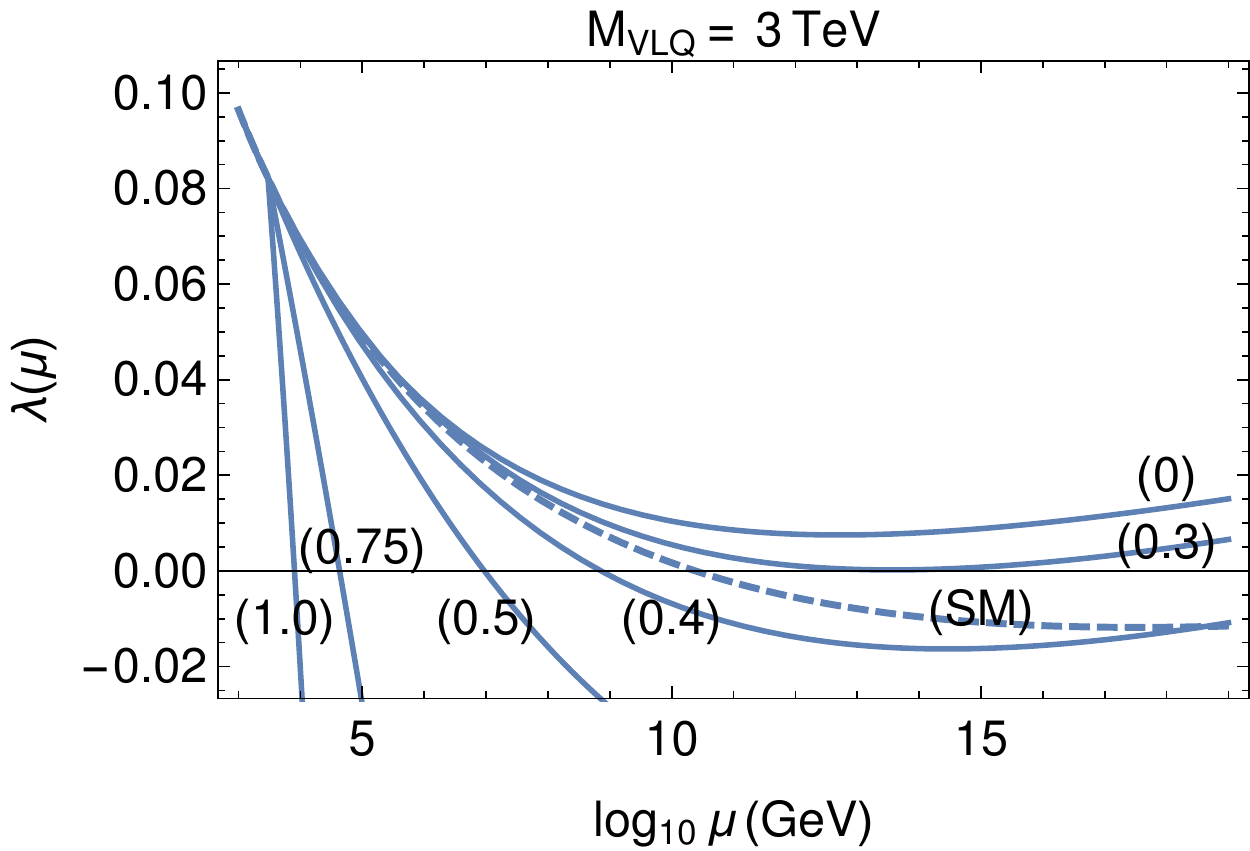} \hspace*{0.25cm}
    \includegraphics[width=0.4\textwidth] {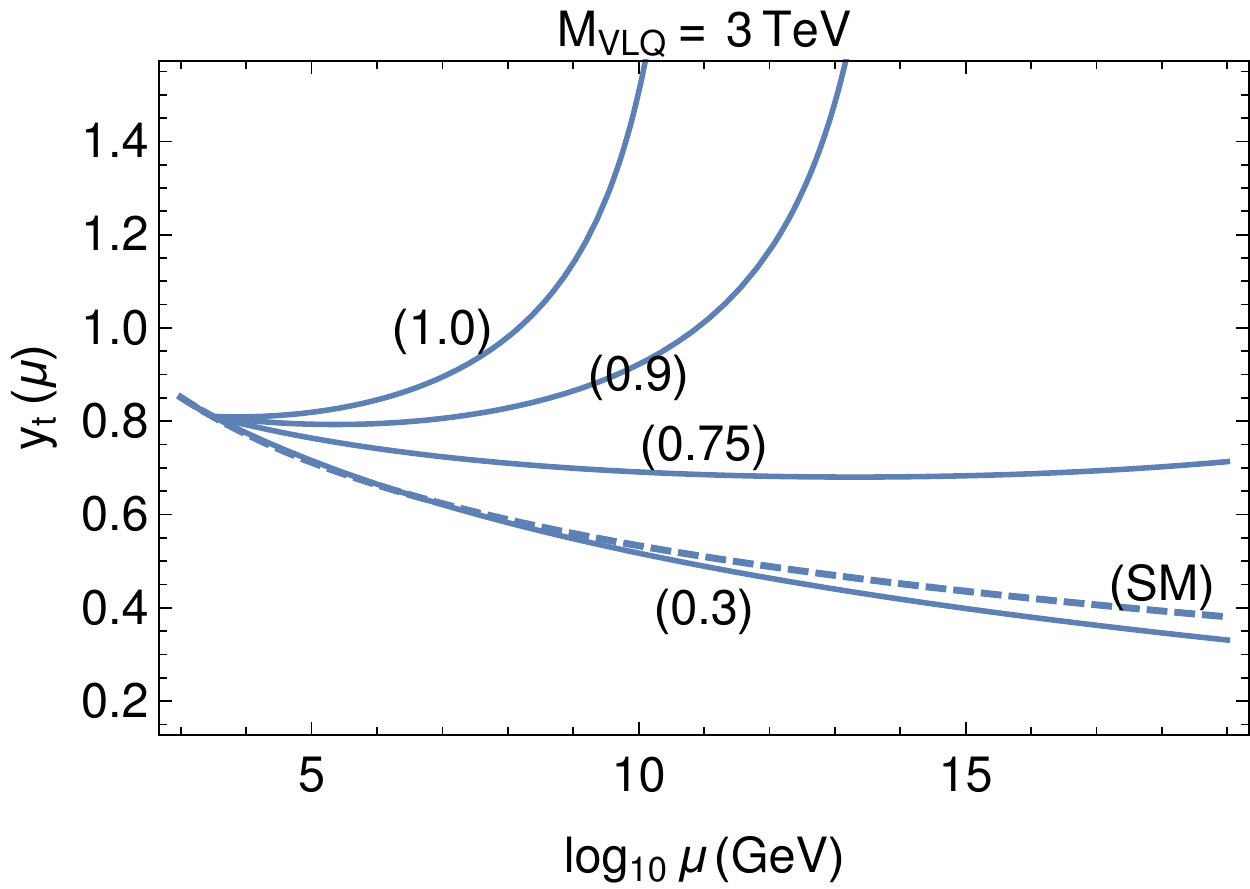} \\
    \vspace*{0.25cm} 
    \includegraphics[width=0.4\textwidth] {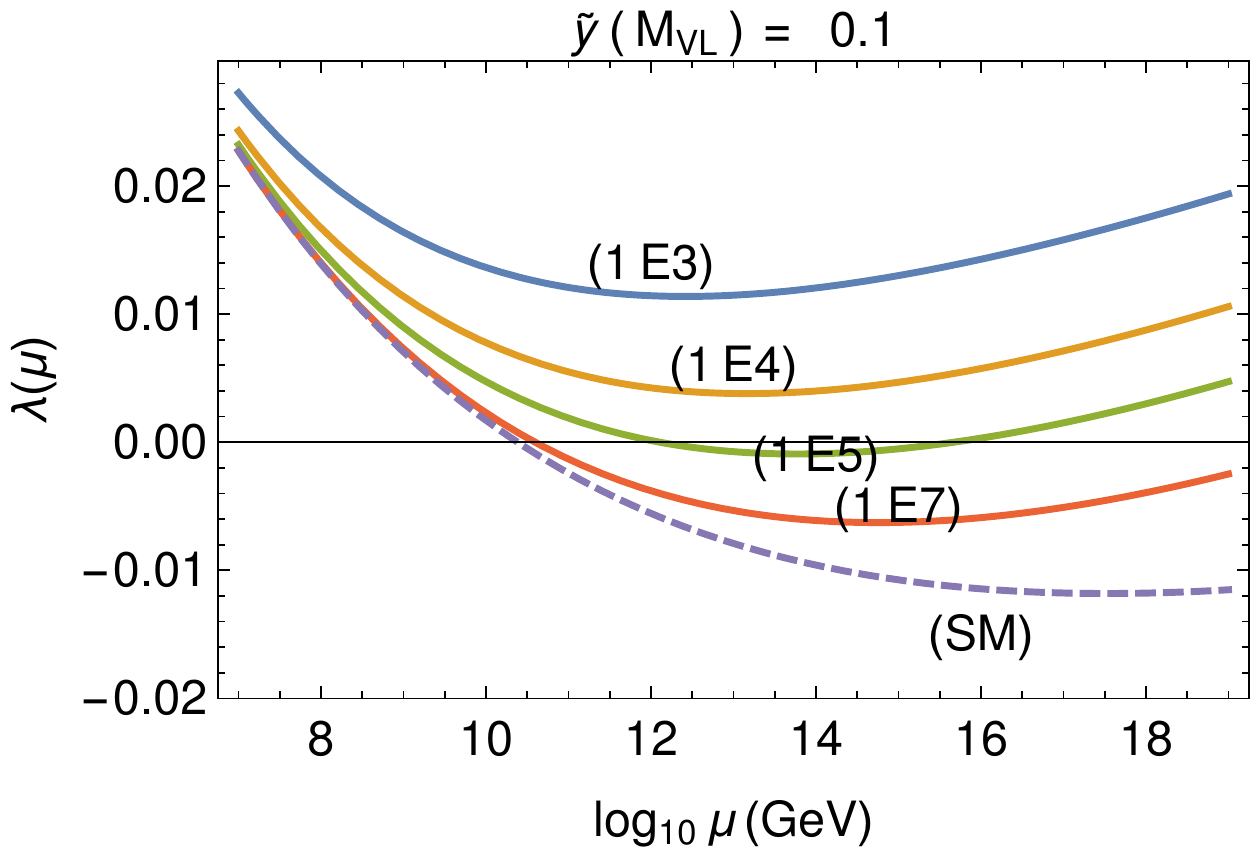} \hspace*{0.25cm}
    \includegraphics[width=0.4\textwidth] {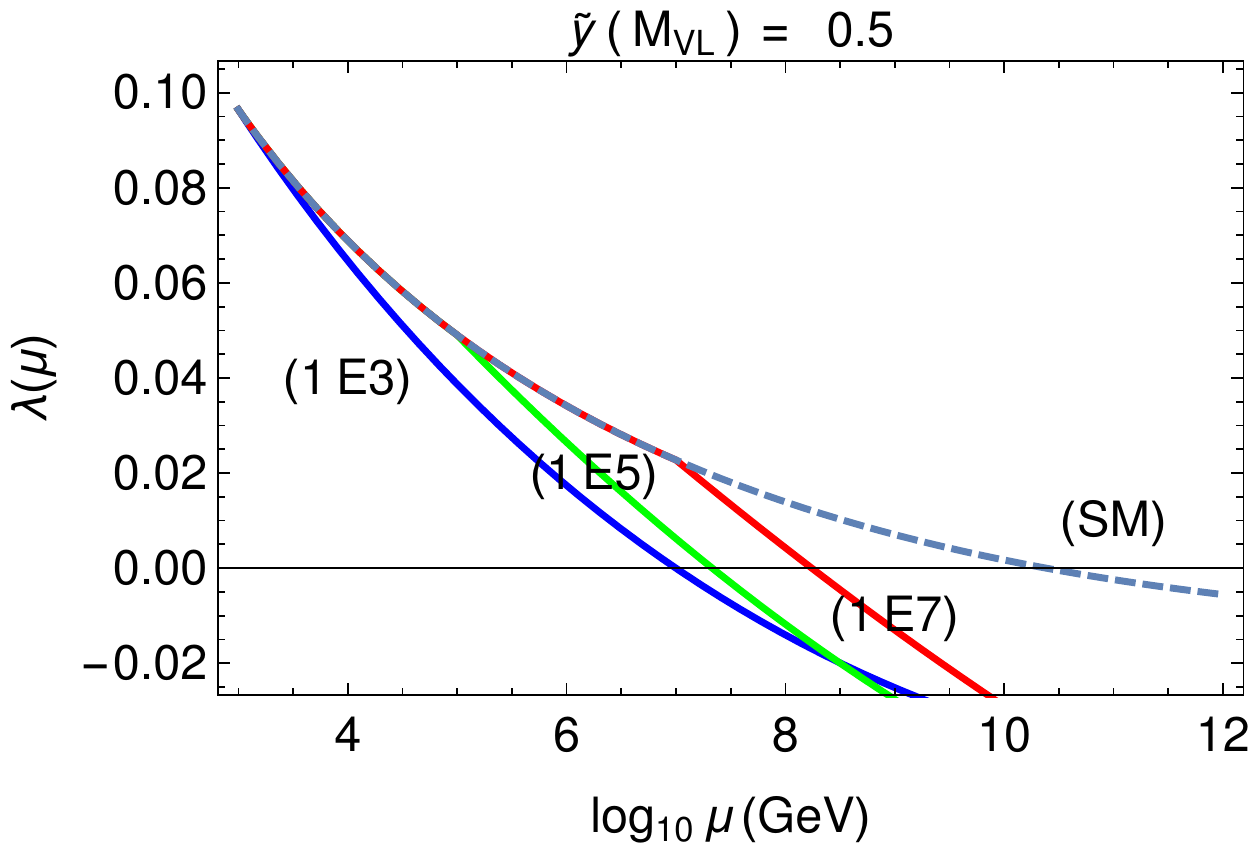} \\
    \vspace*{0.25cm}
    \includegraphics[width=0.4\textwidth] {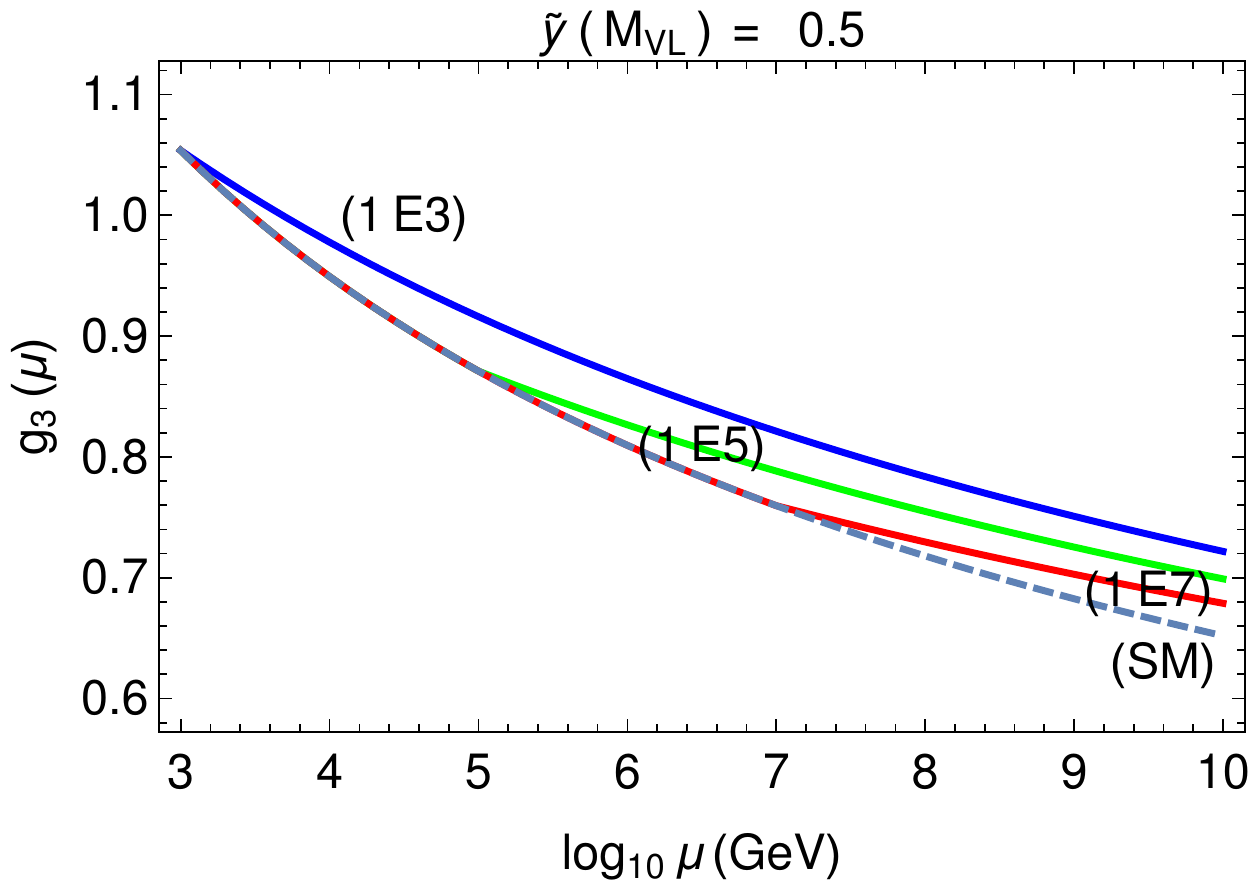} \hspace*{0.25cm}
    \includegraphics[width=0.4\textwidth] {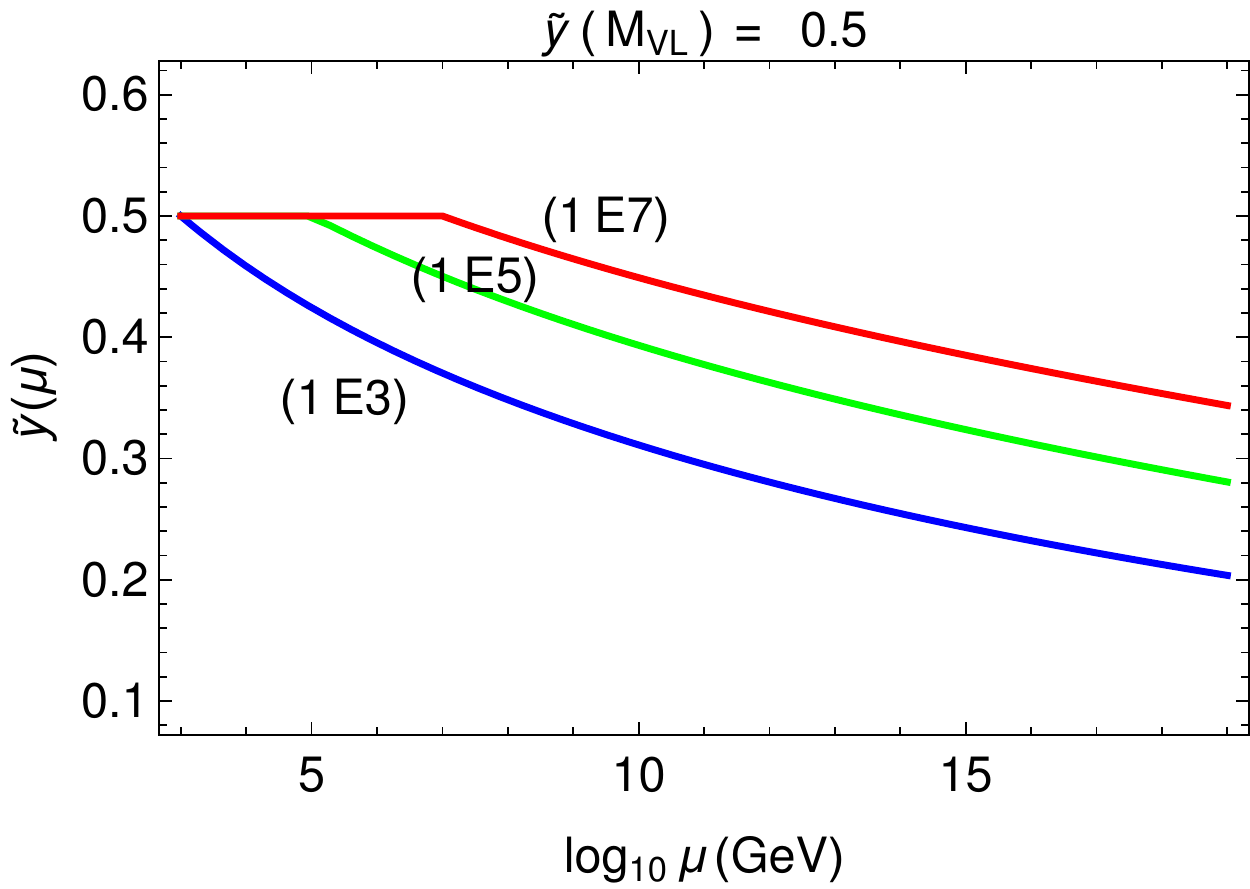}
    \caption{The evolution of $\lambda$, $y_t$, $g_3$ and $\tilde{y}$ with Higgs field value $\mu$
      for a family of one $SU(2)$ doublet VLQ and one singlet VLQ for various $M_{VL}$ and $\tilde y$.
      The top row is for $M_{VLQ} = 3~$TeV, while the bottom row is for different $M_{VL}$ (in GeV)
      for $\tilde{y}(M_{VL}) = \{0.1, 0.5\}$. 
      \label{VLQ2211.FIG}}
  \end{center}
\end{figure}

In Figs.~\ref{VLQsing.FIG}-\ref{VLQ2211.FIG} we show the evolution of the different couplings
for the SM and also for some representative VLF cases. 
In these figures, the dashed lines are for the SM with only the SM particle content with no VLFs,
while the solid lines are for various VLF cases. 
As is evident, in the SM all the couplings decrease with field value $h\equiv \mu$.
The Higgs quartic coupling $\lambda$ becomes zero and goes negative at about $\mu \sim 10^{10.5}~$GeV,
while all the other couplings stay positive all the way up to $M_{\rm Pl}$. 
It is interesting that $\beta_\lambda$ approaches zero for large field values ({\it cf} Fig.~\ref{SBIgnd.FIG}).
In the following, we discuss the evolution of the couplings in the presence of VLFs. 

In Fig.~\ref{VLQsing.FIG} we show the evolution of the couplings with field value $h\equiv \mu$ for $(n)$ degenerate $SU(2)$ singlet VLQs for various $M_{VL}$, 
where the $M_{VL}$ values are shown in the notation $(rEm) \equiv r\times 10^{m}~$GeV. 
When 3 or more degenerate singlet VLQs of mass 3~TeV are added, interestingly, $\lambda$ never goes negative, unlike in the SM. 
When we add only singlet VLQs, $SU(2)$ invariance forbids a coupling of the Higgs to such VLQs,
(we do not turn-on mixed Yukawa couplings between SM fermions and VLFs as we noted earlier). 
However, these VLQs contribute to $\beta_{g_3}$, and also to $\beta_{g_1}$ if the VLQ has hypercharge, and
because of the coupled nature of the RGEs, $\lambda(h)$ does see the effect of the VLQ.
In particular, even if $\tilde y$ is very small, the VLQ contribution to $\beta_{g_3}$ given in Eq.~(\ref{betag3VLF.EQ}) still remains,
which, being positive, results in $g_3(\mu)$ being larger for larger $\mu$ as compared to the SM.
A larger $g_3$ means that the second term in Eq.~(\ref{RGEbeta1ygaSM.EQ}) is more negative,
causing the $y_t(\mu)$ to be smaller in comparison to the SM case.
A smaller $y_t$ implies a less negative contribution to $\beta_\lambda$
from the third term of Eq.~(\ref{RGEbeta1lamSM.EQ}), which means that the $\lambda(\mu)$ is larger with VLQ present. 
For large enough $n_3$ this results even in a turn-around to a positive $\beta_\lambda$ allowing for the possibility of $\lambda$ never going negative.  
We discuss the implications of this to vacuum stability in Sec.~\ref{AbsltStab.SEC}. 

In Fig.~\ref{VLQdoub.FIG} we show the evolution of the couplings with field value $h\equiv \mu$ for $(n)$ degenerate $SU(2)$ doublet VLQs for various $M_{VL}$. 
We observe that when we add one or more doublet VLQs of mass 3~TeV, $\lambda$ never goes negative for the same reasons as above.
We also see that adding a doublet VLQ with mass up to about $10^5$~GeV will have this feature.
We discuss in Sec.~\ref{AbsltStab.SEC} the implications to vacuum stability of $\lambda$ remaining positive.
If we add five or more doublets with 3~TeV mass, we find that
due to the large positive VLF contribution to $\beta_{g_2}$ given in Eq.~(\ref{betag2VLF.EQ}),
$g_2$ grows and becomes non-perturbative at $\mu \approx 10^{16}~$GeV, invalidating this perturbative analysis at around that scale.
In Fig.~\ref{VLQdoub.FIG}, we have restricted the five doublet curves to the region $g_2 < 10$ so that our perturbative analysis
is reliable. 
The negative contribution proportional to $g_2^2$ in $\beta_{y_t}$ given in Eq.~(\ref{RGEbeta1ygaSM.EQ})
becomes significant as $g_2$ becomes large and leads to a smaller $y_t$.
Also, $\beta_{g_3}$ gets a large positive VLF contribution from Eq.~(\ref{betag3VLF.EQ}) causing $g_3$ to increase with $\mu$. 

In Fig.~\ref{VLL2211.FIG} we show the evolution of the couplings with field value $h\equiv \mu$ for a degenerate family of one $SU(2)$ doublet VLL and one singlet VLL
for various $M_{VL}$ and $\tilde y$.
The VLL Yukawa coupling values are shown as $(\tilde{y})$ and the $M_{VL}$ values are shown in the notation $(rEm) \equiv r\times 10^{m}~$GeV. 
For $\tilde y(M_{VL})\! =\! 1$, we see that $\tilde y$ increases as $\mu$ increases.
$y_t$ eventually starts increasing at large $\mu$, which is a behavior unlike in the SM. 
We notice that the scale at which $\lambda$ becomes negative decreases as $\tilde y$ increases, or as $M_{VL}$ decreases. 
 
In Fig.~\ref{VLQ2211.FIG} we show the evolution of the couplings with field value $h\equiv \mu$ for a degenerate
family of one $SU(2)$ doublet VLQ and one singlet VLQ, for various $M_{VL}$ and $\tilde y$.
We see that for $\tilde y(M_{VL})=0.5$, $\tilde y$ decreases as $\mu$ increases.  
For $M_{VL} = 3~$TeV, if $\tilde y > 0.35$, the scale at which $\lambda$ becomes negative decreases as
$\tilde y$ increases or as $M_{VL}$ decreases and is lesser than in the SM, 
while if $\tilde y < 0.35$, the scale at which $\lambda$ becomes negative is larger than in the SM.
In fact, for $\tilde y < 0.3$, $\lambda$ stays positive all the way up to $M_{Pl}$.
We see that when $\tilde y = 0.1$ for example, $\lambda$ stays positive all the way up to $M_{Pl}$
for $M_{VL}$ up to about $10^5$~GeV. 

These examples illustrate a range of effects on the evolution of the couplings due to VLFs. 
For the cases when $\lambda$ does go negative, the EW vacuum is not the absolute minimum but is meta-stable.
There is then a non-zero probability that the EW vacuum will tunnel quantum mechanically
away to those (large) field values where $V_{\rm eff} < 0$.
We turn next to an analysis of this possibility and a computation of the tunneling probability.

\section{Tunneling away from EW vacuum}
\label{TUNL.SEC}

In this section, we compute the tunneling probability from the metastable electroweak Higgs vacuum
into a deeper true vacuum via quantum mechanical barrier penetration.
We do this by computing the {\em Euclidean action} for the {\em bounce} configuration of the Higgs field
(for a review see for example Ref.~\cite{Coleman:1985rnk}). 
We compute the bounce configuration using the running couplings that we presented in Sec.~\ref{RGE.SEC}. 
From the bounce action $S_B$, we compute the tunneling probability $P_{tunl}$.

\subsection{Method of computing the tunneling probability}

We briefly review here how to compute the bounce configuration and the tunneling probability
(for details see for example Refs.~\cite{Sher:1988mj,Coleman:1985rnk} and references therein).
In Sec.~\ref{compSB.SEC} we discuss in detail why we do not use the approximation commonly used in the literature,
but resort to actually solving the bounce EOM numerically as described in this section.

Let us recall that the Lagrangian density ${\cal L}$ and the {\em action} $S$
for the Higgs field in Minkowski coordinates are
\beq
    {\cal L} = \frac{1}{2} \partial^\mu h \partial_\mu h - \Veff(h) \ ; \qquad S = \int d^4 x \, {\cal L} \ ,
\eeq
with the effective potential defined in Eq.~(\ref{VeffhDefn.EQ}). 
We define the Euclidean time $\tau = i t$, the Euclidean coordinates
$\rho^i = \{\tau,x^1,x^2,x^3\}$ with $i=\{1,2,3,4\}$, and
the invariant $\rho^2 = \tau^2 + (x^1)^2 + (x^2)^2 + (x^3)^2$, and write the Euclidean action as
\beq
S_E = \int d^4 \rho \left[ \frac{1}{2} (\partial_i h)^2  + \Veff(h) \right] \ ,
\eeq
where $\partial_i$ is with respect to the Euclidean coordinates $\rho^i$.

As mentioned earlier, taking $\Veff(v) = 0$ at the EW minimum,
the central question of interest in our work here is whether the EW vacuum is absolutely stable or
if there is a possible transition to other field values, which would be possible only if $\Veff(h) < 0$ for some $h$ typically much larger than $v$.
If there exists field values for which $\Veff(h) < 0$, the EW vacuum at $h=v$ is typically separated from this by a (large) barrier and a vacuum transition can only
occur via quantum tunneling. In such a situation we would like to know the time-scale of the tunneling in comparison to the age of the Universe and see if we can gain
an understanding of why the Universe has not tunneled away to the true vacuum with $h \gg v$, but is in the EW vacuum today. 
If for some large field value, $h\equiv\sigma$ say, suppose $\Veff(\sigma) = 0$, and suppose $\Veff$ is negative for $h\gtrsim \sigma$.
($\Veff$ may turn around and have a second minimum (or not) for $h>\sigma$ depending on other BSM contributions in the RGE.)
Equivalently, from our definition of the field dependent coupling in Eq.~(\ref{Veffh4.EQ}), 
suppose $\lambda(\sigma) = 0$, and that $\lambda(h)$ is negative for $h \gtrsim \sigma$. 
The vacuum configuration defined to have total energy $E=0$ at $h=v$ can quantum mechanically tunnel to
$h \geq \sigma$ with $\Veff(h) < 0$.
If the vacuum were to tunnel so, the field then runs down the potential classically toward large field values $h>\sigma$.
The tunneling probability is given in terms of the bounce configuration~\cite{Coleman:1985rnk}
which satisfies $\delta S = 0$, starting with $h(t\! =\! -\infty) = v$, attaining a value
$h(t\! =\! 0) \equiv h_0 \geq \sigma$ and returning to $h(t\! =\! \infty) = v$.
This configuration is a solution of the equation of motion (EOM).
In Euclidean coordinates, the EOM reads
\beq
\partial_i^2 h = \frac{\partial \Veff}{\partial h} \ . 
\eeq
We look for an $O(4)$ symmetric solution~\cite{Coleman:1977th}, which implies that it depends on $\rho$, i.e. $h(\rho^i) = h(\rho)$. 
The EOM then reads
\beq
\frac{d^2 h}{d \rho^2} + \frac{3}{\rho} \frac{d h}{d \rho} = \frac{\partial \Veff}{\partial h} \ , 
\label{EucEOM.EQ}
\eeq
with the boundary conditions (BC) $(dh/d\rho)(\rho\! =\! 0) = 0$ and $h(\rho\! \rightarrow\! \infty) = v$.
We must also have $h(\rho\! =\! 0) = h_0 \geq \sigma$ for this to represent tunneling.
This EOM is identical to that of a classical particle moving in a potential $-\Veff$ with a ``friction'' term present (second term in Eq.~(\ref{EucEOM.EQ})) that dies-off as $1/\rho$
as $\rho$ increases.

In Euclidean space-time, the bounce configuration $h_B(\rho)$, will have the feature of a fairly sharp transition in $\rho$ from $h_0$ to $v$.
In Minkowski space-time, this configuration looks like an expanding bubble
with the bubble-wall separating a region of true-vacuum inside and the false EW vacuum outside.
The bubble nucleation probability per unit 4-volume is given by~\cite{Sher:1988mj} $\Delta \Ptunl/\Delta V_4 = M^4 \exp{(-S_B)}$,
where we have included a prefactor of $M^4$ on dimensional grounds with $M$ an appropriate mass scale,
$\Delta V_4$ is a unit 4-space-time volume, 
and $S_B$ is the Euclidean action for the bounce configuration $h_B(\rho)$ given by
\beq
S_B = 2 \pi^2 \int_0^\infty d\rho \, \rho^3 \left[\frac{1}{2} \left( \frac{dh_B}{d\rho} \right)^2 + \Veff(h_B) \right] \ .
\label{SBIntegral.EQ}
\eeq
We make the choice $M^4=\Veff(h_0)$ since $h_0$ is typically the largest scale in the problem and gives the largest tunneling rate,
and hence the most conservative bound on the allowed VLF parameter-space from vacuum tunneling.

If a bubble bigger than the critical size is nucleated anywhere in our past light-cone it would have engulfed us by now
and we would not find ourselves in the EW vacuum now.
The (dimensionless) volume of our past light cone is about $V_4 \sim (1/m^4_t)\exp{(404)}$,
which is nothing but our Hubble 4-volume in $1/m^4_t$ units,
and we choose this unit since our starting point for the running is at the $m_t$ scale.
Thus, the total probability that we would have nucleated a bubble in our Hubble volume and tunneled into the true vacuum by now is
$\Ptunl \sim (d\Ptunl/dV_4) V_4$, which gives~\cite{Sher:1988mj}
\beq
\Ptunl = (h_0/m_t)^4\, e^{(404 - S_B)} \ .
\label{PtunlSB.EQ}
\eeq
If $\Ptunl \ll 1$ for the given $\Veff$, we deem this as an acceptable situation.
In other words, if $\Ptunl \gtrsim 1$, we take this to mean that the probability that we would have tunneled into the true vacuum due to a bubble nucleating in our past light-cone
is essentially unity, and therefore the model that generated that $\Veff$ we consider is disfavored.
Evidently, the larger $S_B$ is, the smaller $\Ptunl$ is, and the latter is exponentially suppressed by $S_B$.
In the following section, we solve numerically the bounce EOM to get the bounce configuration $h_B(\rho)$,
compute $S_B$ for this $h_B$, and compute $\Ptunl$.
We do this for the SM and some VLF extensions.

\subsection{Tunneling probability numerical results}
\label{tunProbNumRes.SEC}

Here we describe the method we use to solve the bounce EOM numerically, obtain the bounce configuration $h_B(\rho)$,
the bounce action $S_B$ and the tunneling probability $\Ptunl$ for the SM and various VLF extensions.

The value of $S_B$ largely depends on the behavior of $\Veff$ at large field values $h$ where the $h^4$ term dominates, which is why we included only that term in $\Veff$.  
Nevertheless, for completeness, we insert an EW minimum at $v$ by including it in $\lambda(h)$ for $h\sim v$ as follows.
Keeping the $h^2$ term, we have for $h\sim v$ the potential $V = -(\mu_h^2/2) h^2 + (\lambda/4) h^4$,
which we define to be $\Veff = (\lambda(h)/4) h^4$ as in Eq.~(\ref{Veffh4.EQ}).
This definition implies that we have the effective quartic coupling given by $\lambda(h) =  \lambda(v) (1-2v^2/h^2) $ for $h \sim v$.
Slightly above the scale $v$, we match this to the $\lambda(h)$ obtained by solving the RGE.
In this way we effectively obtain a minimum at $h=v$ while the larger field value evolution is governed by the RGE.
We add a constant term to $\Veff$ and make $\Veff(v) = 0$. 

We obtain a solution of the EOM in Eq.~(\ref{EucEOM.EQ}) numerically, subject to the BC $h(\rho\!\! =\!\! 0) = h_0$, $(dh/d\rho)(\rho\!\! =\!\! 0)=0$.
$h_0$ is unknown and so we iteratively search for that $h_0$ that will lead to $h(\rho_{\rm end}) = v$ and $(dh/d\rho)(\rho_{\rm end})=0$.
Although in theory $\rho_{\rm end} \to \infty$, in practice it can be picked finite but large enough that the bounce has completed the transition from $h_0$ to $v$.  
In our numerical implementation, we work with the dimensionless quantities
$\rhoh \equiv \tilde{m}_t \rho$, $\hh \equiv h/\tilde{m}_t$ and $\Veffh(h) = \Veff(h)/{\tilde{m}_t}^4$. 

The ``friction'' term that goes like $1/\rho$ in Eq.~(\ref{EucEOM.EQ}) will be problematic numerically near $\rho \rightarrow 0$,
and we therefore obtain an analytical solution in this region, valid for $\rhoh \in (0,\epsilon)$ for $\epsilon \ll 1$,
and match this onto a numerical solution of the EOM for $\rhoh \geq \epsilon$.
We now give the solution valid in $\rhoh \in (0,\epsilon)$.
For small $\rhoh$, we expand as $(3/\rhoh) d\hh/d\rhoh \equiv s(\rhoh) = s_0 + s_1 \rhoh + (s_2/2) \rhoh^2 + {\cal O}(\rhoh^3)$,
and require all the $s_i$ to be finite so that the friction term is finite as $\rhoh \to 0$.
Integrating this, we find $\hh(\rhoh) = \hh_0 + (s_0/6) \rhoh^2 + {\cal O} (\rhoh^3)$.
Differentiating the earlier equation, we have $d^2\hh/d\rhoh^2 = s_0 + (4s_1/3)\rhoh + (5s_2/6)\rhoh^2 + {\cal O} (\rhoh^3)$.
We find $\lambda(\hh(\rhoh)) = \lambda_0 + ({\beta_\lambda}_0/\hh_0) (d^2\hh/d\rhoh^2)_0 \rhoh^2/2 + {\cal O}(\rhoh^3)$, and 
$\beta_\lambda(\hh(\rhoh)) = {\beta_\lambda}_0 + ({\partial\beta_\lambda}/\partial\hh)_0 (d^2\hh/d\rhoh^2)_0 \rhoh^2/2 + {\cal O}(\rhoh^3)$,
where $\lambda_0 \equiv \lambda(h_0)$, ${\beta_\lambda}_0 \equiv \beta_\lambda(h_0)$, $(d^2\hh/d\rhoh^2)_0 \equiv (d^2\hh/d\rhoh^2)(\rhoh = 0)$ and
$({\partial\beta_\lambda}/\partial\hh)_0 \equiv ({\partial\beta_\lambda}/\partial\hh)({\hh=\hh_0})$.
Also, $\partial\Veffh/\partial\hh = [\lambda(\hh) + \beta_\lambda(\hh)/4] \hh^3$. 
Substituting these into the EOM in Eq.~(\ref{EucEOM.EQ}), we get by matching powers of $\rhoh$,
$s_0 = (\lambda_0 + {\beta_\lambda}_0/4) \hh_0^3/2$, $s_1 = 0$ and
$s_2 = (3/8) (d^2\hh/d\rhoh^2)_0 [ 3 (\lambda_0 + {\beta_\lambda}_0/4) \hh_0 + ({\beta_\lambda}_0/\hh_0 + ({\partial\beta_\lambda}/\partial\hh)_0/4 ) \hh_0^3 ]$.
This is the solution valid for $\rhoh \in (0,\epsilon)$, and the solution at $\rhoh = \epsilon$ is got by substituting $\rhoh=\epsilon$ in this.

Taking the $\hh(\epsilon)$ and $(d\hh/d\rhoh)|_\epsilon$ obtained as above at the point $\epsilon$ as a BC,
we numerically integrate the EOM in Eq.~(\ref{EucEOM.EQ}) for $\rhoh \in (\epsilon,\rhoh_{\rm end})$ and obtain $\hh_B(\rhoh)$ over this domain.
The large values of the fields and the presence of the friction term complicates the numerical implementation. 
A further challenge is that satisfying the required end-condition requires an extremely sensitive tuning of the starting value $\hh_0$.
By an iterative search algorithm we are able to obtain the bounce configuration $h_B(\rhoh)$ using Mathematica.

Piecing together the analytical solution above and the numerical solution, we obtain the bounce configuration over the complete domain $\rhoh \in (0,\rhoh_{\rm end})$. 
Following this procedure, we present below the bounce configuration, the bounce action evaluated for this bounce,
and the tunneling probability for the SM and various VLF extensions. 

\begin{figure}
  \begin{center}
    \includegraphics[width=0.4\textwidth]{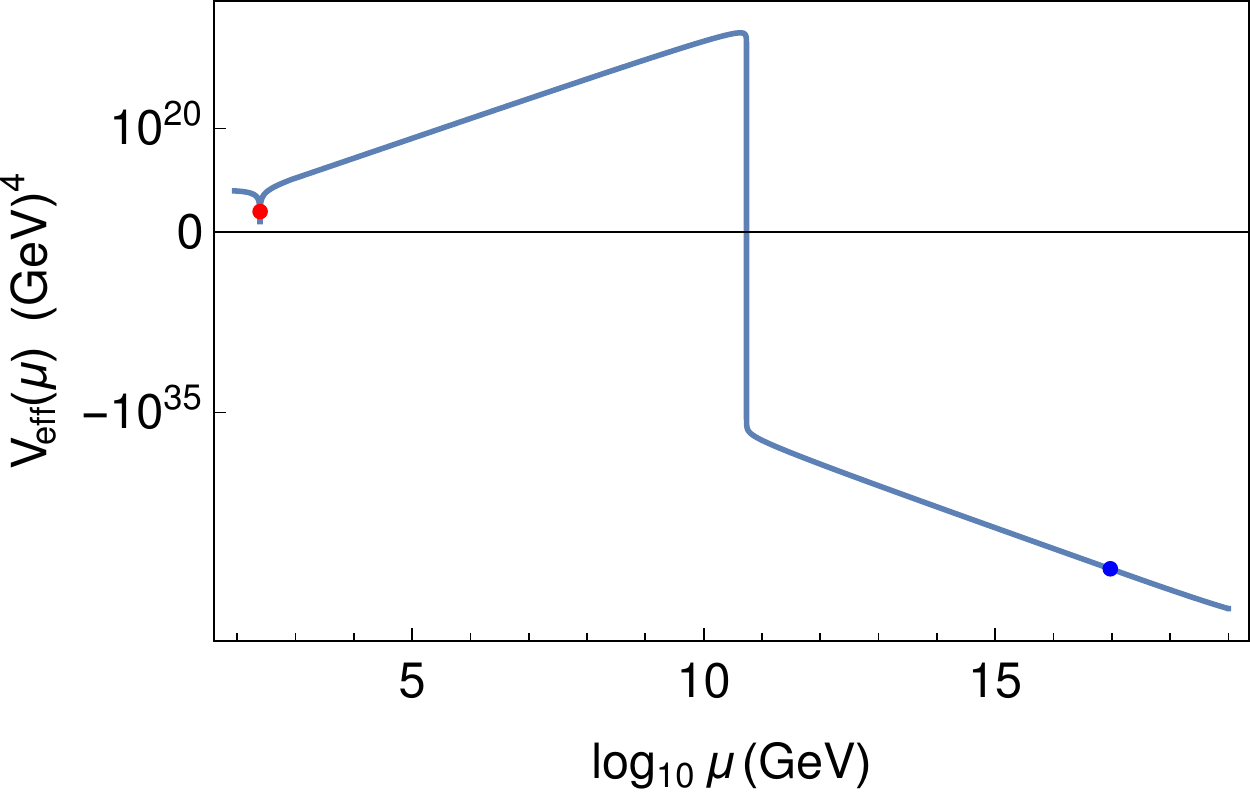} \hspace*{0.25cm}
    \includegraphics[width=0.4\textwidth]{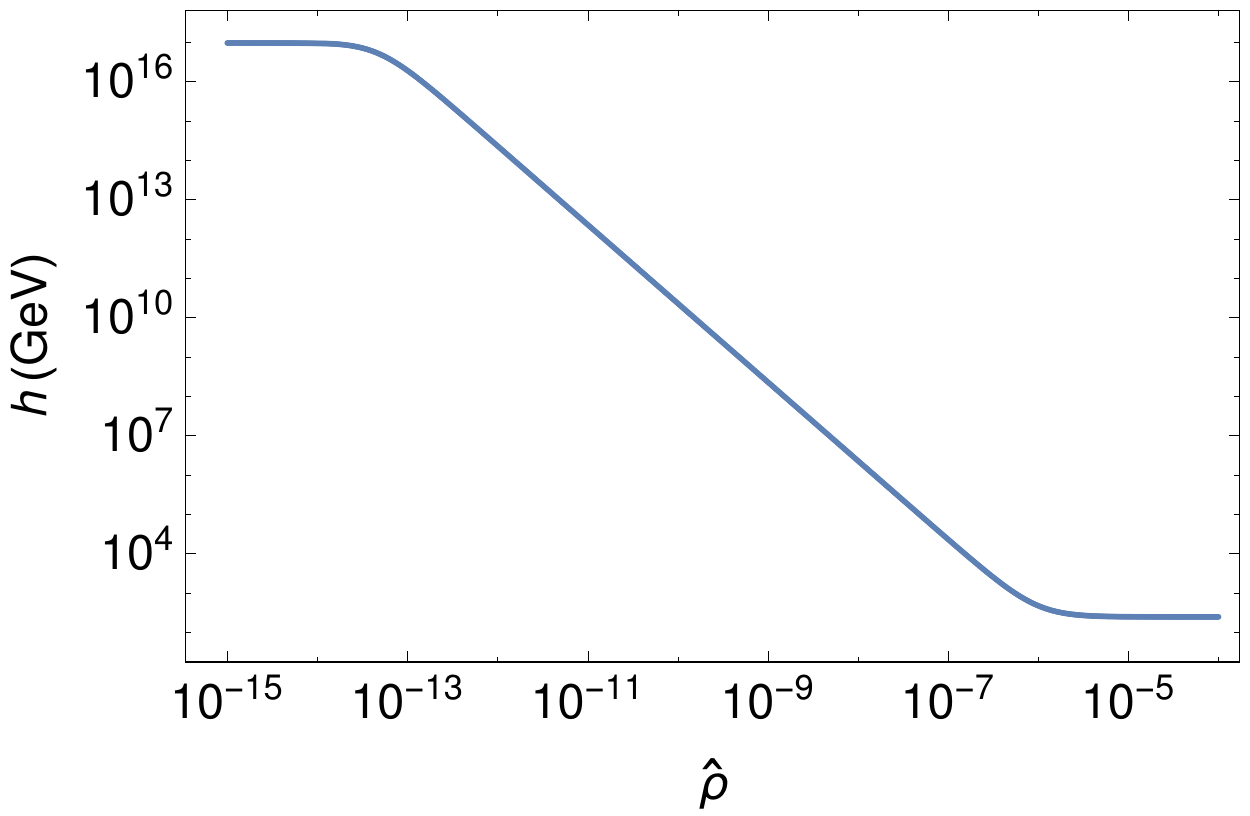}
    \caption{For the SM, 
      the effective potential as a function of the field value $h=\mu$,
      and the bounce configuration $h_B(\hat\rho)$.
      The blue (red) dot shows the starting (ending) value of the bounce. 
      \label{SMTunl.FIG}}
  \end{center}
\end{figure}

For the SM, the $\Veff (h\!\equiv\! \mu)$
and the bounce configuration $h_B(\rho)$ obtained numerically are shown in Fig.~\ref{SMTunl.FIG}. 
The $\Veff(\mu)$ is positive for smaller $\mu$, crosses zero at about $\mu \approx 10^{10.75}~$GeV,
and is negative for larger $\mu$.
The blue dot shows the starting field value ($h_0$) of the bounce and the red dot the ending field value ($v$).   
For this bounce, we find by numerical integration of Eq.~(\ref{SBIntegral.EQ}) that the value of the
Euclidean bounce action is $S_B = 2866$ (in $\hbar=1$ units).
From this, we compute the tunneling probability into the true vacuum in our Hubble volume from Eq.~(\ref{PtunlSB.EQ}) to be
$\Ptunl \sim 10^{-1013}$, which is an incredibly small probability.
This, and many other comparisons we have done for the SM, are in excellent agreement with the results obtained in Ref.~\cite{Buttazzo:2013uya}.

Next, we solve the bounce EOM, and compute the $S_B$ and $\Ptunl$ for various VLF representations.
We start with a (color singlet) VLL family with SM-like hypercharge assignment present,
i.e. an $SU(2)$ singlet with hypercharge $-1$ and an $SU(2)$ doublet VLL with hypercharge $-1/2$
both present, for various common mass $M_{VL}$ and various $\tilde y$. 

\begin{figure}
  \begin{center}
    \includegraphics[width=0.4\textwidth]{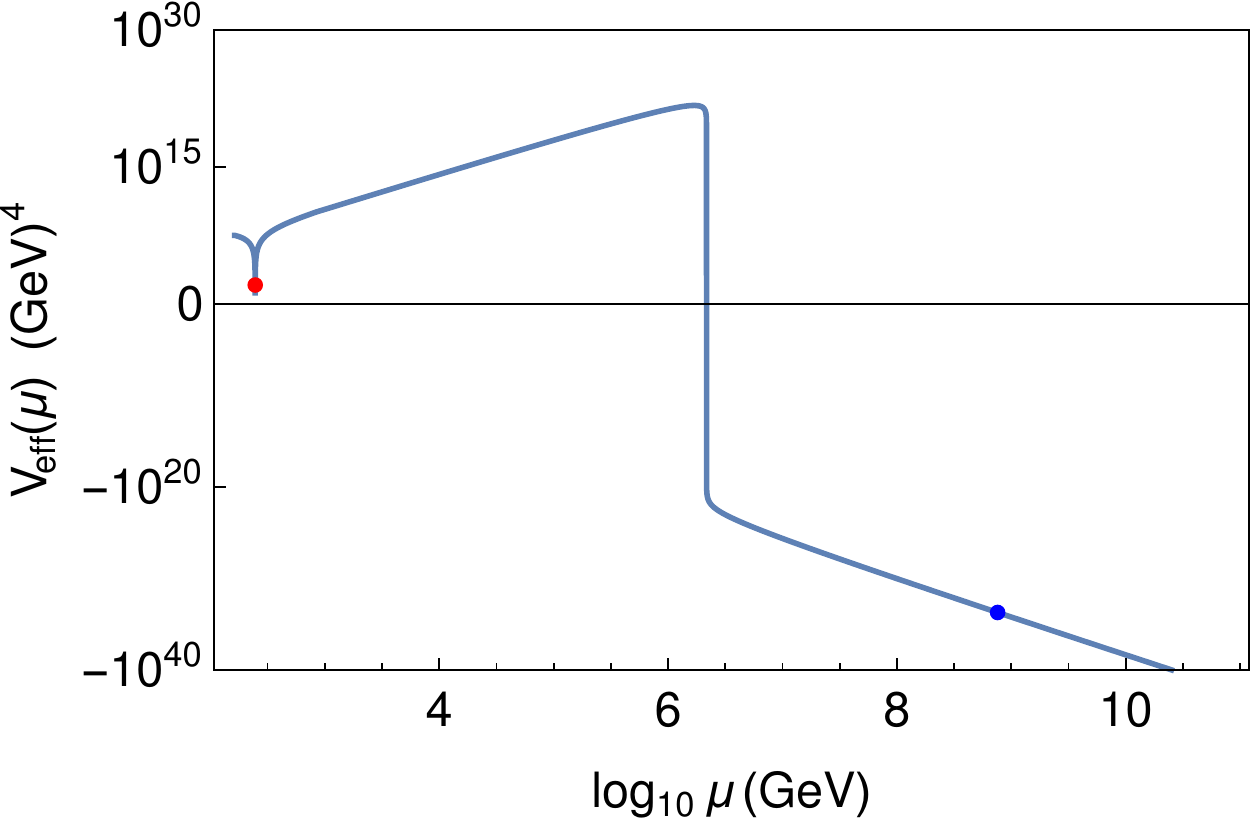} \hspace*{0.25cm}
    \includegraphics[width=0.4\textwidth]{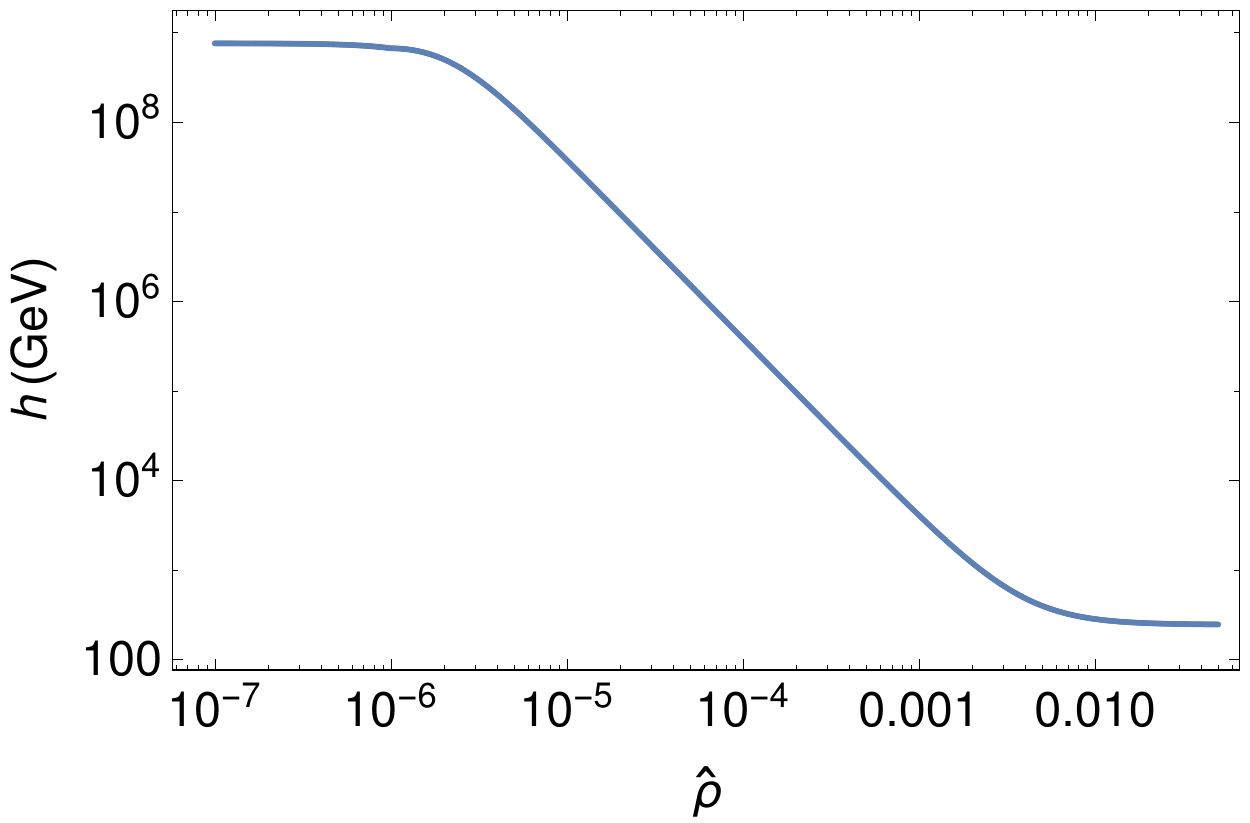} \\
    \vspace*{0.25cm}
    \includegraphics[width=0.4\textwidth]{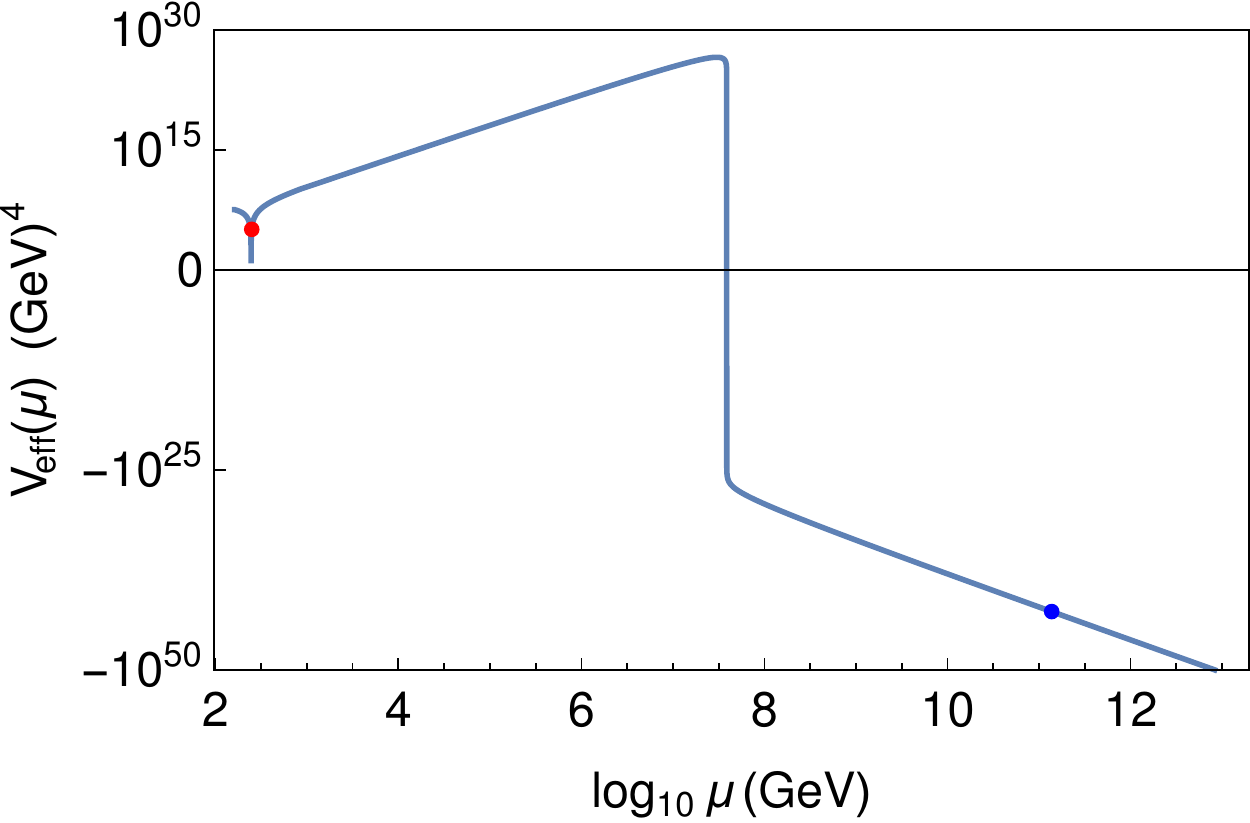} \hspace*{0.25cm}
    \includegraphics[width=0.4\textwidth]{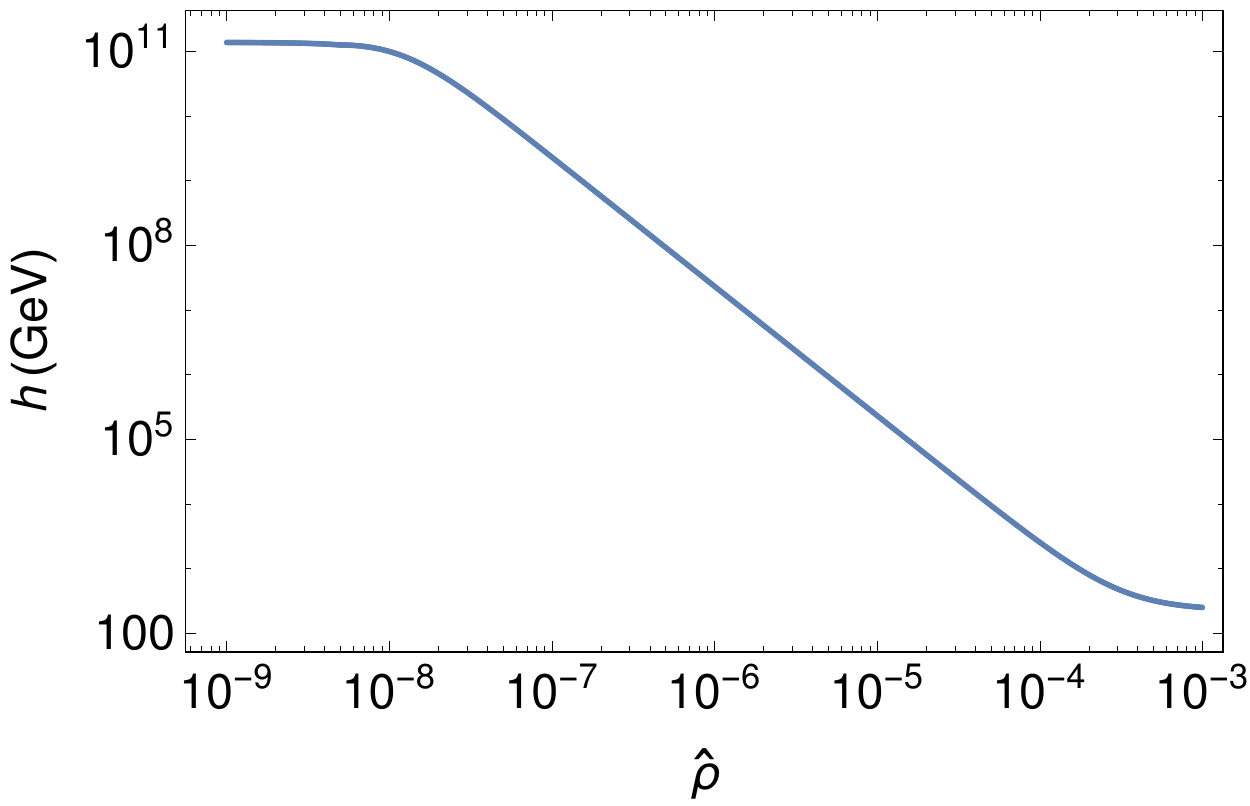} \\
    \vspace*{0.25cm}
    \includegraphics[width=0.4\textwidth]{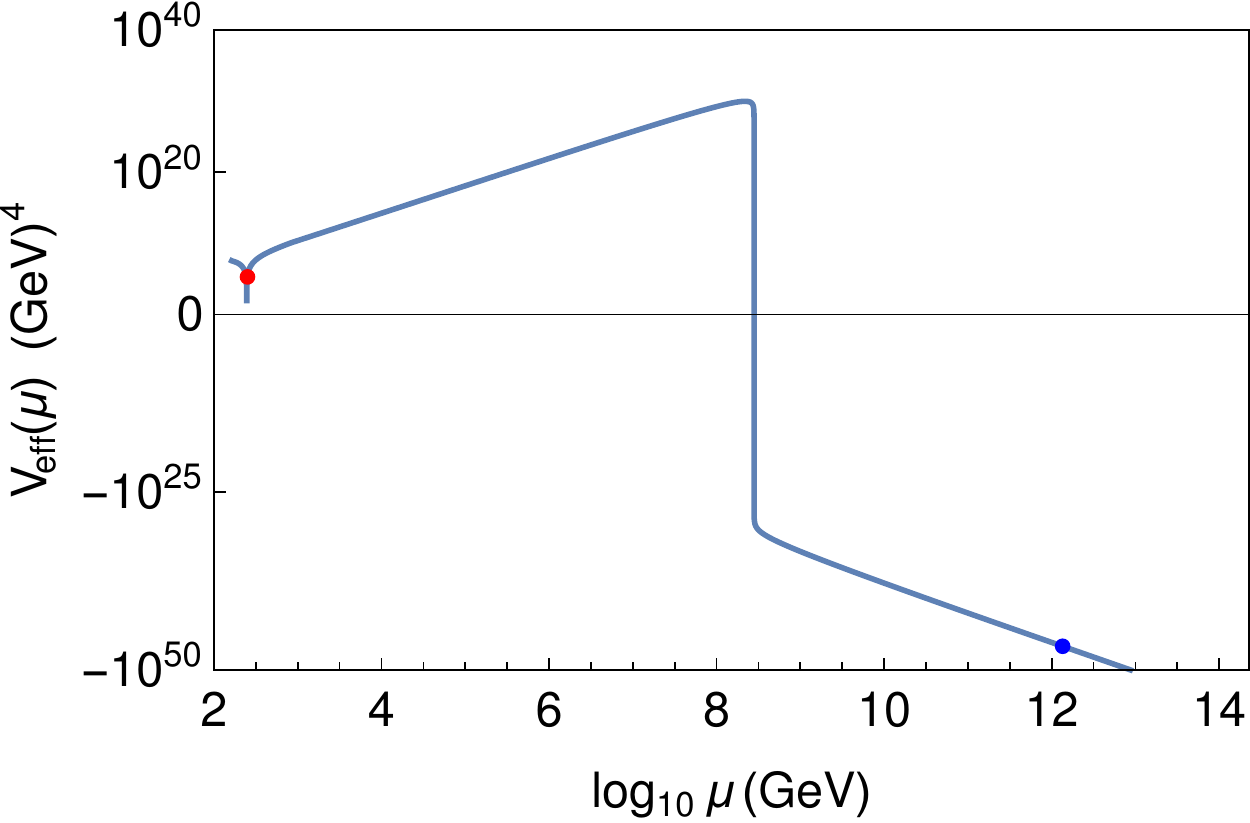} \hspace*{0.25cm}
    \includegraphics[width=0.4\textwidth]{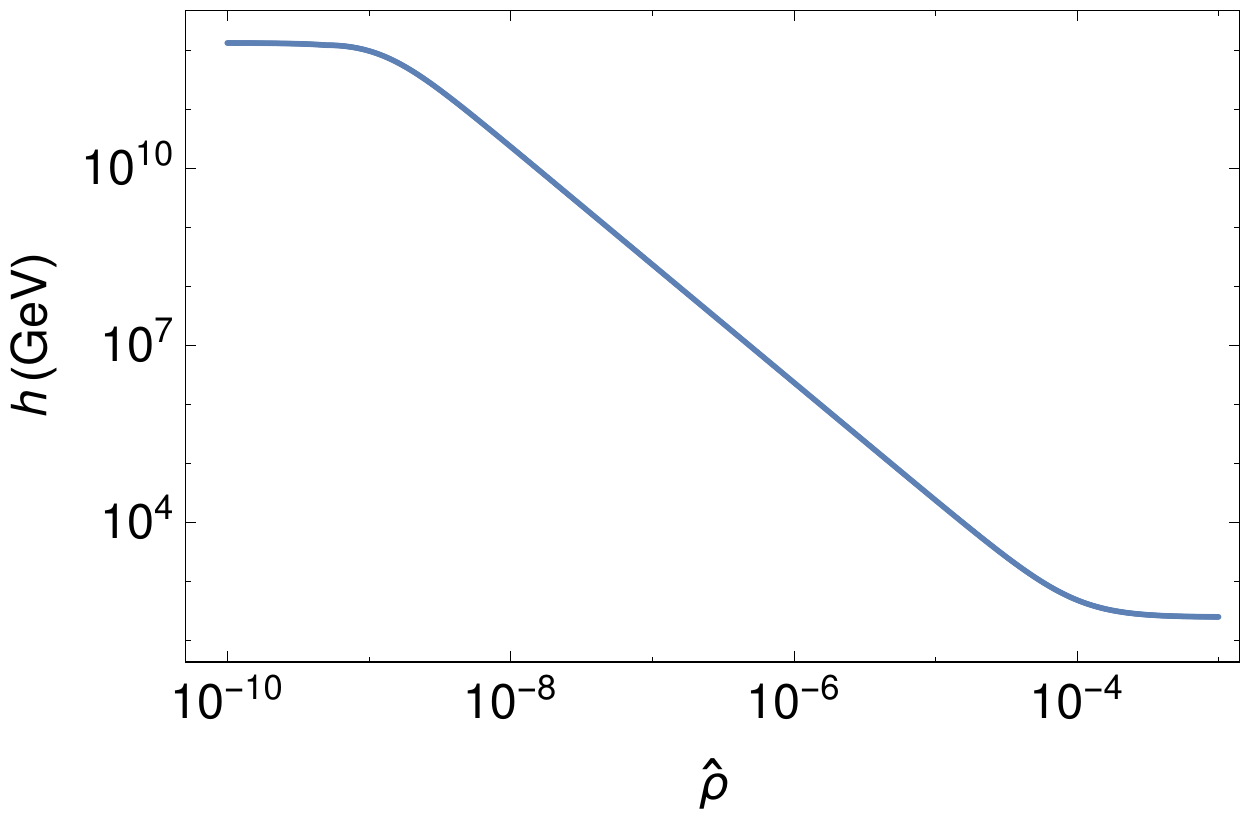}    
    \caption{For a VLL family with $M_{VL} = 10^3~$GeV and $\tilde y = 0.6$ (top row),
      $M_{VL} = 10^5~$GeV and $\tilde y = 0.57$ (middle row), and
       $M_{VL} = 10^7~$GeV and $\tilde y = 0.6$ (bottom row),
      the effective potential as a function of the field value $h=\mu$,
      and the bounce configuration $h_B(\hat\rho)$.
      The blue (red) dot shows the starting (ending) value of the bounce.
      \label{VLL-Tunl.FIG}}
  \end{center}
\end{figure}

For a VLL family with $M_{VL} = 10^3~$GeV and $\tilde y = 0.6$, the $\Veff (h\!\equiv\! \mu)$, and
the bounce configuration are shown in Fig.~\ref{VLL-Tunl.FIG} (top row). 
The $\Veff(\mu)$ is positive for smaller $\mu$, crosses zero at about $\mu \approx 10^{6.5}~$GeV, and is negative for larger $\mu$.
The blue dot shows the starting field value ($h_0$) of the bounce and the red dot the ending field value ($v$).
For this bounce configuration, we find $S_B = 472$ and $\Ptunl \sim 10^{-6}$.
This parameter-space point is thus acceptable as the tunneling probability into the true vacuum is sufficiently small
for us to understand why the electroweak vacuum has still not tunneled away into the true vacuum
within the age of the Universe.
That is, for this model with VLL present,
the probability of a true vacuum bubble having nucleated in our past light-cone is sufficiently small,
although this probability is much much larger than in the SM.
We see that the presence of a VLL increases the tunneling probability dramatically compared with the SM.
As another example, we consider a VLL family with $M_{VL} = 10^3~$GeV and $\tilde y = 0.61$,
for which we find $S_B = 422$ and $\Ptunl \sim 10^{17}$.
This large value implies that the EW vacuum could have tunneled into the true vacuum in our Hubble volume essentially with unit probability. 
Therefore, this parameter-space point can be considered severely disfavored.
These two examples also show that $\Ptunl$ is extremely sensitive to $\tilde y$, with a small change of $0.01$ in $\tilde y$ between the two cases
resulting in a change of $S_B$ of 50, which in turn results in a $\Ptunl$ 23 orders of magnitude different because of its exponential dependence on $S_B$. 
As another example, for a VLL family with $M_{VL} = 10^5~$GeV and $\tilde y = 0.57$, the bounce configuration is shown in
Fig.~\ref{VLL-Tunl.FIG} (middle row).
For this bounce configuration, we find $S_B = 498$ and $\Ptunl \sim 10^{-7}$ which is acceptable.
For a VLL family with $M_{VL} = 10^7~$GeV and $\tilde y = 0.6$, the bounce configuration is shown in
Fig.~\ref{VLL-Tunl.FIG} (bottom row). 
For this bounce configuration, we find $S_B = 500$ and $\Ptunl \sim 10^{-4}$ which is acceptable.

The hidden sector Higgs-portal dark matter model of Ref.~\cite{Gopalakrishna:2009yz} essentially behaves
like a VLL family considered above, for the following reason.
Although in the model of Ref.~\cite{Gopalakrishna:2009yz}, the VLF dark matter is a singlet and does not couple directly to the Higgs,
due to the Higgs mixing with a hidden-sector scalar, a coupling with the Higgs is induced with size
$\tilde y \equiv \kappa s_h$,
where the right-hand-side is in the notation of that paper and involve the parameters of that model.
As can be inferred from the analysis in Ref.~\cite{Gopalakrishna:2009yz},
we require $\tilde y \ll 1$ to keep the direct-detection rate small in order to honor experimental constraints.
Thus, from the results above, we infer that EW vacuum stability constraints are not too severe in such models. 

\begin{figure}
  \begin{center}
    \includegraphics[width=0.5\textwidth]{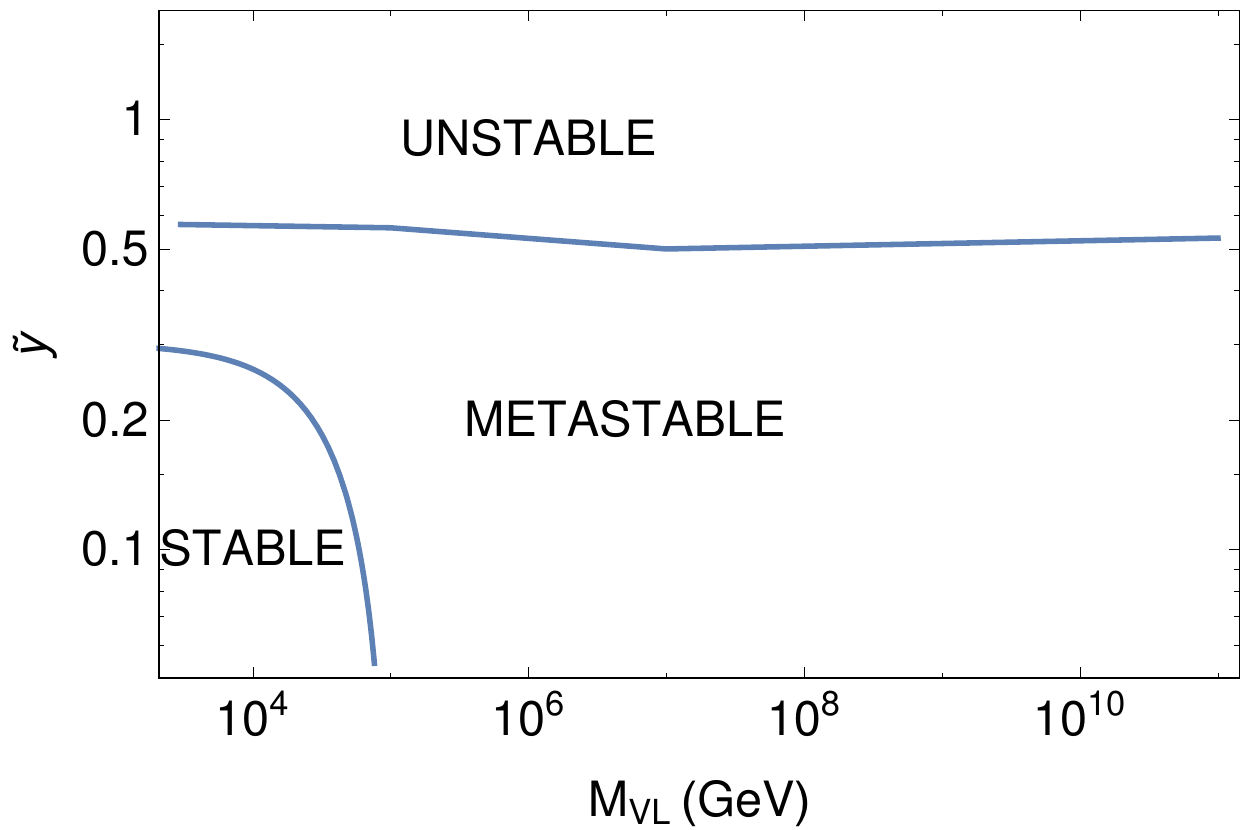} 
    \caption{With the addition of a VLQ family, the regions of stability, meta-stability and instability as a function of $M_{VL}$ (in GeV) and
      $\tilde y$.  
      \label{VLQstabReg.FIG}}
  \end{center}
\end{figure}

Next, we compute $S_B$ and $\Ptunl$ with a color triplet VLQ family with SM-like hypercharge assignment present,
consisting of an $SU(2)$ singlet VLQ with hypercharge $2/3$ and an $SU(2)$ doublet VLQ with hypercharge $1/6$ both present,
for various common mass $M_{VL}$ and various $\tilde y$.
With the addition of a VLQ family, in Fig.~\ref{VLQstabReg.FIG} we show the regions of stability, meta-stability and instability
as a function of $M_{VL}$ (in GeV) and $\tilde y$.
In the region marked ``stable'' the Higgs electroweak minimum is the absolute minimum and is discussed further in
Section~\ref{AbsltStab.SEC};
in the region marked ``metastable'' there is a lower minimum at large field values with $\Ptunl \lesssim {\cal O}(1)$,
and in the region marked ``unstable'' $\Ptunl \gtrsim {\cal O}(1)$. 
We find that for $\tilde y \gtrsim 0.5$ the $\Ptunl \gtrsim {\cal O}(1)$ quite independently of $M_{VL}$. 
This parameter-space leads to an unstable vacuum and we consider this region disfavored from the vacuum stability point of view.

\subsubsection{Second minimum in $V_{eff}$}

\begin{figure}
  \begin{center}
    \includegraphics[width=0.4\textwidth] {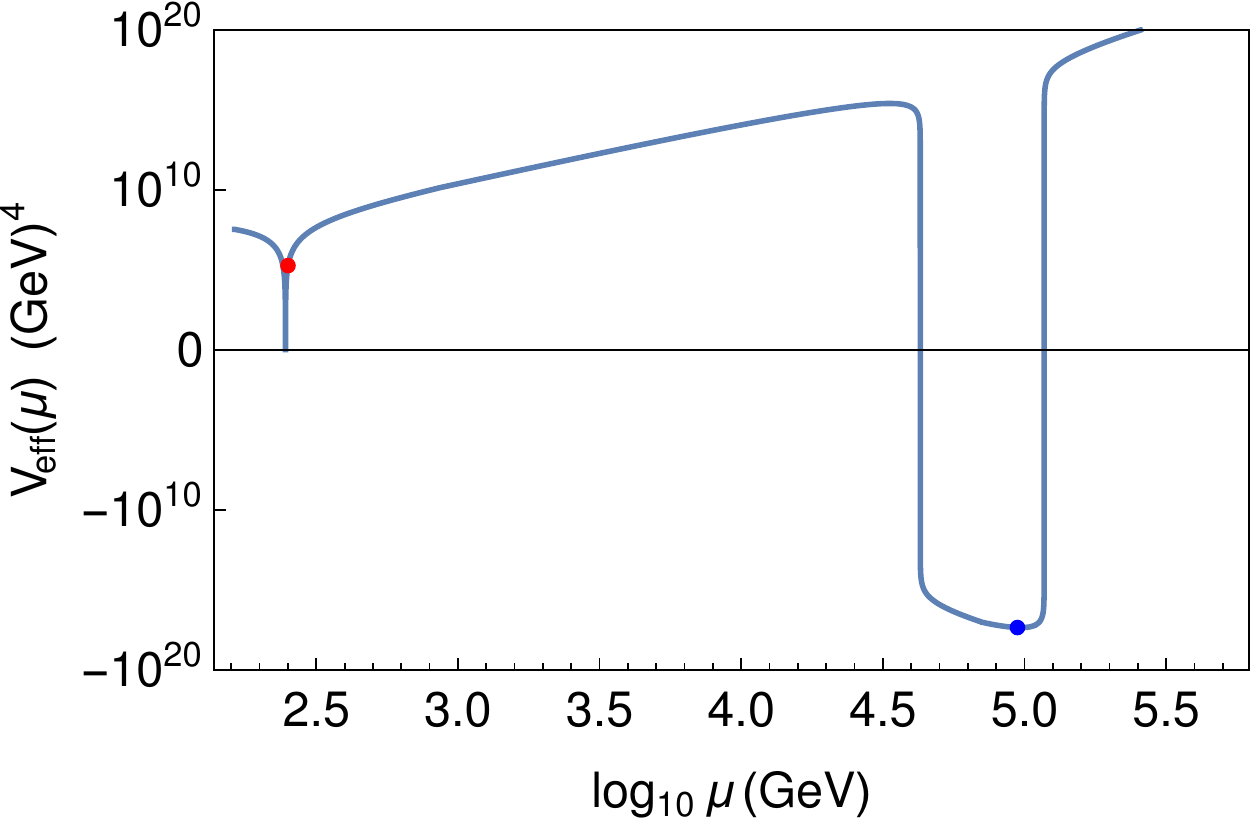} \hspace*{0.25cm}
    \includegraphics[width=0.4\textwidth] {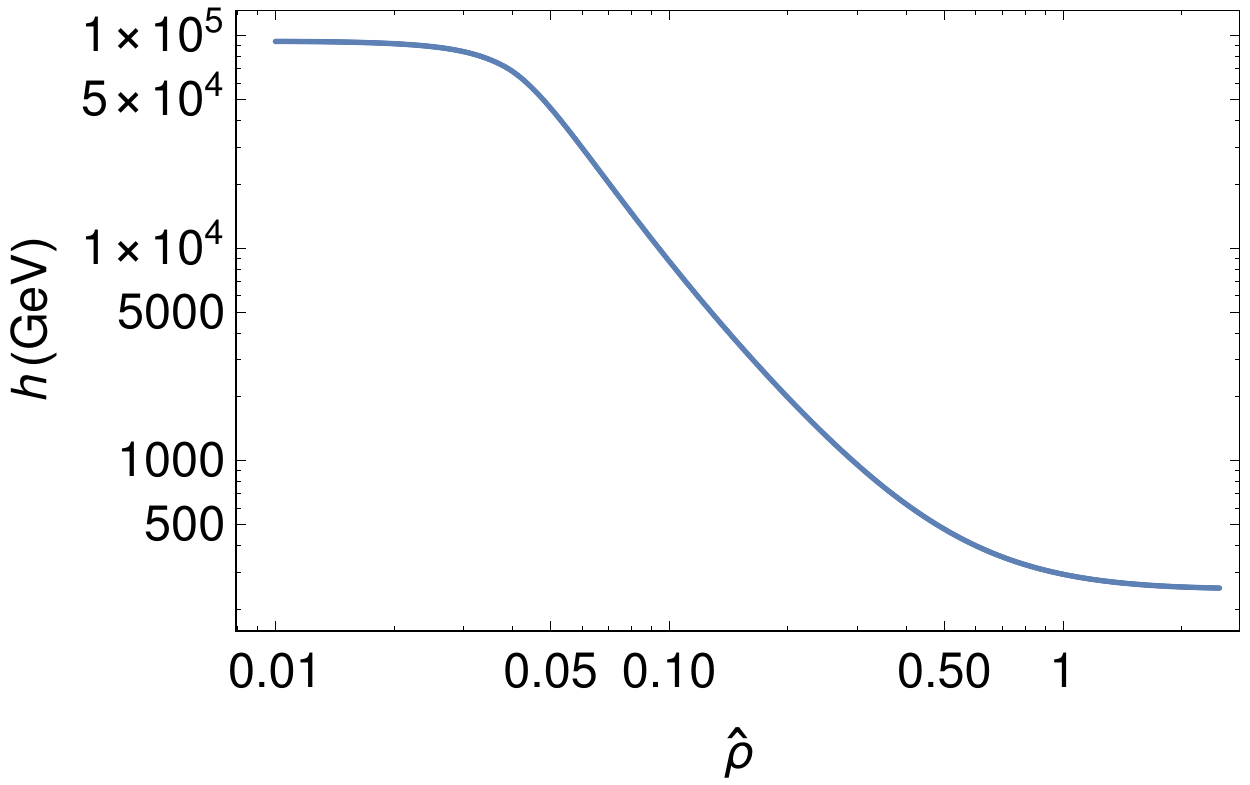}
    \caption{For a VLQ family with $M_{VL} = 3\times 10^3~$GeV and $\tilde y = 0.75$,
      with a second minimum in the effective potential, 
      the effective potential as a function of the field value $h=\mu$,
      and the bounce configuration $h_B(\hat\rho)$.
      The blue (red) dot shows the starting (ending) value of the bounce.
      \label{VLQPy0p75M3E3scm-Tunl.FIG}}
  \end{center}
\end{figure}

Thus far, we have investigated the situation when only the VLF is present and the effective potential
has only a minimum at $v$ and no second minimum at large field values but rather runs off in a bottomless manner.
If the VLF is accompanied by other states, presumably in a UV completion that it is a part of,
one can contemplate the possibility of the potential being turned around due to the contributions of the extra states and the appearance of a
second minimum at large field values.
We encode this possibility by adding a second minimum in the effective potential as shown in
Fig.~\ref{VLQPy0p75M3E3scm-Tunl.FIG} for the case of a VLQ family with $M_{VL} = 3\times 10^3~$GeV and $\tilde y = 0.75$.
The $\Veff (h\!\equiv\! \mu)$, and the
bounce configuration for this modified potential are shown in Fig.~\ref{VLQPy0p75M3E3scm-Tunl.FIG}. 
The $\Veff(\mu)$ is positive for smaller $\mu$, crosses zero at about $\mu \approx 10^{4.65}~$GeV and becomes negative,
obtains a minimum at about $\mu \approx 10^{5}~$GeV, crosses zero again and becomes positive for larger $\mu$.
The blue dot shows the starting field value ($h_0$) of the bounce and the red dot the ending field value ($v$).   
For this bounce configuration, we find $S_B = 3071$ and $\Ptunl \sim 10^{-1150}$, which is an incredibly
tiny tunneling probability and very comfortably acceptable.

The reason why this parameter-space point which was excluded if no second minimum was present as found earlier, 
is now allowed if a second minimum present is as follows. 
In the bounce configuration for this situation, the field value starts close to the second minimum,
and stays there for a substantial amount of time (i.e. $\rho$) near the minimum since $d\phi/d\rho$ is small there, 
and as a result the friction term starts reducing in significance due to its $1/\rho$ behavior.
Since the friction term becomes small, the field value can overcome the barrier and reach $v$.
(It can even overshoot $v$ leading to an imaginary solution if the initial value of the field is chosen too big.)
Therefore, the starting field value is much lesser compared to the earlier case without the second minimum,
and we find that the resulting $S_B$ is much larger and $\Ptunl$ much smaller, now allowing a parameter-space point that was excluded earlier.

\section{Absolute stability of the EW vacuum}
\label{AbsltStab.SEC}

We have seen that the EW vacuum is meta-stable in the SM
as there is a deeper minimum below the EW vacuum albeit shielded by a potential barrier,
and due to the tunneling probability being incredibly small, the life-time of the metastable vacuum is extremely large compared to the age of the Universe.
In Section~\ref{TUNL.SEC} we added VLFs and analyzed regions of $\tilde y$ and $M_{VL}$ parameter-space for which there again is a deeper minimum making the
EW vacuum metastable. We computed the tunneling probability and found that in some regions of parameter space, $\Ptunl$ is acceptably small
while in others it is unacceptably large.
In this section, we highlight VLF cases where the addition of VLFs makes the EW vacuum the global minimum, rendering it absolutely stable. 

Consider first adding some number of either $SU(3)$ singlet VLQs or doublet VLQs, but not both.
For instance, we showed in Section~\ref{RGERes.SEC}, Fig.~\ref{VLQsing.FIG},
that when 3, 4 or 5 $SU(2)$ singlet VLQs all with 3~TeV mass are added, $\lambda(h)$ never goes negative,
implying that the EW minimum is the global minimum and absolutely stable, unlike the SM situation.
The reason for this behavior is explained in detail in Section~\ref{RGERes.SEC}. 
As we show in Fig.~\ref{VLQdoub.FIG}, the same conclusion holds also when we add
one to four $SU(2)$ doublet VLQs with a 3~TeV mass, or one doublet with mass less than about $10^5$~GeV.
When both singlet and doublet VLFs are present, i.e. when a VLF family is added,
the situation changes since a Yukawa coupling ($\tilde y$) with the Higgs can be written down.
Nevertheless, when $\tilde y$ is small, the behavior is similar to the above two cases.
For a VLQ family with one singlet and one doublet VLQ added, as can be seen in Fig.~\ref{VLQ2211.FIG}, 
for a small $\tilde y = 0.1$ and for $M_{VL} \lesssim 10^5$~GeV the EW minimum becomes absolutely stable. 
Thus, as we see in these examples, the presence of either $SU(2)$ singlet VLQs, or doublet VLQs,
or a full family with a small enough $\tilde y$,
allows the intriguing possibility that the EW vacuum is rendered absolutely stable.

For example, the hidden-sector dark matter model in Ref.~\cite{Gopalakrishna:2017zku} contains a singlet VLQ mediating
loop-level couplings between the hidden-sector dark matter and the SM.
Such models can also be written down with a doublet VLQ.  
For proper choices of the number of VLQs and masses,
it is interesting that the Higgs vacuum could be absolutely stable in such models,
unlike in the SM in which it is meta-stable.

\section{Comparison with the analytical approximation of $S_B$}
\label{compSB.SEC}

Here we compare our numerical results for $S_B$ obtained in Sec.~\ref{tunProbNumRes.SEC}
with an analytical approximation developed in Refs.~\cite{Lee:1985uv,Arnold:1989cb}, 
which is, 
\beq
S_B^{\rm approx} = \frac{8 \pi^2}{3 (-\lambda(t))} \ ,
\label{SBapprox.EQ}
\eeq
where $t$ is a typical scale at which the bounce makes the transition from large field values to $v$.
This approximation can yield a reasonably good estimate of $S_B$
when the bounce transition happens at a fairly constant value of $\lambda(t)$,
i.e. when $h_0$ is close to where $\beta_\lambda(h_0) \approx 0$.
Furthermore, when $S_B$ is so large that errors due to the transition not happening at a constant $\lambda(t)$ are small
compared to $S_B$, this approximation yields a good enough estimate.
When these conditions are not realized, one has to be cautious in using the expression in Eq.~(\ref{SBapprox.EQ}).
We elaborate on this statement below with many examples. 

\begin{figure}
  \begin{center}
    \includegraphics[width=0.4\textwidth]{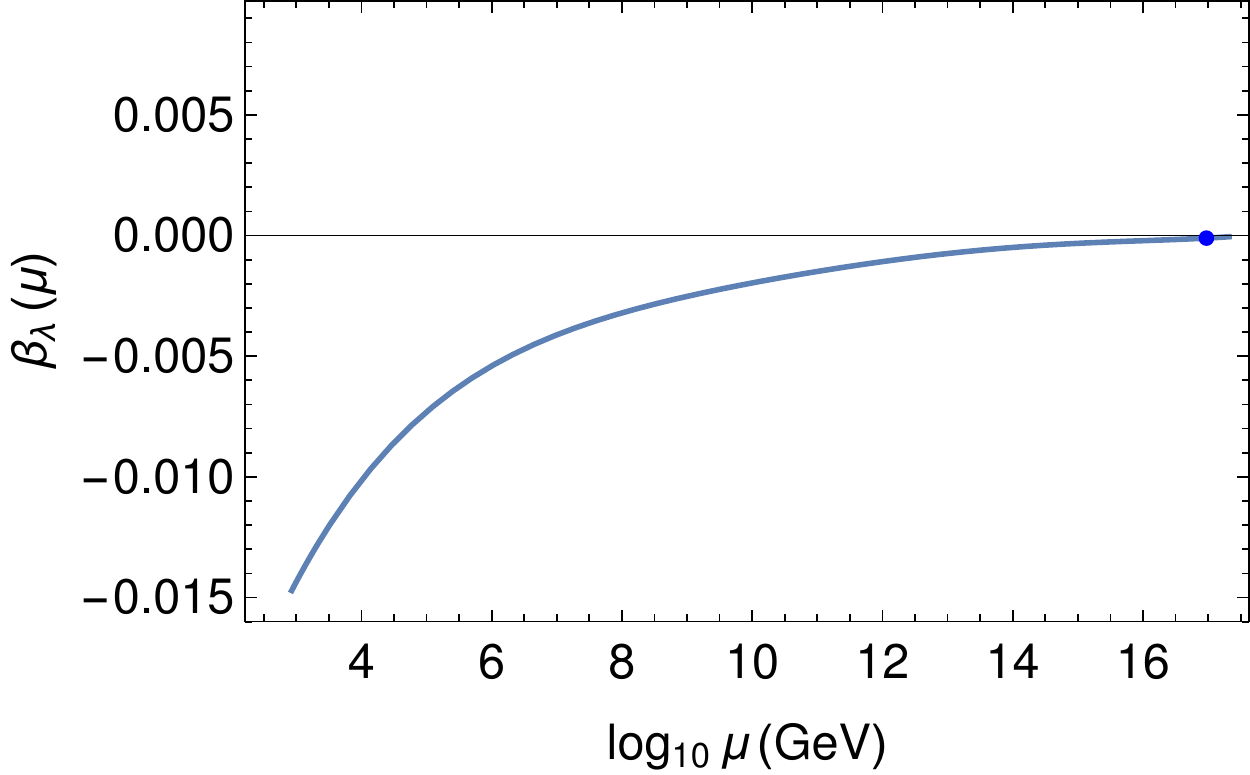} \hspace*{0.25cm}
    \includegraphics[width=0.4\textwidth]{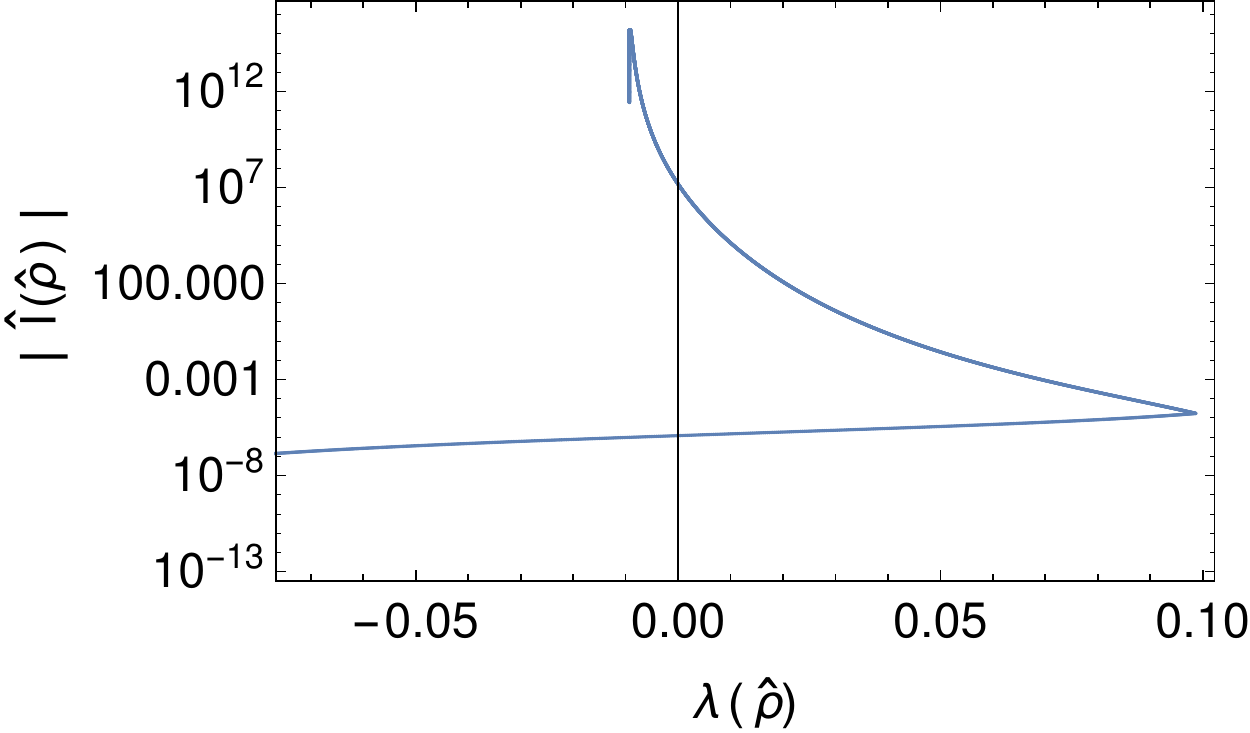} \\ \vspace*{0.25cm}
    \includegraphics[width=0.4\textwidth]{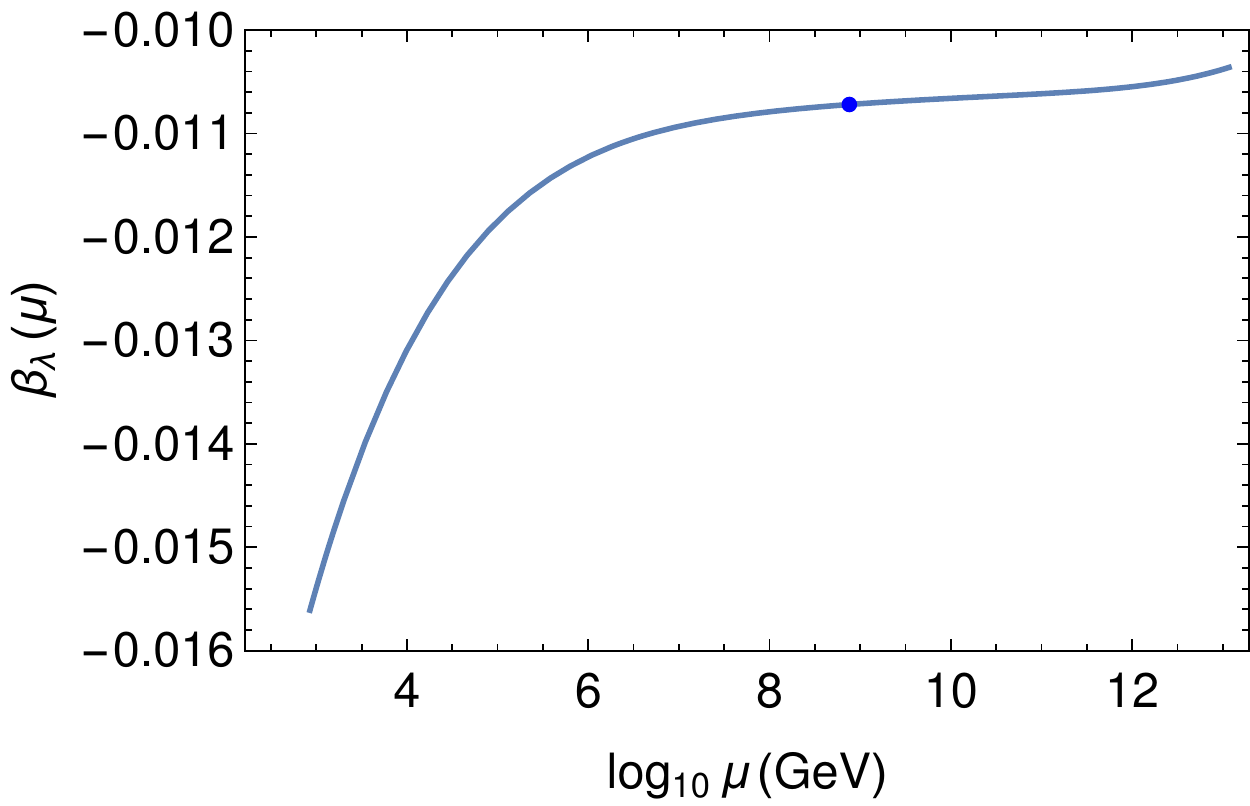} \hspace*{0.25cm}  
    \includegraphics[width=0.4\textwidth]{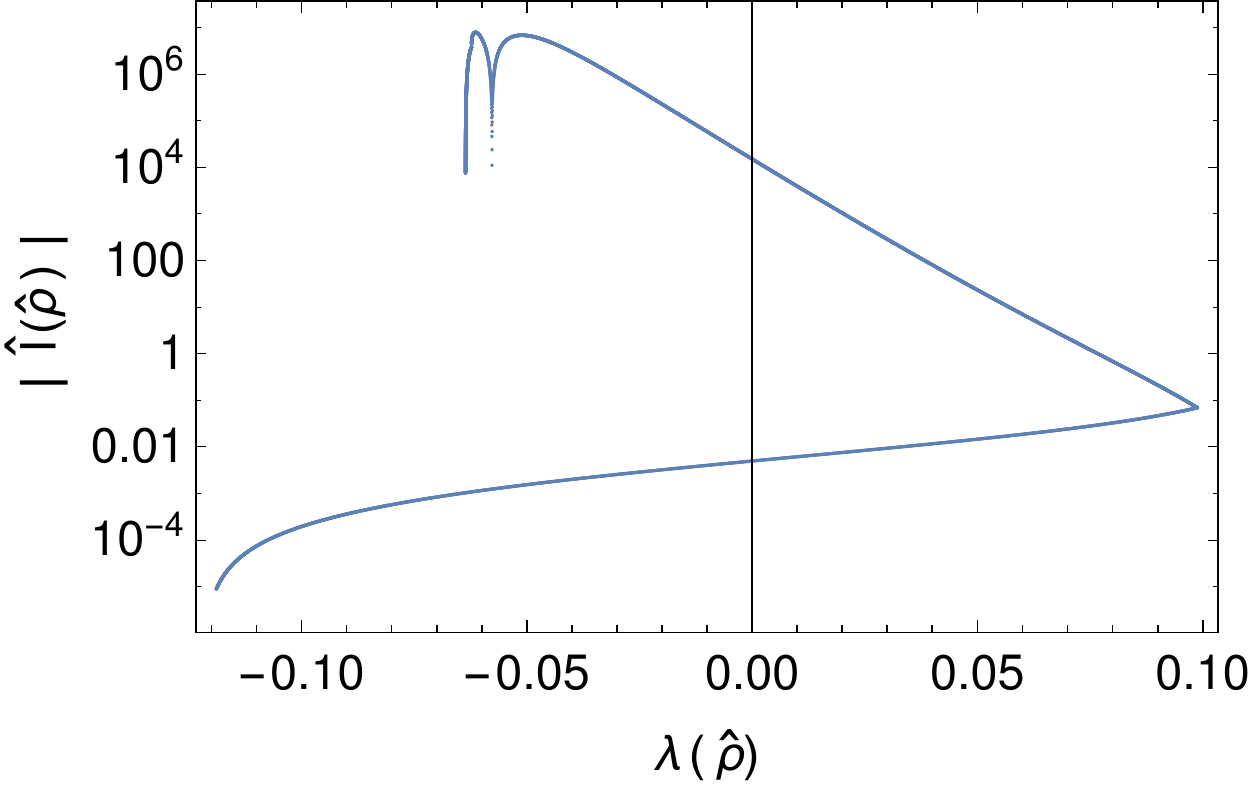} \\ \vspace*{0.25cm}
    \includegraphics[width=0.4\textwidth]{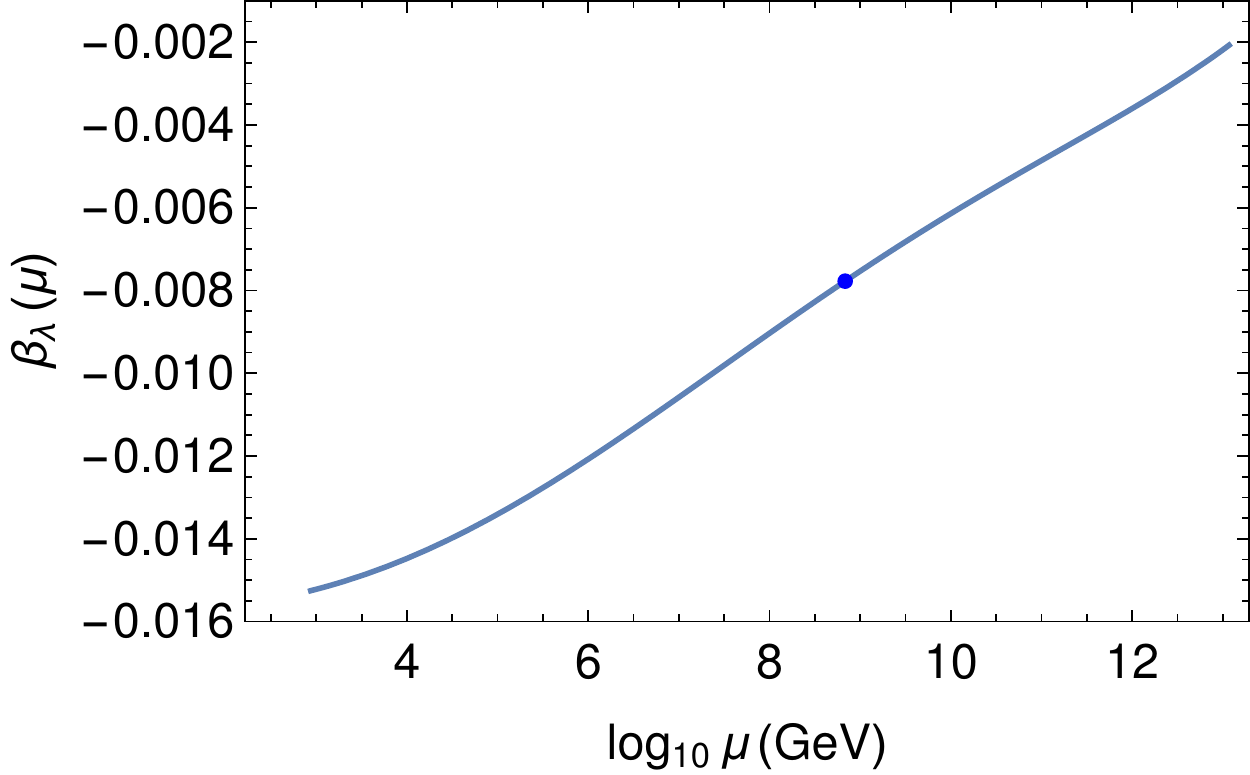}   \hspace*{0.25cm}
    \includegraphics[width=0.4\textwidth]{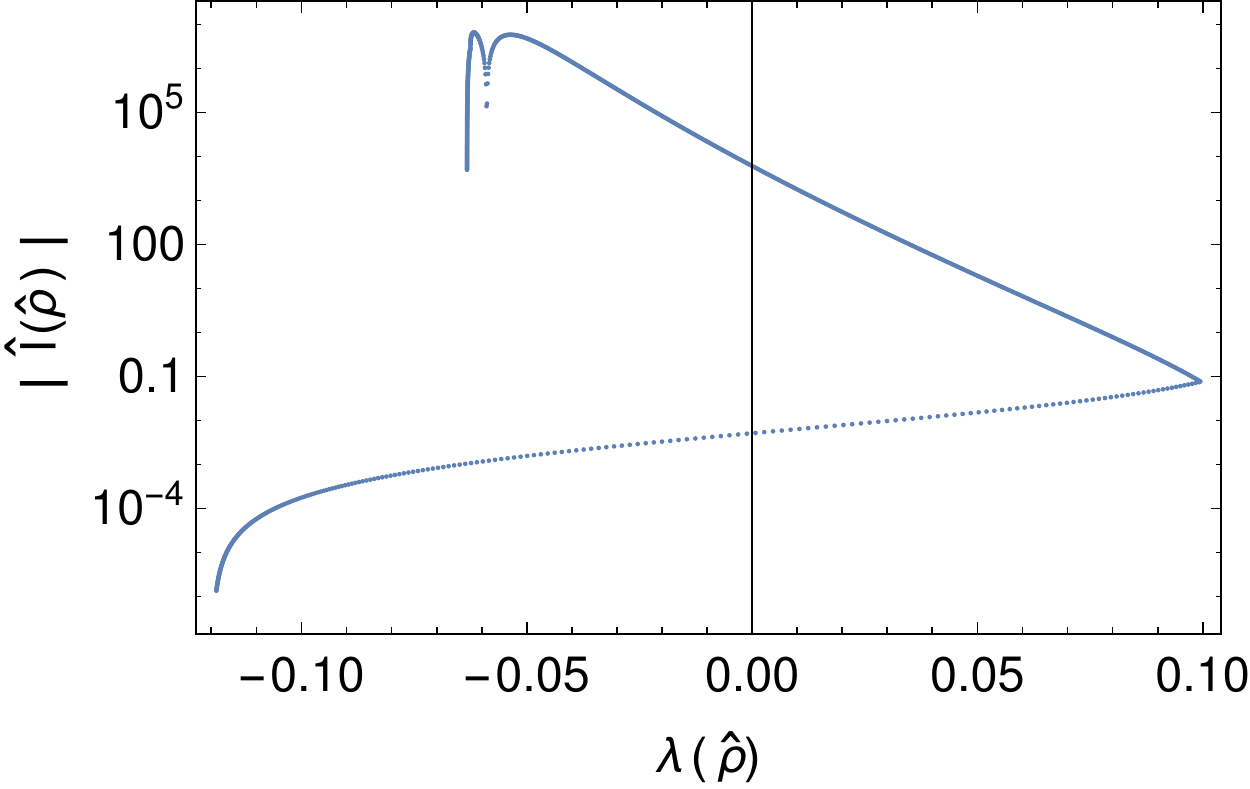} \\ \vspace*{0.25cm}
    \includegraphics[width=0.4\textwidth] {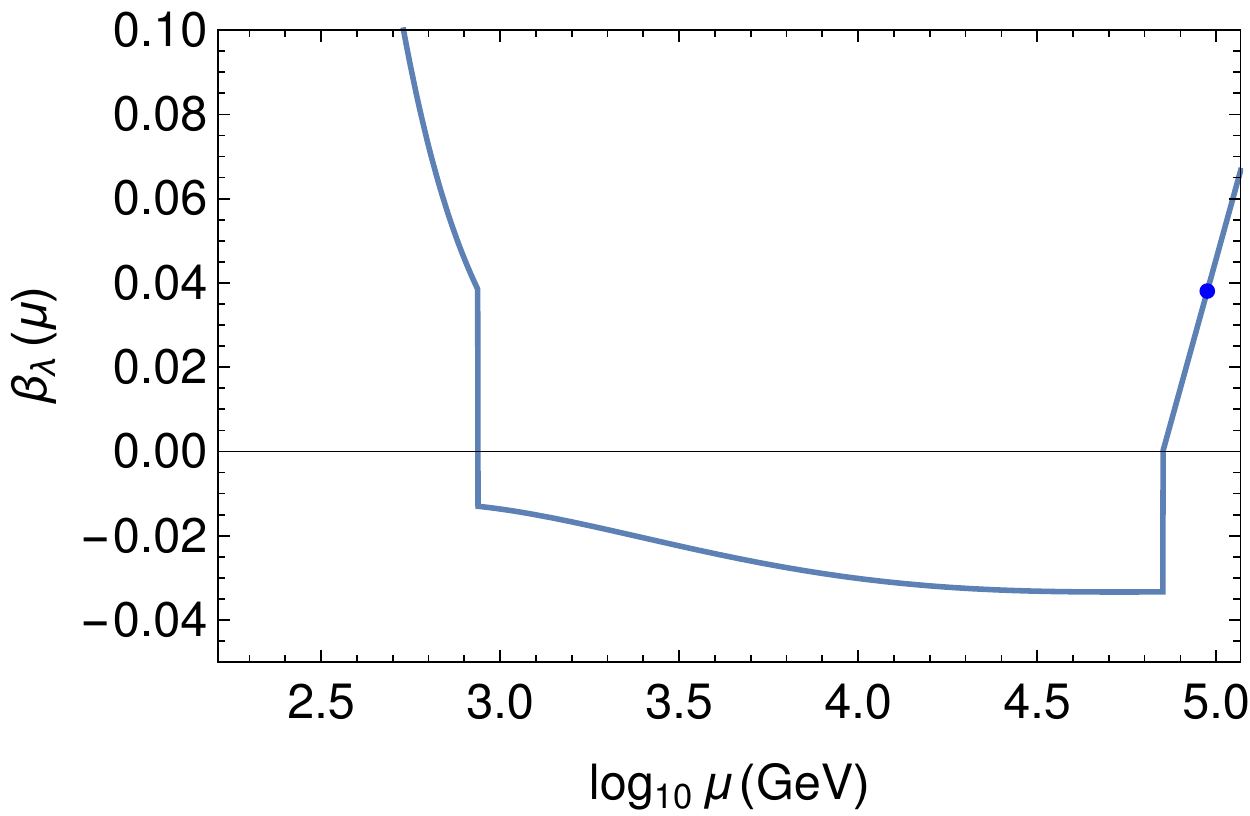}  \hspace*{0.25cm}    
    \includegraphics[width=0.4\textwidth]{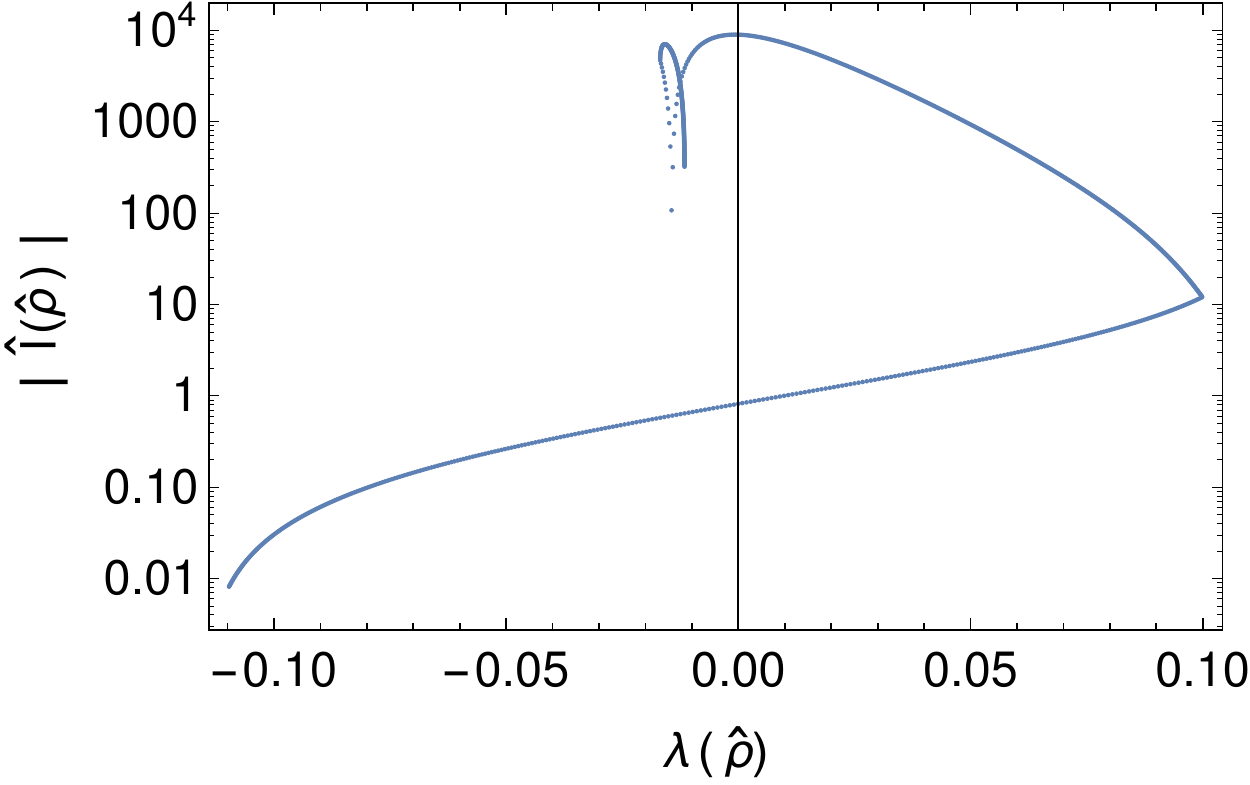}
    \caption{The $\beta_\lambda(\mu)$ as a function of the field value $h(\rho)=\mu$ (left column) and the integrand
      of the bounce action integral Eq.~(\ref{SBIntegral.EQ}) vs. $\lambda(h(\rho))$ (right column),
      for the SM ({\it cf} Fig.~\ref{SMTunl.FIG}) (top row),
      and the remaining for the different cases with a VLF family as follows:
      a VLL family with $M_{VL} = 10^3~$GeV and $\tilde y = 0.6$ ({\it cf} Fig.~\ref{VLL-Tunl.FIG}) (top row),
      a VLQ family with $M_{VL} = 3\times 10^3~$GeV and $\tilde y = 0.57$, and, 
      a VLQ family with $M_{VL} = 3\times 10^3~$GeV and $\tilde y = 0.75$ with a second minimum ({\it cf} Fig.~\ref{VLQPy0p75M3E3scm-Tunl.FIG}).
      The blue dot shows the starting value $h_0$ for the bounce configuration. 
    }
    \label{SBIgnd.FIG}
  \end{center}
\end{figure}

In Fig.~\ref{SBIgnd.FIG} in the left column we show $\beta_{\lambda}(\mu)$ vs. $\mu$ where $\mu \equiv h(\rho)$. 
In the right column we show the (absolute value of) integrand of Eq.~(\ref{SBIntegral.EQ})
made dimensionless by multiplying the integrand by $1/m_t^4$ and denoted as $|\hat{I}(\hat\rho)|$,
vs. $\lambda(h(\rho))$, with $\rho$ being the parameter (not shown).
As shown in the topmost row in Fig.~\ref{SBIgnd.FIG}, for the SM,
it is evident that most of the contribution to the integral comes from when $\lambda$ takes a specific value.
For the SM, we can compare the $S_B$ computed numerically in Sec.~\ref{tunProbNumRes.SEC}, which is $2866$,
with the $S_B$ got from the approximation in Eq.~(\ref{SBapprox.EQ}) with the $t$ taken to be at the scale at which
$\beta_\lambda = 0$ where $\lambda = -0.009$, which gives $S_B^{\rm approx} = 2848$.
This is in excellent agreement with our numerical computation of $S_B$, and as discussed earlier, this is because $\beta_\lambda = 0$
does get satisfied for the SM, presenting a natural choice for $t$.
That this approximation works is also borne out by the plot showing $|\hat{I}(\hat\rho)|$ for the SM,
where most of the contribution to the $S_B$ integral is indeed coming for $\lambda = -0.009$,
where the bounce spends most of its time ($\hat\rho$).
Indeed, Eq.~(\ref{SBapprox.EQ}) was put-forth for the SM, where it can be safely applied. 

As we see from the last three rows in Fig.~\ref{SBIgnd.FIG}, with VLF present,
$\beta_\lambda$ is not close to zero anywhere, and thus there is no clear choice of $t$ that is suggested. 
In such a situation, we cannot use Eq.~(\ref{SBapprox.EQ}) but have to compute $S_B$ numerically. 
Indeed as the $|\hat{I}(\hat\rho)|$ for these cases show,
the integral gets its contributions for a range of $\lambda$.
These show the inadequacy of the approximate formula, and that a numerical evaluation is necessary. 
We have therefore computed the bounce EOM numerically and the bounce action for it,
from which we computed the tunneling probability.

\section{Conclusions}
\label{Concl.SEC}

We study the stability of the electroweak vacuum in the presence of new vector-like fermions. 
We work with the renormalization group improved Higgs effective potential,
identifying the Higgs field value $h\equiv \mu$. 
We first review the computation of the beta-functions in the SM, paying particular attention to the
SM fermion contributions.
We use dimensional regularization for our computation.
We then derive the VLF contributions to the 1-loop beta-functions which can be applied to
various $SU(3)$ and $SU(2)$ representations, namely VLQs and VLLs.
We apply this to a few example cases with singlet VLFs, doublet VLFs,
and a family consisting of one doublet VLF and one singlet VLF coupled to the Higgs via the Yukawa coupling $\tilde y$.
We include the significant 2-loop contributions in the beta-functions. 
We numerically integrate the RGE to determine the scale at which $\lambda(\mu)$ becomes zero and goes negative.

The Higgs effective quartic-coupling $\lambda(\mu)$ becoming negative signals that the EW vacuum is a false vacuum
and is unstable, 
and can tunnel away quantum mechanically via barrier penetration to (large) field values that have a lower effective potential.
We compute the probability $\Ptunl$ that the EW vacuum would have tunneled away by a true-vacuum bubble nucleating in our
Hubble four-volume. 
Computing $\Ptunl$ requires computing the bounce configuration in Euclidean space-time,
and the value of the Euclidean action $S_B$ for the bounce configuration. 
We solve the bounce configuration EOM numerically and compute $S_B$ for it. 

We compare our numerical evaluation with the approximation commonly used for $S_B$,
which is written in terms of $\lambda$ at a single scale where $\beta_\lambda(\mu)$ is approximately zero.
This is because the bounce transition is mostly completed when $\lambda(\mu)$ has this value. 
For the SM, there is such a scale which is about $10^{16}$~GeV, and we verify
by comparing with our numerical evaluation that the approximation is perfectly adequate. 
When VLFs are present, there is no scale at which $\beta_\lambda(\mu)$ is close to zero,
and so the approximation cannot be applied.
A numerical evaluation is then required, which we resort to.

We take example cases where a single VLL or VLQ family is added and show the bounce transition, compute $S_B$ for it
and obtain $\Ptunl$.
We find that $\Ptunl$ is extremely sensitive to $\tilde y$ as it exponentially depends on $S_B$. 
Interestingly, we find that for some VLF representations and parameters, adding only singlet VLFs,
a doublet VLFs, or a full family with a small enough $\tilde y$,
$\lambda$ stays positive to arbitrarily large scales, i.e. the EW vacuum is rendered absolutely stable,
unlike in the SM in which it is meta-stable.
For other parameters, adding VLFs still keeps the EW vacuum meta-stable,
either with a larger $\Ptunl$ than in the SM, or a smaller $\Ptunl$. 

In summary, our work here helps us get an idea of what the impact of VLFs is on the stability of the
Higgs electroweak vacuum.

\medskip
\noindent {\it Acknowledgements:}
We thank Romesh Kaul for valuable discussions.

\appendix

\section{SM $\beta$-functions (fermion contributions)}
\label{betaFcnSM.SEC}

Here, we review the calculation of the fermion contributions to the 1-loop $\beta$-functions in the SM so that we can extend this to include VLF contributions in the next section. 
Since we are interested in field values $h$ much larger than the particle masses, we neglect the particle masses. 

We expand the SM Lagrangian density shown in Eq.~(\ref{LSMth.EQ}) by writing the Higgs doublet as
$H=(1/\sqrt{2})((\phi^1 + i \phi^2)\ \ (v + h + i\phi^3))^T$,
where $v\approx 246$~GeV is the EW VEV, $h$ is the physical Higgs boson and $\phi^i$ are the Goldstone bosons.
The Lagrangian density in terms of the bare fields $\{\tilde{h}, t_0\}$ and bare coupling $y_0$ is
\bea
    {\cal L} &\supset& \bar{t}_0 i \slashchar{D} t_0
                      + \half \left[ (D_\mu \tilde{h})^2 + (D_\mu \phi^i_0)^2 \right] \nonumber \\
    && - \left[ \frac{y_0}{\sqrt{2}} \left(  \bar{t}_0 (\tilde{h}+i\gamma^5\phi^3_0) t_0 - \bar{t}_0 (\phi^1_0 + i \phi^2_0) P_L b_0 \right) + h.c. \right] \ , 
\eea
where for notational brevity we denote $y_t$ just as $y$, and the covariant derivatives are in the usual notation.
In terms of the renormalized fields and counter-terms, we have for the $\{t,h\}$ sector
\bea
    {\cal L} &\supset & \bar{t} i \slashchar{D} t + \half (D_\mu h)^2
    - \left( \frac{y}{\sqrt{2}} h \bar{t} t  + h.c. \right) \nonumber \\ 
    && + \half (Z_h - 1) (D_\mu h)^2 + (Z_{t_{L,R}} - 1) \bar{t}_{L,R} i \slashchar{D} t_{L,R}
    - \left[\frac{y}{\sqrt{2}} \left(Z_y \sqrt{Z_h Z_{t_{L}} Z_{t_{R}}} - 1 \right) \, h \bar{t} t  + h.c. \right]
    \ ,
\eea
where the renormalized fields $h,t$ are defined by
$\tilde{h} = \sqrt{Z_h} h$, ${t_0}_{L,R} = \sqrt{Z_{t_{L,R}}} t_{L,R}$,
and the renormalized Yukawa coupling $y$ is defined by $y_0 = Z_y y$.
Expanding as a perturbation series in $y$, we define $a$'s to leading order as
$(Z_h-1) \equiv a_h y^2/2 $, $(Z_{t_{L,R}}-1) \equiv a_{t_{L,R}} y^2/2$, $(Z_y - 1) \equiv a_y y^2/2$,
and we also define 
$(\hat Z_y - 1) \equiv ( Z_y \sqrt{Z_h Z_{t_{L}} Z_{t_{R}}} - 1 ) = (a_y + a_h/2 + a_{t_L}/2 + a_{t_R}/2) y^2/2 \equiv \hat{a}_y y^2/2$.
Similarly the renormalized Lagrangian density for the Goldstone fields can be written down.

The Feynman vertices in momentum-space are as follows:
\begin{itemize}
\item[] propagator $(t_{L,R}t_{L,R})(p)$ is $i\slashchar{p}/(p^2 + i \epsilon)$
        with the counter-term $i a_{t_{L,R}} \slashchar{p} y^2/2$,
\item[] propagator $(hh)(p)$ is $i/(p^2 + i \epsilon)$
        with the counter-term $i a_h p^2 y^2/2$,
\item[] Yukawa coupling $ht \bar{t}$ is $-iy/\sqrt{2}$
  with the counter-term $-i \hat{a}_y y^3/(2\sqrt{2})$,
\item[] for the Goldstone bosons, the Yukawa coupling $\phi^3 t \bar{t}$ is $-y\gamma^5/\sqrt{2}$ with the counter-term $\hat{a}_y y^3 \gamma^5/(2\sqrt{2})$; 
  and $\phi^{1,2} b \bar{t}$ is $iyP_L(1,i)/\sqrt{2}$ with the counter-term $-i \hat{a}_y y^3 P_L(1,i)/(2\sqrt{2})$.    
\end{itemize}

We give next the one-loop corrections involving the $t$, which we compute using dimensional regularization.
We present only the SMF $t$ contributions,
since our goal is to use these to derive the VLF contributions to the $\beta$-functions later.

The 1-loop correction to the Higgs 2-point function (wave-function renormalization of the $h$) due to the fermion is
$i \Sigma_h(p_h) = i (N_c y^2/(8 \pi^2)) p_h^2 \left( 1/\epsilon - \ln{p_h} - i \pi/2 + \ln\sqrt{4\pi} - \gamma/2 + {\cal O}(\epsilon)  \right)$,  
after including a factor of $(-1)$ for the fermion loop, where $\gamma \approx 0.5772$ is the Euler-Mascheroni constant.
To cancel this divergence, we fix the counter-term from the condition
$i\Sigma_h(p_0) + i a_h p_0^2 y^2/2 = 0$ at the subtraction scale $p_0$.
As mentioned above, since our goal is to derive the $t$ contributions to the $\beta$-functions,
we show only the $y$ dependent terms in the counter-terms, and we omit other terms.  
This yields
$a_h = - N_c/(4 \pi^2) \left( 1/\epsilon - \ln{p_0} - i \pi/2 + ... \right)$, 
and we have 
\beq
i \Sigma_h(p_h)  + i a_h \frac{y^2}{2} p_h^2 = -i \frac{N_c y^2}{8\pi^2} p_h^2 \ln{\left(\frac{p_h}{p_0}\right)} \ .
\label{Sigmah.EQ}
\eeq
The 1-loop correction to the fermion 2-point function (wave-function renormalization of the $t_{L,R}$)
proportional to $y$ is
$i \Sigma_{t_{L,R}}(p_t) = i y^2/(32\pi^2) \slashchar{p}_t \left( 1/\epsilon - \ln{p_t/2} + \ln\sqrt{4\pi} - \gamma/2 + {\cal O}(\epsilon) \right)$,
and to cancel the divergence we fix the counter-term from the condition
$i\Sigma_t(p_0) + i a_t \slashchar{p}_0 y^2/2 = 0$, 
which yields
$a_{t_{L,R}} = -1/(16\pi^2) \left( 1/\epsilon - \ln{p_0/2} + \ln\sqrt{4\pi} - \gamma/2 \right)$,
and we have 
$i\Sigma_{t_{L,R}}(p_t) + i a_{t_{L,R}} \slashchar{p}_t y^2/2 = -i y^2/(32\pi^2) \slashchar{p}_t \ln{p_t/p_0}$. 
The vertex 1-loop correction proportional to $y$ is
$$i V(p_h) = i y^3/(16\sqrt{2} \pi^2) \left( 1/\epsilon - \ln{p_h/2} + \ln\sqrt{4\pi} - \gamma/2 + 1/2 + {\cal O}(\epsilon) \right) \ ,$$ 
where we take the Higgs momentum as $p_h$, and the fermion momenta as $-p_h/2$ and $p_h/2$.
To cancel the divergence, we fix the vertex counter-term from the condition
$iV(p_0) - i\hat{a}_y = 0$,
which yields
$iV(p_h) - i \hat{a}_y y^3/(2\sqrt{2}) = -i y^3/(16\sqrt{2}\pi^2) \ln{p_h/p_0}$, 
and we have 
$$a_y = (2N_c + 3)/(16\pi^2) \left( 1/\epsilon - \ln{p_0/2} + \ln\sqrt{4\pi} - \gamma/2 + 1 - \ln{2}/(2N_c + 3) \right) \ .$$ 

We discuss next the Goldstone boson contributions proportional to $y$. 
Starting with the self-energy corrections, we have  
the $\phi^3$ contribution to $\Sigma_{t_{L,R}}$ is equal to the $h$ contribution,
the $\phi^{1}$ and $\phi^{2}$ contributions to $\Sigma_{t_{R}}$ are equal to the $h$ contribution, and, 
the $\phi^{1}$ and $\phi^{2}$ contributions to $\Sigma_{t_{L}}$ are proportional to $y_b$ which we drop and take to be zero. 
Turning next to the vertex corrections, we have 
the $\phi^3$ contribution to the $h t_L \bar{t}_R$ vertex ($V_{LR}$) is negative of the $h$ contribution to this vertex, 
the $\phi^3$ contribution to the $h t_R \bar{t}_L$ vertex ($V_{RL}$) is again negative of the $h$ contribution to this vertex, and, 
the $\phi^{1,2}$ contribution to $V_{LR,RL}$ is proportional to $y_b$ and hence we take it to be zero.

One way to extract the $\beta$-function is from the divergent part of the bare coupling.~\footnote{We briefly summarize here
   the method to obtain the $\beta$-function from the bare coupling, following 't\!~Hooft's method as described in Ref.~\cite{Weinberg:1996kr}. 
   With $\kappa_0$ the bare-coupling, we write in $d=4-\epsilon$ dimensions,
   $\kappa_0 \mu^{-\Delta(d)} = \kappa(\mu,d) - b(\kappa(\mu,d))/\epsilon$, $\Delta(d)\equiv \Delta - \rho \epsilon$,  
   $\kappa_0$ being $\mu$ independent, and $\kappa(\mu,d)$ is the renormalized coupling.
   Then, we write $d\kappa(\mu,d)/d\ln{\mu} \equiv \beta_\kappa - \alpha_\kappa \epsilon$ with $\beta_\kappa$ being the $\beta$-function,
   and by matching powers of $\epsilon$, we obtain
   $\beta_\kappa = -\Delta \kappa - \rho b + \rho \kappa \partial b/\partial \kappa$.
   We generalize this to a system of many couplings $\kappa_i$ by writing
   ${\kappa_0}_i \mu^{-\Delta_i(d)} = \kappa_i(\mu,d) - \sum_j b_{ij}(\kappa(\mu,d))/\epsilon$ with $\Delta_i(d)\equiv \Delta_i - \rho_i \epsilon$.
   Then, the $\beta$-functions are $\beta_{\kappa_i} = -\Delta_i \kappa_i - \sum_j [\rho_i b_{ij} - (\partial b_{ij}/\partial \kappa_j) \rho_j \kappa_j]$.
   For the couplings encountered here, we have: $\Delta_y = 0$, $\rho_y = -1/2$; $\Delta_\lambda = 0$, $\rho_\lambda = -1$;
   and $\Delta_{g_a} = 0$, $\rho_{g_a} = -1/2$ (for $a=\{1,2,3\}$). 
}
From the contributions computed above, we find the contributions proportional to $y$ to be  
$y_0 \supset y + (y^3/(16\pi^2)) ((3 + 2N_c)/2) (1/\epsilon)$,
from which we obtain the fermionic contribution to $\beta_y$ as 
\beq
\beta_y \supset \frac{y^3}{16\pi^2} \left[ \frac{(3 + 2N_c)}{2} \right] \ ,
\label{betaySM.EQ}
\eeq
after including the $t,h,\phi^{1,2,3}$ contributions.
This is in agreement with the results in Refs.~\cite{Cheng:1973nv,Buttazzo:2013uya}, for example. 
Interestingly, the $\phi^{1,2,3}$ contribute zero after including all their contributions. 

The $y g_3^2$ and $y g_2^2$ contributions to $\beta_y$ can be written as~\cite{Buttazzo:2013uya}
\beq
\beta_y \supset \frac{y}{16 \pi^2} \left( - 8 g_3^2 - \frac{9}{4} g_2^2 \right) \ , 
\label{betayg3ng2SM.EQ}
\eeq
which are included in Eq.~(\ref{RGEbeta1ygaSM.EQ}).
To derive the $y g_1^2$ contribution, we start by extracting the relevant Feynman rules for the hypercharge gauge boson $B_\mu$ interactions. 
With all momenta going into the vertex,
with $Y_{L,R}$ being the hypercharges of $\psi_{L,R}$ and $Y_H = 1/2$ being the hypercharge of the Higgs doublet,
we have the Feynman rules: 
\begin{itemize}
\item[] $\phi^3(p_3) h(p_h) B_\mu$: $-g' Y_H (p_3^\mu - p_h^\mu)$; \ \ \ \ $hhB_\mu B_\nu$: $2 i g'^2 Y_H^2 g_{\mu\nu}$;  \ \ \ \ $hB_\mu B_\nu$ : $2 i g'^2 Y_H^2 v g_{\mu\nu}$;
\item[] $\psi_{L,R} \bar\psi_{L,R} B_\mu$: $ig' Y_{L,R} \gamma^\mu$.   
\end{itemize}
Computing the $B_\mu$ contribution at 1-loop order in the 't\!\!~Hooft-Feynman $\xi\!\! =\!\! 1$  gauge,
we obtain the following divergent pieces:
the $\psi_R \bar\psi_L h$ vertex correction due to $B_\mu$ exchange gives $iV^{(B_\mu)} \supset -i 8 Y_L Y_R y g'^2/(\sqrt{2}\, 16\pi^2 \epsilon)$; 
the Higgs 2-point function correction due to $\phi^3 \-- B_\mu$ exchange gives $i\Sigma_h^{(B_\mu)} \supset -i 4 g'^2 Y_H^2 p_h^2/(16\pi^2 \epsilon)$; and 
the fermion 2-point function corrections due to $B_\mu$ exchange gives $i\Sigma_{\psi_L,\psi_R}^{(B_\mu)} \supset i 2 g'^2 Y_{L,R}^2 \pslash/(16\pi^2 \epsilon)$. 
We include in the counter-terms a piece to cancel these divergences at the subtraction scale $p_0$, i.e.,
$i(Z_h-1) p_0^2/2 \supset - i\Sigma_h^{(B_\mu)}(p_0)$;
$i(Z_{\psi_L,\psi_R}-1) \pslash_0 \supset - i\Sigma_{\psi_L,\psi_R}^{(B_\mu)}(p_0)$; and
$i(y/\sqrt{2})(\hat Z_y - 1) \supset iV^{(B_\mu)}$. 
From these we determine $(Z_y - 1) = -g'^2(8 Y_L Y_R + 4 Y_H^2 - Y_L^2 - Y_R^2) /(16\pi^2\epsilon)$.
Thus, since the bare coupling is $y_0 = Z_y y$, we get the contribution
\beq
\beta_y \supset \frac{-y g'^2}{16 \pi^2} \left( 8 Y_L Y_R + 4 Y_H^2 - Y_L^2 - Y_R^2 \right)
           = \frac{y}{16 \pi^2} \left[ - \frac{9}{5} g_1^2 \left(Y_H^2 + 2 Y_L Y_R \right) \right] \ , 
\label{betayg1SM.EQ}
\eeq
where we make use of $Y_H = Y_R - Y_L$ required for $U(1)_Y$ invariance of the Yukawa term in the Lagrangian, and $g'=\sqrt{3/5} g_1$. 
For the top, using $Y_L = 1/6$, $Y_R = 2/3$, we obtain the contribution shown in the last term of Eq.~(\ref{RGEbeta1ygaSM.EQ}). 

We compute the fermion loop contribution to the Higgs 4-point vertex that is proportional to $y$,
from which we can write the bare coupling as
$\lambda_0 \supset \lambda - N_c y^4/(8\pi^2) \left( 1/\epsilon + {\rm finite} \right)$. 
From this, we infer that this contribution leads to 
\beq
\beta_\lambda \supset \frac{1}{16\pi^2} \left[ - 2 N_c y^4 \right] \ .
\label{betalamy4SM.EQ}
\eeq

Writing the Higgs four-point vertex as $i \lambda_{\rm eff}$,
the contribution to its evolution, i.e. $\beta_\lambda$, due to the fermion loop in the $h$-leg is just four times $\sqrt{\Sigma_h}$.
Thus, for this contribution we have
$i\lambda_{\rm eff} \supset (-i\lambda) (i/p^2) (4\times i\Sigma_\phi/2)$,
and from Eq.~(\ref{Sigmah.EQ}), we have 
$\lambda(\mu) = \lambda(M) + N_c \lambda y^2/(4\pi^2) \ln(\mu/M) + ... $, 
where $\mu$ is the renormalization scale, and $M$ is a subtraction scale.
From this, and since $\beta_\lambda = d\lambda(\mu)/d\ln\mu$, we have
\beq
\beta_\lambda \supset \frac{1}{16\pi^2} \left[ 4 N_c y^2 \lambda \right] \ .
\label{betalamy2lam.EQ}
\eeq

We turn next to the $\beta$-functions of the gauge couplings $g_a = \{g_3,g_2,g_1\}$, focusing on the SM fermion contribution.
We recall the definition $\beta_a = g_a^3 b_a/(16\pi^2)$.
For $\beta_{g_3}$ we have the well-known result (see for example Ref.~\cite{Weinberg:1996kr})
\beq
\beta_{g_3} = \frac{g_3^3}{16\pi^2} \left( -\frac{11}{3} N_c + \frac{2}{3} n_3 \right) \ ,
\label{betag3SM.EQ}
\eeq
where the second term is due to fermions, with $n_3$ as the number of colored fermions in the fundamental representation of $SU(3)$.
Note that the top-quark is vector-like with respect to the $SU(3)$.
In the SM, at large $\mu$, we have $n_3 = 6$ for three generations of quarks, which implies $b_3 = -7$.
Similarly, for $\beta_{g_2}$ we have
\beq
\beta_{g_2} = \frac{g_2^3}{16\pi^2} \left( -\frac{11}{3} (2) + \half\times \frac{2}{3} n_2 + \frac{1}{6} \right) \ ,
\label{betag2SM.EQ}
\eeq
where we have taken $N=2$ for $SU(2)$ in the first term, the second term is the fermion $SU(2)$ doublet contribution with $n_2$ being the number of doublet fermions, and the last term is the Higgs doublet contribution.
Since the SM fermions are chiral under $SU(2)$ with only the $L$ chirality contributing, we include an extra factor of $1/2$ in the second term (since we are neglecting the effects of masses).
Thus in the SM, at large $\mu$, $n_2 = (3N_c+3)$ for the three generations of quark and lepton doublets, which yields $b_2 = -19/6$. 
Lastly, for $\beta_{g_1}$ we have
\beq
\beta_{g_1} = \frac{g_1^3}{16\pi^2} \left(\frac{2}{5} \sum_f Y_f^2 + \frac{1}{5} \sum_\phi Y_\phi^2 \right) \ ,
\label{betag1SM.EQ}
\eeq
where the sum in the first term is over all fermions $f$ with hypercharge $Y_f$,
and the sum in the second term is over all complex scalars $\phi$ with hypercharge $Y_\phi$.
We recall that we use $SU(5)$ normalization for $g_1$,
i.e. the SM hypercharge gauge coupling $g'$ is related to $g_1$ by $g_1 = \sqrt{5/3} g'$.
Thus in the SM, at large $\mu$, for three generations, $\sum_f Y_f^2 = 3\times (10/3) = 10$, and for one Higgs doublet containing two complex fields with $Y_H = 1/2$, $\sum_\phi Y_\phi^2 = 2 (1/2)^2 = 1/2$,
we get $b_1 = 41/10$. This agrees with, for example, Ref.~\cite{Mihaila:2012fm}. 

We complete our derivation of the SM fermion contributions to the 1-loop $\beta$-functions.
After adding the other contributions, the complete 1-loop $\beta$-functions are as given in Eqs.~(\ref{RGEbeta1lamSM.EQ})-(\ref{RGEbeta1gaSM.EQ}).

\section{VLF contributions to the RGE}
\label{betaFcnVLF.SEC}

Here we extend the $\beta$-functions derived in Sec.~\ref{SMRGE.SEC} and App.~\ref{betaFcnSM.SEC} to include VLF contributions, which we denote as $\beta^{VLF}_\kappa$. 

We first derive the $\beta_{g_a}^{VLF}$. 
The $\beta_{g_3}^{VLF}$ is got easily from Eq.~(\ref{betag3SM.EQ}). Since the SM quark is vector-like with respect to $SU(3)$, we have an identical contribution for a VLQ,
and we obtain the result shown in Eq.~(\ref{betag3VLF.EQ}). 
For obtaining $\beta_{g_2}^{VLF}$, we note that this is similar to the $SU(3)$ contribution owing to the fact that for a VLF $SU(2)$ is also vector-like just as the SMF was for $SU(3)$.
Thus taking twice the second term in Eq.~(\ref{betag2SM.EQ}) will give us $\beta_{g_2}^{VLF}$ as given in Eq.~(\ref{betag2VLF.EQ}). 
Since for a VLF both $L$ and $R$ chiralities contribute, we take twice the first term in Eq.~(\ref{betag1SM.EQ}) to obtain $\beta_{g_1}^{VLF}$ as given in Eq.~(\ref{betag1VLF.EQ}); 
the $2n_2$ is just the number of fermions in $n_2$ doublets having hypercharge $Y_\chi$. 

Next, we derive the contributions present only when a full family is added, i.e. when the $\tilde y$ operator of Eq.~(\ref{LagrVLF.EQ}) can be written down. 

Let us recall that the SM top (and bottom) sector have the following Feynman rules for the couplings with the Higgs-doublet fields $\{h,\phi^{1,2,3} \}$
(all vertices have an overall $(-i y_t/\sqrt{2})$) written in terms of the Dirac spinors $t=(t_L\ t_R)^T$ and $b=(b_L\ b_R)^T$:
\begin{itemize}
\item[] $\{h,\phi^3 \} t \bar{t}$ : $\{1,-i\gamma^5\}$ and $\{\phi^1, \phi^2\} b \bar{t}$ : $\{-P_L, -iP_L\}$.
\end{itemize}  
Now, when a full VLF family is present, we have the $SU(2)$ doublet VLF $\chi=(\chi_1\ \chi_2)^T$ and a singlet $\xi$.
We can assemble the following Dirac spinors:  $\psi_1 = ({\chi_1}_L\ \xi_R)^T$, $\psi_2 = (\xi_L \ {\chi_1}_R)^T$, $\xi=(\xi_L\ \xi_R)^T$ and $\chi_2 = ({\xi_2}_L \ {\xi_2}_R)^T$.
Using these, we can write the Feynman rules with the Higgs-doublet fields $\{h,\phi^{1,2,3} \}$ as follows (all vertices have an overall $(-i\tilde{y}/\sqrt{2})$):
\begin{itemize}
\item[] $\{h,\phi^3 \} \psi_1 \bar\psi_1$ : $\{1,-i\gamma^5\}$ and  $\{\phi^1, \phi^2\} \chi_2 \bar\psi_1$ : $\{-P_L, -iP_L\}$, 
\item[] $\{h,\phi^3 \} \psi_2 \bar\psi_2$ : $\{1,i\gamma^5\}$ and $\{\phi^1, \phi^2\} \chi_2 \bar\psi_2$ : $\{-P_R, -iP_R\}$. 
\end{itemize}
We write it this way to bring forth the analogy between the SMF and VLF, with the realization that for the VLF we have two Dirac sets that each have a similarity with the SM couplings.
The first Dirac fermion $\psi_1$ has identical couplings, while the second Dirac fermion $\psi_2$ has couplings that is similar but not identical,
with a change $i\to -i$ in the $\phi^3$ couplings, and $P_L \to P_R$ in the $\phi^{1,2}$ couplings. 
We observe that all the diagrams that contribute to the $\beta$-functions are immune to both of these changes, and therefore each of them give the same SM contribution as for the $t$ quark.
Furthermore, the Goldstone bosons with each Dirac fermion contributes zero to the $\beta$-function as in the SM. 
Thus, to obtain the VLF contributions to $\beta_\lambda^{VLF}$, we just multiply the SM contribution after combining Eqs.~(\ref{betalamy4SM.EQ})~and~(\ref{betalamy2lam.EQ}) by a factor of 2
and obtain Eq.~(\ref{betalamVLF.EQ}). 
Next, the VLF contribution to $\beta_{y_t}$ is due to only the wavefunction renormalization contribution to $h$, i.e. $\Sigma_h$,
and this contribution can be got from the second term in Eq.~(\ref{betaySM.EQ}), but multiplied by 2 since there are two VLF Dirac sets as argued above,
and changing the coupling to $y_t \tilde{y}^2$ instead of $y_t^3$, which then gives us the $\beta_{y_t}^{VLF}$ in Eq.~(\ref{betaytVLF.EQ}). 
Next, consider the evolution of either the $h\psi_1 \bar\psi_1$ coupling, or the $h\psi_2 \bar\psi_2$ coupling, either of which is $\tilde y$.
The VLF contribution to $\beta_{\tilde y}$ is due to these three contributions:
(i) the vertex contribution proportional to $3\tilde{y}^3$ as in the first term in Eq.~(\ref{betaySM.EQ}), 
(ii) the VLF contributions in $\Sigma_h$ which yields twice the second term in Eq.~(\ref{betaySM.EQ}) proportional to $2\times 2N'_c\tilde{y}^3$ since each of the $\psi_1$ and $\psi_2$ contribute as in the SM,
and (iii) the top-quark contribution in $\Sigma_h$ which yields $2N_c y_t^2 \tilde{y}$ as in the second term in Eq.~(\ref{betaySM.EQ}).
Adding these three contributions then gives the first part of $\beta_{\tilde y}$ in Eq.~(\ref{betatwyVLF.EQ}).
We write the $\tilde{y} g_a^2$ contributions to $\beta_{\tilde{y}}$ following Eqs.~(\ref{betayg3ng2SM.EQ})~and~(\ref{betayg1SM.EQ}),
which gives the last part in Eq.~(\ref{betatwyVLF.EQ}). 
We thus complete the derivation of the VLF contributions to the $\beta$-functions given in Eqs.~(\ref{betag3VLF.EQ})-(\ref{betatwyVLF.EQ}). 



\end{document}